\documentclass[12pt,twocolumn]{aastex62}

\usepackage{hyperref}
\usepackage{amsmath,mathtools,mathrsfs,amssymb}
\usepackage{graphicx}
\usepackage{bm}
\usepackage[percent]{overpic}

\defcitealias{kadowaki_etal_2020}{KGM20}

\graphicspath{{./}{figures/}}

\received{September, 2020}
\submitjournal{ApJ}

\shorttitle{Particle acceleration by magnetic reconnection in relativistic jets}
\shortauthors{Medina-Torrejon et al.}

\usepackage{multirow}

\begin{document}
\title{Particle acceleration by relativistic magnetic reconnection driven by kink instability turbulence in Poynting flux dominated jets}

\correspondingauthor{Elisabete M. de Gouveia Dal Pino
}
\email{dalpino@iag.usp.br}

\author[0000-0003-4666-1843]{Tania E. Medina-Torrej\'{o}n}
\affiliation{Universidade de S\~{a}o Paulo, Instituto de F\'{i}sica, S\~{a}o Paulo, Brasil}
             
\author[0000-0001-8058-4752]{Elisabete M. de Gouveia Dal Pino}
\affiliation{Universidade de S\~{a}o Paulo, Instituto de Astronomia, Geof\'{i}sica e Ci\^{e}ncias Atmosf\'{e}ricas, Departamento de Astronomia, 1226 Mat\~{a}o Street, S\~{a}o Paulo, 05508-090, Brasil}

\author[0000-0002-6908-5634]{Luis H.S. Kadowaki}
\affiliation{Universidade de S\~{a}o Paulo, Instituto de Astronomia, Geof\'{i}sica e Ci\^{e}ncias Atmosf\'{e}ricas, Departamento de Astronomia, 1226 Mat\~{a}o Street, S\~{a}o Paulo, 05508-090, Brasil}

\author[0000-0002-0176-9909]{Grzegorz Kowal}
\affiliation{Escola de Artes, Ci\^encias e Humanidades - Universidade de S\~ao Paulo,
Av. Arlindo B\'ettio, 1000 -- Vila Guaraciaba, CEP: 03828-000, São Paulo - SP, Brazil}

\author[0000-0002-7782-5719]{Chandra B. Singh}
\affiliation{South-Western Institute for Astronomy Research, Yunnan University, \\ University Town, Chenggong District, Kunming 650500, People's Republic of China}

\author[0000-0002-8131-6730]{Yosuke Mizuno}
\affiliation{Tsung-Dao Lee Institute and School of Physics and Astronomy, Shanghai Jiao Tong University, 200240, People's Republic of China}
\affiliation{Institut f\"{u}r Theoretische Physik, Goethe Universit\"{a}t, 60438 Frankfurt am Main, Germany}




\begin{abstract}

Particle acceleration in magnetized relativistic jets still puzzles theorists. In this work we investigate the acceleration of particles injected in a three-dimensional relativistic magnetohydrodynamical jet subject to current driven kink instability (CDKI). We find that, once turbulence driven by CDKI fully develops, achieving nearly stationary state, the amplitude of excited wiggles along the jet spine attains maximum growth, causing disruption of the magnetic field lines and the formation of several sites of fast reconnection. Low energy protons injected in the jet at this state experience exponential acceleration, mostly in  directions parallel to the local magnetic field, up to  maximum energies $E\sim 10 ^{16}$  eV, for   $B \sim 0.1$ G, and $E\sim10 ^{18}$   eV, for $B \sim 10$ G. The Larmor radius of the particles attaining these energies corresponds to the size of the acceleration region ($\sim$ diameter of the perturbed jet). There is a clear association of the accelerated particles with regions of fast reconnection. 
In the early non-linear growth stage of the CDKI, when there are  no sites of fast reconnection yet, injected particles with  initial much larger energy are accelerated by magnetic curvature drift.  We have also obtained the acceleration time due to reconnection with a dependence on the particles energy, $t_A \propto E^{0.1}$. The energy spectrum of the accelerated particles develops a power-law index $p \sim$ -1.2, in the beginning, in agreement with earlier works. Our results provide multi-dimensional framework for exploring this process in real systems and explain their emission patterns, specially at very high energies and associated neutrino emission recently detected in some blazars.

\end{abstract}



\section{Introduction} \label{sec:intro}

Relativistic, collimated  jets are ubiquitous in extreme astrophysical sources  like  microquasars (BH XRBs),  active galactic nuclei (AGNs), and  gamma-ray bursts (GRBs). An important property common to all these classes of objects, is the required presence of magnetic fields to allow the formation of these jets
\citep[e.g.,][]{blandford1977, blandford1982}.  Observed 
polarized non-thermal emission in all wavelengths also evidences that these jets are highly magnetized, specially near the launching region at the source
 \citep[e.g.,][]{laurent_etal_11, doeleman_etal_12, marti-vidal_etal_15}. In particular, it has been reported  evidences of a helical magnetic field feature in  the  M87 jet 
 \citep[][]{harris_etal_03}.
General relativistic magnetohydrodynamic (GRMHD) simulations with accretion disks around the spinning black hole of these sources  are compatible with the observations and the  proposed mechanisms for jet formation. 
They reveal the formation of a Poynting flux dominated jet spine with large Lorentz factor, surrounded by a mildly relativistic (matter-dominated) sheath, and a sub-relativistic wind  \citep[e.g.,][]{mckinney2006, hardee_etal_07, abramowicz_fragile_13, yuan_narayan_14}. 
This spine-sheath structure has indeed been inferred in VLBI observations from nearby FRI and FRII radiogalaxies like M87 \citep{kovalev_etal_07}, 3C84 
 \citep{nagai_14}, and Cyg A \citep{boccardi_etal_15}, as well as  in blazars like Mrk501 \citep{giroletti_etal_04} and 3C273 \citep{lobanov_zensus_01}.

Another  important constraint from  the observations is the fact that at distances large enough from the sources
these jets should  become kinetically dominated and as such,  they should convert most of the magnetic energy into kinetic.
A mechanism that could favor such conversion in jets is magnetic reconnection \citep[e.g.,][and references there in]{giannios_10,dalpino_kowal_15,dalpino_etal_2018,werner_etal_2018}. 
Lately, this process has been largely investigated in the framework of compact sources
 like pulsar nebulae \citep[e.g.,][]{lyubarsky_etal_2001,clausen-brown_2012,cerutti_etal_2014,sironi_spitkovsky_2014}, gamma-ray bursts (GRBs) \citep[e.g.,][]{drenkhahn_2002,giannios_spruit_2007, zhang_yan_11}, as well as  
jets and accretion flows around black holes \citep[e.g.,][]{dalpino_lazarian_2005,giannios_etal_2009,dalpino_etal_2010a,dalpino_etal_2010b,
giannios_10,mckinney_2012,kadowaki_etal_15,
singh_etal_15,khiali_etal_15,singh_etal_16,sironi_etal_2015,dalpino_etal_2018,
kadowaki_etal_18a,kadowaki_etal_2018b,
rodriguezramires_etal_18,christie_etal_19,fowler_etal_2019, giannios2019, nishikawa_etal_2020, nathanail_etal_20, davelaar_etal_2020} 

Among all classes of relativistic jets, reconnection can be particularly important in blazars,  which are a sub-class of AGNs with  jets making very small angles with our line of sight.  
This geometry is particularly favorable since relativistic effects, like Doppler boosting with apparent enhancement of the observed emission, are maximized. 
Blazars produce  usually highly variable, non-thermal emission in all wavelengths, which  is generally attributed  to relativistic particles (protons and electrons) accelerated stochastically in recollimation shocks along the jet and in their head \citep[e.g.]{mizuno_etal_15,hovatta_lindfors19,matthews_etal_2020}. However,  there is increasing evidence   that shock acceleration may not be always  as  efficient in the magnetically dominated regions of these jets, particularly to explain the very high energy emission \citep{sironi_etal_2013,dalpino_kowal_15,bell_etal_2018}.  
This may be the case, for instance, of the  blazars  PKS 2155-304 \citep{aharonian_etal_07} and MRK501 \citep{albert_etal_07}
(see also 3C 279 \citep{ackermann_etal_2016} and 3C 54.3 \citep{britto_elal_2016}). They produce very short duration
gamma-ray flares, of minutes,  at  the TeV band, which imply extremely compact acceleration/emission regions ($< R_S/c$, where $R_S$ is  the Schwartzschield radius) with
Lorentz factors much larger than the typical jet bulk values  in these sources (which are $\Gamma \simeq$ 5--10). This is
the only way to prevent the re-absorption of the  gamma-rays  within the source due to electron-positron pair creation  
\citep[e.g.,][]{begelman_etal_08}. The only  mechanism  that seems to be able to drive both, the high variability and
compactness of the TeV emission is fast magnetic reconnection involving misaligned current sheets inside the jet
\citep{giannios_etal_2009,giannios_2013,kushwaha_etal_17}.  A similar mechanism has been also invoked to explain the 
prompt  emission in gamma-ray-bursts \citep[e.g.][]{giannios_2008,zhang_yan_11}. Moreover, the  recent simultaneous detection of gamma-rays and  high-energy neutrinos from the blazar TXS 0506 +056 \citep{aartsen_etal_2018}, have evidenced the presence of high-energy protons interacting with ambient photons, producing  pions and a subsequent decay in gamma-rays and neutrinos.  It has been argued that if these protons are produced in magnetically dominated regions of the jet near the core, then they are probably accelerated by fast magnetic reconnection \citep[e.g.,][]{dalpino_etal_2018}.


Magnetic reconnection is produced from the merging of two magnetic fluxes of opposite polarity. This causes partial breaking and rearrangement of the field topology, and it is $fast$ when its rate $V_R$ is a substantial fraction of the local Alfvén speed, $V_A$, \citep[e.g.,][]{zweibel_yamada_2009, lazarian12, lazarian20}. The ubiquitous turbulence in astrophysical flows is acknowledged as one of the main driving mechanisms of fast reconnection due to the wandering of the magnetic field lines that allows for many simultaneous events of reconnection and the broadening of the outflow region removing the reconnected flux more efficiently. These two factors result in the reconnection rate to be independent of the small microscopic magnetic resistivity \citep{lazarian_vishiniac_99,eyink_etal_2011,kowal_etal_09,takamoto_etal_15, santoslima_etal_2010, santos-lima_etal20}. 

The break and rearrangement of the reconnected lines causes the conversion of magnetic energy  into kinetic energy and particle acceleration.
\citet{dalpino_lazarian_2005}  realized that particles could be accelerated in a fast magnetic reconnection site in a similar way as in diffusive shock acceleration \citep{blandford_eichler_1987,bell1978}. In other words,   particles bouncing  back and forth between  two converging magnetic fluxes of opposite polarity in a reconnection discontinuity (or current sheet),  gain energy  due to head-on collisions with  magnetic fluctuations at a rate $\Delta E /E \propto V_R /c$, which implies  a first order Fermi process. 
This has been  successfully tested numerically \citep[see e.g.][]{drake_etal_2006,kowal_etal_2011,kowal_etal_2012}.
\citet*{kowal_etal_2011},  in particular, have demonstrated  by means of two- and three-dimensional  (2D and 3D) MHD simulations with test particles, the equivalence between this process and that of particles being accelerated 
within two-dimensional merging magnetic islands (or plasmoids)  excited, e.g., by tearing mode instability.
It is important to remark that magnetic islands are actually the 2D  cross sections of 3D flux tubes and in real systems, reconnection is 3D and thus particle acceleration in reconnection sites \citep[e.g.][]{dalpino_kowal_15, kowal_etal_2011}.

The realization that  reconnection acceleration can be a fundamental  mechanism to explain observed non-thermal, highly variable emission, specially at very-high-energies (VHEs), in magnetically dominated sources, has motivated intensive study mainly through 2D and 3D  particle-in-cell (PIC) simulations of  current sheets in a slab geometry  in collisionless plasmas 
\citep[e.g.][]{drake_etal_2006,zenitani_H_2001,zenitani_H_2007,zenitani_H_2008,lyubarsky_etal_2008,drake_etal_2010,clausen-brown_2012,cerutti_etal_2012,li_etal_2015,lyutikov_etal_2017,werner_etal_2018,werner_etal_2019,sironi_spitkovsky_2014,guo_etal_2015,guo_etal_2016}.
These studies have probed the kinetic scales of the process. To assess the stochastic Fermi mechanism at the macroscopic scales of collisional flows present in most astrophysical systems, the tracking of test particle distributions in such flows is also a very useful and complementary tool to help in the understanding of the overall process across  the scales. Such studies have been  performed combining  2D and 3D MHD simulations  with the injection of thousands of test particles in the reconnection domain \citep{kowal_etal_2011,kowal_etal_2012,delvalle_etal_16, beresnyak_etal_2016, dalpino_etal_2018,dalpino_etal_2020}. 
In particular, \citet*{kowal_etal_2012}  have distinguished the first order Fermi process that occurs inside large scale current sheets with embedded  turbulence  driving fast reconnection, from a second order Fermi process occurring in pure turbulent environments \citep*[see also][]{brunetti_lazarian2011, brunetti_vazza_2020}.\footnote{It is worth to note that in a more recent PIC study, \citet{comisso18, comisso19} also considered a pure turbulent, magnetically-dominated system, but with no guide field and found that particles are initially exponentially accelerated in small scale reconnecting current sheets and then undergo further acceleration by stochastic interactions with the turbulent fluctuations at a slower rate. This is, in principle, similar to the process described above, specially  in \citet[][see their Figure 1, bottom panel]{kowal_etal_2012}. 
But in an MHD environment, the presence of turbulence naturally leads to fast reconnection with the formation of 3D current sheets in the entire turbulent domain 
\citep[Lazarian-Vishniac theory,][]{lazarian_vishiniac_99} and to stochastic particle acceleration in these sites, as described previously. A key difference between \cite{comisso18, comisso19} simulations  and those of \cite{kowal_etal_2012} (and the ones in the present work)  is that in the latter there is the underlying large scale magnetic field that favors the confinement of the particles in the regions where the current sheets are formed and thus the exponential acceleration can be sustained longer and not only during the initial stages.}
Both,  PIC and test particle+MHD approaches  have probed  the efficiency of the stochastic reconnection acceleration, particularly  in 3D dimensions,  with an exponential growth of the particle energy in time, implying a power-law  energy dependence of the acceleration rate, and the production of  an extended, non-thermal distribution of relativistic particles with a power-law tail \citep*[e.g.,][]{delvalle_etal_16}.

The results above  are applicable  to  magnetized astrophysical flows in general and specially to relativistic jets in regions near the source where they are possibly magnetically dominated.  The investigation of magnetic reconnection acceleration in these objects is the aim of the present work.

The presence of instabilities in the  jet can drive turbulence and thus fast magnetic  reconnection 
\citep[e.g.,][]{spruit_etal_2001, barniol_eltal_2017,dalpino_etal_2018, gill_etal_2018}.
In particular, jets with helical magnetic field structure, can be subject  to current-driven kink (CDK)  instability \citep[e.g.,][]{begelman_1998,giannios_spruit_2006,mizuno_etal_09,mizuno_etal_11,mizuno_etal_12,mizuno_etal_14,das_begelman2019}, and a  
number of recent numerical  works 
have revealed that this instability  can operate  in the jet spine without  disrupting  the entire jet structure, 
convertig magnetic into kinetic energy, and driving  reconnection \citep{porth_etal_2015,singh_etal_16, bromberg_etal_2016, Tchekhovskoy_etal_2016, striani_etal_2016,bromberg_etal_2019,davelaar_etal_2020}. In particular, in their  3D relativistic MHD (RMHD) simulations of  Poynting flux  dominated, rotating jets with helical fields,  \citet*{singh_etal_16}  verified that the CDK-induced-turbulence triggers  the formation of current-sheets with  fast reconnection rates $\sim  0.05V_A$. 

In a companion work to the present one
\citep[][hearafter \citetalias{kadowaki_etal_2020}]{kadowaki_etal_2020},
we have  expanded  upon the previous studies above,
applying a magnetic reconnection search-algorithm, developed in \citet*{kadowaki_etal_18a}
\citep[see also ][]{zhdankin_etal_13}, to a simulated 3D relativistic MHD (RMHD) 
jet with helical field \citep*[as in][]{singh_etal_16}. With this study, we have been able to identify several sites of reconnection and
obtain robust values of the reconnection rates and the magnetic power of every reconnection events inside the jet, in different snapshots, as well as the topological
characteristics of each reconnection region \citep*[see also][]{kadowaki_etal_2018b}.  
In the present work, in  order to obtain a fully understanding on how this magnetic energy released by the CDK instabiliy can be  channeled into energetic nonthermal particles in the fast reconnection regions, we present a study  of $in$  $situ$  particle  acceleration by injecting  hundreds to thousands of test particles in the same 3D  RMHD jet model we employed in 
\citetalias{kadowaki_etal_2020}.

We should remark that preliminary results of this study have been presented in \citet{dalpino_etal_2018,dalpino_etal_2020}. Other recent studies based on PIC simulations of relativistic jets have also explored particle acceleration in relativistic jets \citep[e.g.,][]{alves_etal_2018,nishikawa_etal_2020,davelaar_etal_2020}.  In \citet{alves_etal_2018}, the authors investigated particle acceleration in the early non-linear stage of the CDK instability, before the development of turbulence,  and verified that particles are accelerated  by  magnetic curvature drift. 
We have also identified this process in  our simulations in the early stages of the CDK growth, but it is soon replaced by reconnection acceleration when turbulence settles in the system (see below). \citet{davelaar_etal_2020}, on the other hand, in their simulations with high magnetization, detected only the acceleration by magnetic reconnection.
\citet{nishikawa_etal_2020} 
have explored particle acceleration in a 3D PIC relativistic  jet  interacting with the environment, 
considering different driving mechanisms of turbulence inside the jet, such as Weibel, kinetic Kelvin-Helmholtz, and  mushroom instabilities,
and they also conclude that magnetic reconnection is the dominant acceleration process.

In the next sections we organize the paper as follows. In section \ref{sec:numethod}, we describe the numerical method and  our setup both for  the RMHD jet simulations and the test particle method; in Section \ref{sec:results} we present our numerical results of the development of reconnection and the injection of test particles into the entire jet domain subject to the CDK instability, and obtain the  properties of the acceleration such as the particles energy growth with time, the  acceleration rates, and the particles spectrum, as well as the connections of the accelerated particles with the reconnection regions (identified in \citetalias{kadowaki_etal_2020}). Finally, in Section \ref{sec:discut} we discuss our findings and draw our conclusions. 

~\\

\section{Numerical Method and Setup} \label{sec:numethod}

Following \citet*{kowal_etal_2011,kowal_etal_2012} and \citet*{delvalle_etal_16} \citep[see also][]{dalpino_kowal_15}, we inject test particles (100 - 10,000 protons) into  frozen-in-time 3D MHD domains, in order to test particle acceleration by magnetic reconnection. However, instead of taking a large scale single current-sheet with embedded controlled weak stochastic turbulence, as in these former works, we consider here $in-situ$ particle acceleration  in a relativistic 3D MHD jet where turbulence and fast magnetic reconnection are naturally induced by current-driven-kink (CDK) instability \citet*{singh_etal_16}. As in \citet{kowal_etal_2011,kowal_etal_2012}, we can 
neglect the macroscopic MHD dynamical time variations (see further justification for this in section \ref{sec:discut}) and,
once the instability reaches saturation and turbulence is fully developed with the appearance of several sites of fast reconnection in the jet,  we can inject test particles into a  snapshot to  follow their acceleration. In fact, we show that particles  undergo acceleration mostly in the reconnected regions once this condition is fulfilled.

\subsection{3D MHD jet simulation setup} \label{sec:MHDsetup}

As in \cite{singh_etal_16}, 
we  perform RMHD simulations using  the  three-dimensional GRMHD code \texttt{RAISHIN} \citep{mizuno_etal_2006,mizuno_etal_11,mizuno_etal_14}. A pre-existing jet is established across the computational domain (tower jet). We use a similar parametrization as in model $D2$ of    \cite{singh_etal_16} which has a rotating jet with an initial force-free helical magnetic field and decreasing radial density profile.

Table \ref{tablejet} gives the initial conditions for the jet model. The computational domain is $6L \times 6L \times 6L$ in a Cartesian (x, y, z) coordinate system, where $L$ is the length scale unit of the computation domain. We consider two different grid resolutions in the three directions:  $\Delta L=L/40$, corresponding to 240 cells in each direction (model j240 in Table \ref{tablejet}), and $\Delta L=L/80$, corresponding to 480 cells in each direction (model j480 in Table \ref{tablejet}). 
Distinctly from \cite{singh_etal_16}, we impose outflow boundaries only in the transverse directions $x$ and $y$, and adopt
 periodic boundaries in the $z$ direction in a similar way to the setup in \cite{mizuno_etal_12}.

The code unit (c.u.) for the velocity is the speed of light $c$, the magnetic field is in units of $\sqrt{4 \pi \rho_0 c^2}$, the density is in units of $\rho_0$ 
(being $\rho_0 = $ 1 in the code), 
the pressure is in units of $\rho_0 c^2$ and the time is in units of $L/c$. 

The jet   initial angular velocity is given by:
\begin{equation}
\Omega=\left\{\begin{matrix}
\Omega_0 & \mathrm{if}\,\,\, R\leq R_0 \\ 
\Omega_0(R_0/R) & \mathrm{if}\,\, R>R_0\, ,
\end{matrix}\right.
\label{ohm0}
\end{equation}
where $R_0$ is the radius of the jet core. In the simulations $R_0=(1/4)L$, and $\Omega_0=2.0$ $L/c$. 

The initial helical magnetic field has poloidal and toroidal  components, given respectively by:
\begin{equation}
B_z=\frac{B_0}{1+(R/R_0)^2}\, ,
\end{equation}
\begin{equation}
B_{\phi}=-\frac{B_0(R/R_0)[1+(\Omega R_0)^2]^{1/2}}{1+(R/R_0)^2}\, .
\end{equation}

The initial poloidal and toroidal components of the drift velocity are given by
%
\begin{equation}
v_z=-\frac{B_\phi B_z}{B^2}\Omega R \, ,
\end{equation}
\begin{equation}
v_\phi=\left(1-\frac{B_\phi^2}{B^2}\right )\Omega R\, .
\end{equation}

The initial density profile decreases with the radius according to  $\rho=\rho_1 \sqrt{B^2/B^2_0}$, where $\rho_1=0.8\rho_0$ and the magnetic field amplitude is $B_0=0.7$. 

The equation of state is given by $p=(\Gamma-1)\rho e$, where $\Gamma$ is the adiabatic index equal to $5/3$ and $e$ is the specific internal energy density. We assume an initial gas pressure decreasing radially, similar to equation \eqref{ohm0}:
\begin{equation}
p=\left\{\begin{matrix}
p_0 & \mathrm{if}\,\, R\leq R_p \\ 
p_0(R_p/R) & \mathrm{if}\,\, R>R_p\, ,
\end{matrix}\right.
\label{p0}
\end{equation}
with $R_p=(1/2)L$ and $p_0 = 0.02\, \rho_0 c^2$. These values correspond to an initial beta  parameter $\beta = p_0/(B^2_0/8\pi)=$ 0.08 and a magnetization parameter
$\sigma = B_0^2/ \gamma^2 \rho h =$ 0.6 (where $h$ is the specific enthalpy) at the jet axis.
The latter increases to maximum values around unity in more evolved times.
We should remark that, although it is generally believed that the jet is highly magnetized at the launching site,  the location of the region where effective acceleration and radiative dissipation occurs is not well understood yet. Nevertheless, it is reasonable to suppose that an efficient reconnection acceleration and dissipation should occur when the magnetization has decreased to values $\gtrsim 1$. This is, in fact, compatible with
recent proposal by 
\citet{giannios2019}.
We here explore a similar magnetization condition.

The initial profiles above are presented in Figure 1 in \citet{singh_etal_16}, for $\Omega_0 =$2   $L/c$. 

The code setup for spatial development of the CDK instability is the same as in \citet{mizuno_etal_12}. 
In order to drive the instability and induce turbulence, a precession perturbation is applied  by a radial velocity profile given by:
\begin{equation}
\frac{v_R}{c}=\frac{\delta v}{N}\exp{\left ( -\frac{R}{R_a} \right )}\sum_{n=1}^{N}\cos(m\theta)\sin\left ( \frac{\pi n z}{L_z} \right ) \, ,
\end{equation}
where the amplitude of the perturbation is $\delta v = 0.01 c$, $N=8$, the radial width $R_a = 0.25 L$, and we consider the mode $m=1$, in order to induce the CDK instability.

\begin{table*}
\centering
\caption{Parameters for the MHD simulations.}
\label{tablejet}
\begin{tabular}{c c ccccc}
\firsthline
\hline
Model	& Resolution	& $p_0 [\rho_0 c^2]$      & $B_0[\sqrt{4 \pi \rho_0 c^2}]$    & $\Omega_0 [c/L]$    & $R_0 [L]$ & $\sigma$\\
\hline
$j240$    & $240^3$   & 0.02 	& $0.7$     & $2.0$     &$0.25$  & 0.6\\ 

$j480$    & $480^3$	& 0.02	& $0.7$     & $2.0$     &$0.25$  & 0.6\\ 
\lasthline
\end{tabular}
\end{table*}

\begin{table*}
\centering
\caption{Parameters for the test particles.}
\label{tablep}
\begin{tabular}{r c r c c c c}
\firsthline
\hline
Test		& Jet snapshot	& N        & $B_0$        & x \& y boundaries   & Initial distribution  & Jet Resolution\\
\hline
$t25o$      & 25	& 1,000 	& $0.094$ 	& outflow	& Maxwellian    &$240^3$\\ 

$ut25o$   	& 25	& 1,000 	& $0.094$ 	& outflow	& Monoenergetic      &$240^3$\\ 

$t30o$   	& 30	& 1,000 	& $0.094$ 	& outflow	& Maxwellian    &$240^3$\\ 

$t40o$   	& 40	& 5,000 	& $0.094$  	& outflow	& Maxwellian    &$240^3$\\

$t44o$   	& 44	& 5,000 	& $0.094$  	& outflow	& Maxwellian    &$240^3$\\

$t46o$   	& 46	& 1,000 	& $0.094$  	& outflow	& Maxwellian    &$240^3$\\

$t50o$   	& 50	& 10,000 	& $0.094$ 	& outflow	& Maxwellian    &$240^3$\\ 

$t50p$   	& 50	& 1,000 	& $0.094$ 	& periodic	& Maxwellian    &$240^3$\\ 

$9t50p$       & 50	& 100   	& $9.4$ 	& periodic	& Maxwellian    &$240^3$\\

$480t50o$	    & 50	& 1,000 	& $0.094$ 	& outflow	& Maxwellian    &$480^3$\\
\lasthline
\end{tabular}
\end{table*}

\begin{table*}
\centering
\caption{Reconnection velocity values in units of the local Alfv\'en speed.}
\label{tablevrec}
\begin{tabular}{ccccc}
\firsthline
\hline
Jet snapshot	& $\langle V_{rec} \rangle$      & $max(V_{rec})$     & Total counts     & Counts $(V_{rec} \geq \langle V_{rec} \rangle$) \\
\hline
25	 	& 0 	    & 0     & 0     & 0\\ 

30	 	& 0.01  & 0.02    & 25     & 11 (44\%)\\ 

40   	& 0.03  & 0.06    & 130    & 55 (42\%)\\

44	 	& 0.03  & 0.1     & 161    & 58 (36\%)\\

46	 	& 0.04  & 0.2     & 121    & 39 (32\%)\\

50	 	& 0.05 	& 0.1     & 136    & 66 (49\%)\\ 
\lasthline
\end{tabular}
\end{table*}

~\\

\subsection{Setup for Test Particle Acceleration}\label{sec:setuptp}

We inject test particles (protons) into a snapshot of RMHD jet simulations and integrate their trajectories using the \texttt{GACCEL} code \citep{kowal_etal_2011, kowal_etal_2012, delvalle_etal_16}, which solves the relativistic equation of motion of a charged particle:
\begin{equation}
\frac{\mathrm{d} (\gamma m \mathbf{u})}{\mathrm{d} t}=q(\mathbf{E}+\mathbf{u}\times \mathbf{B})\, ,
\label{eqmov0}
\end{equation}
where $\gamma \equiv (1-u^2/c^2)^{-1}$ is the particle Lorentz factor, and $\mathbf{u}$, $m$ and $q$ are the particle velocity, mass and electric charge, respectively. The electric field is generated by the background flow of magnetized plasma and by magnetic resistivity effects, and can be obtained directly from the Ohm`s law equation:
\begin{equation}
\mathbf{E}=-\mathbf{v}\times \mathbf{B}+\eta \mathbf{J}\, ,
\label{Ohm0}
\end{equation}
where $\mathbf{v}$ is the plasma velocity obtained from RMHD simulations. We neglect here the Ohmic resistivity (second term on the RHS) 
in order to study the acceleration provided by the plasma magnetic fluctuations (first term on the RHS) \citet{kowal_etal_2011,kowal_etal_2012}.\footnote{We note that the resistive term can be important only in the initial phases of the particle acceleration in the reconnection regions, but it is soon dominated by the first term, once the particles Larmor radius becomes as large as the size of the magnetic fluctuations \citep[e.g.,][]{kowal_etal_2012}.}
Substituting eq.  \eqref{Ohm0}, the equation of motion \eqref{eqmov0} is written as
\begin{equation}
\frac{\mathrm{d} }{\mathrm{d} t}(\gamma m \mathbf{u})=q[(\mathbf{u}-\mathbf{v})\times \mathbf{B}]\, .
\label{eqmov1}
\end{equation}

The particle equation of motion \eqref{eqmov1} is integrated using the fourth-order Runge-Kutta method and the background plasma velocity $\mathbf{v}$ and magnetic field $\mathbf{B}$ at each step of the integration are obtained through linear interpolation of the values from the  discrete grid of the MHD simulation domain.

We integrate equation \eqref{eqmov1} for 100  to 10,000 protons 
with randomly chosen initial positions and directions of motion in the domain.\footnote{We have found that the results are very similar whether particles are injected in a specific region of the jet with large concentration of reconnection sites or in the entire jet.} 
In most of the simulated models, we assume an initial Maxwellian distribution for the particle velocities corresponding to a temperature of $10^{10}K$ (non-relativistic particles) and a mean kinetic energy of the order of 1 MeV ($\sim 10^{-3} m_p c^2$). Only in one of the models investigated, we have assumed an initial monoenergetic distribution, $ut25o$  (see Table \ref{tablep} and Section \ref{sec:t25}).

In order to check potential boundary effects, the trajectories of the particles are integrated considering two different boundary conditions, periodic in all directions (test particle model names  ending with "p" in Table \ref{tablep}), or periodic in the $z$ direction only with outflow boundaries in the $x$ \& $y$ directions (test particle model names ending with "o" in Table \ref{tablep}) as in the simulated MHD background. In the first case, the particles are re-injected in the domain whenever they cross any jet boundary, while in the second case, they are re-injected only  when they cross the periodic boundary of the jet (in $z$ direction). The results, as we will see in Section \ref{sec:bound}, are very similar in both cases, but the first implies much longer computing time. 

As we have described in \S\ref{sec:MHDsetup}, the background jet simulation is performed in code units (c.u.), but in the \texttt{GACCEL} code, we assume physical units. The adopted time unit is one hour, the velocity unit is the light speed $c$, and for the magnetic field we adopt two possible values that correspond to the initial magnetic field in the jet axis, that is $B_0 = 0.094$ and $9.4$ G.  These values come from converting the code unit for the magnetic field strength, $B_0=0.7$ c.u., or $B_0=0.7 \sqrt{4 \pi \rho_0 c^2}$, assuming $\rho_0= 1.0$ cm$^{-3}$ and 10$^{ 4}$,  
respectively.

The initial and boundary conditions, as well as the number of particles considered for each test particle simulation are presented in Table \ref{tablep}. The table also shows the chosen snapshots of the simulated jet in which the  particles have been injected:  $t =$ 25, 30, 40, 44, 46, and 50 $L/c$,
and the corresponding initial magnetic field at the jet axis ($B_0$). The simulated test particle models in the first column of the  table are named according to the jet snapshot where the test particles were injected and the type of boundary condition of the injection in x and y directions (either outflow or periodic). 

In the current study we do not include radiative losses, so that the test particles can gain (or lose energy) only through the interactions with the moving magnetized plasma and its fluctuations. 

We also note that the test particle simulations with \texttt{GACCEL} in the jet domain are extremely computer-time consuming and for this reason, we have also performed a few runs initially injecting 100 particles only.
In spite of the smaller number, the resulting particle acceleration rates and energy growth are similar to those obtained with 1,000 and 10,000 particles injected. This is due to the fact that we allow the particles to be re-injected into the domain when crossing the boundaries, which naturally increases the number of interactions and thus helps to improve the statistics of the events. The re-injection of the particles in z direction mimics the large extension of a real jet, since in our simulations we only consider a small portion of it.



\section{Results}\label{sec:results}

\begin{figure*}[h] 
\centering
   \includegraphics[scale=0.12]{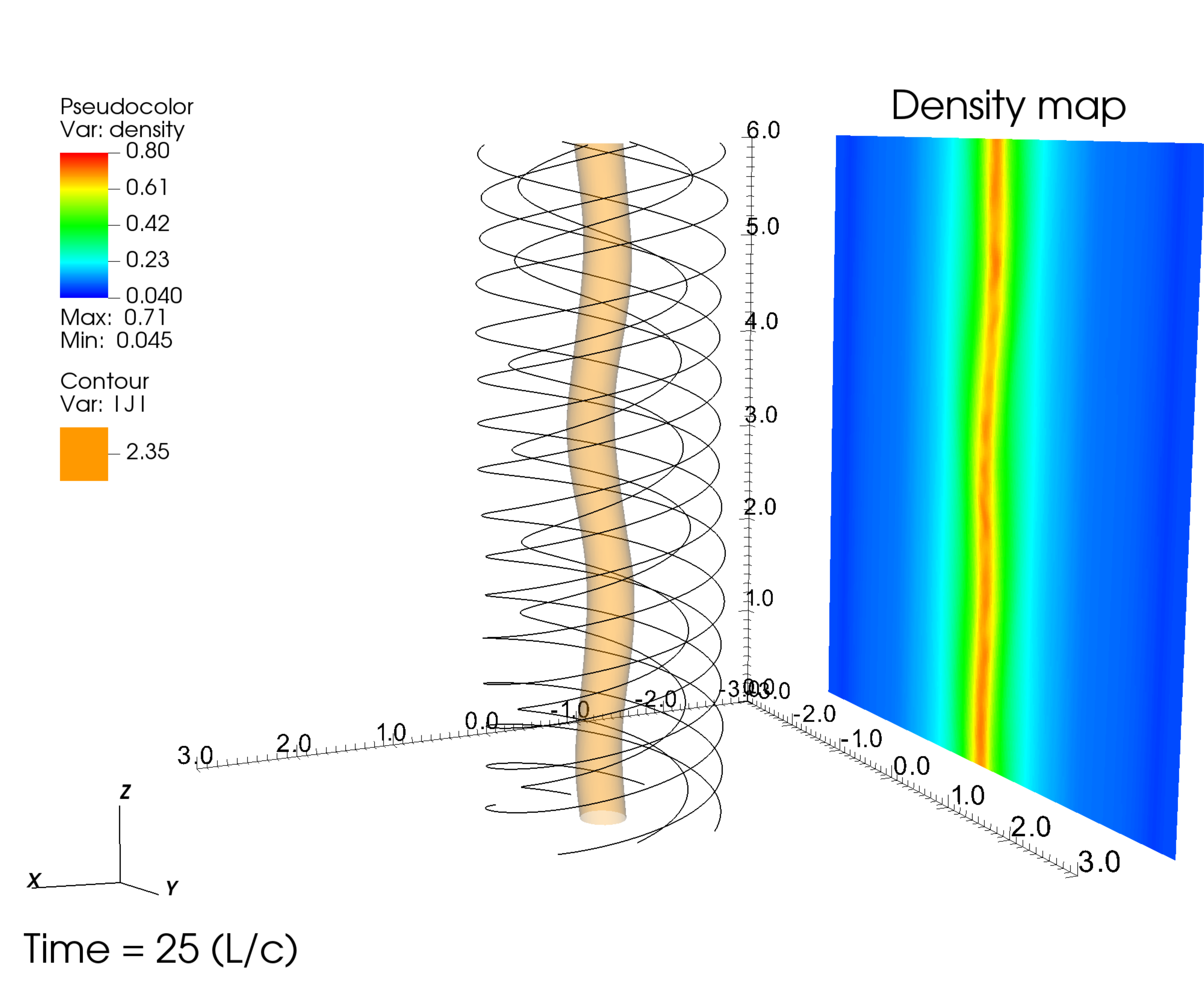}
   \includegraphics[scale=0.12]{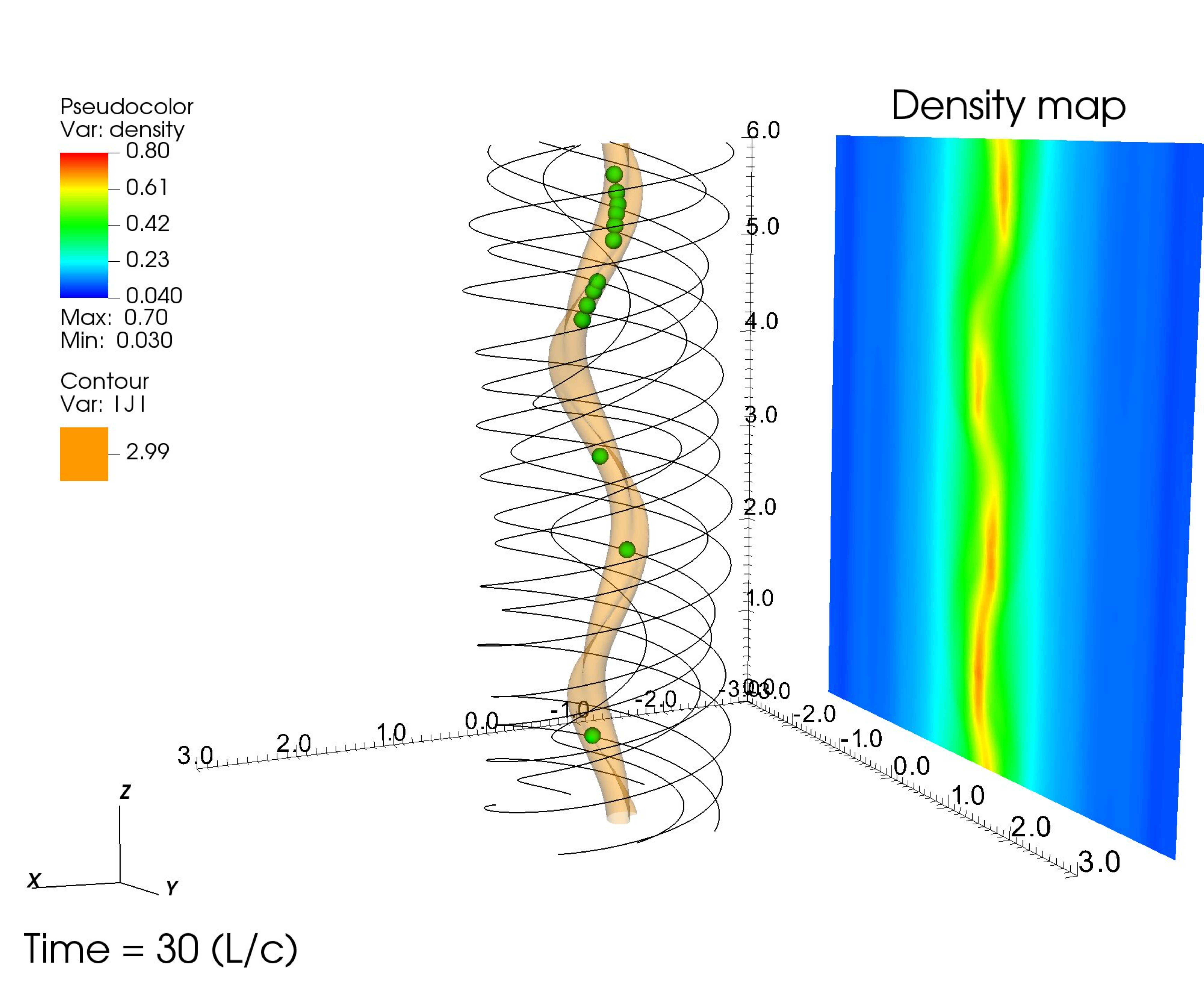}
   \includegraphics[scale=0.12]{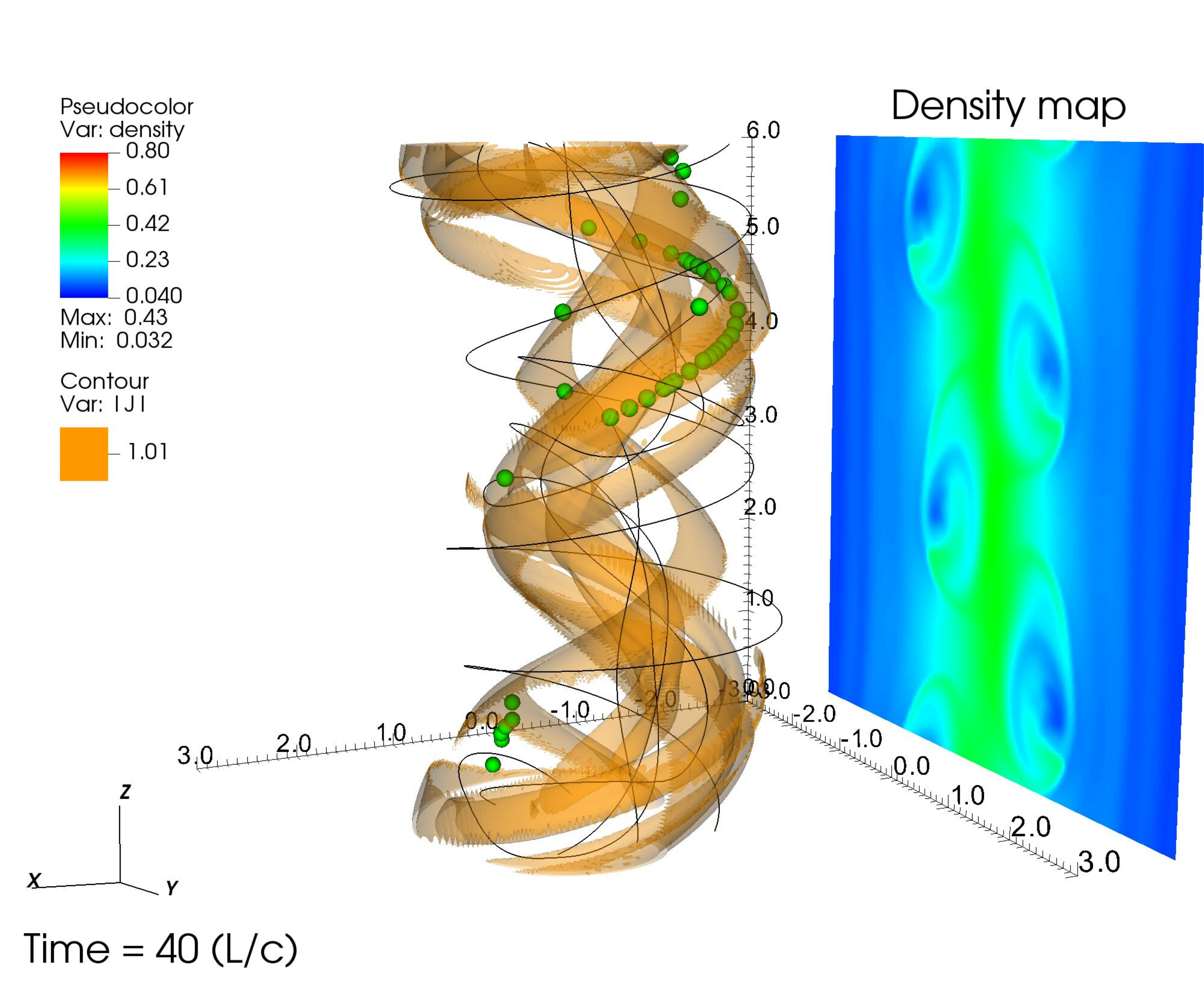}
   \includegraphics[scale=0.12]{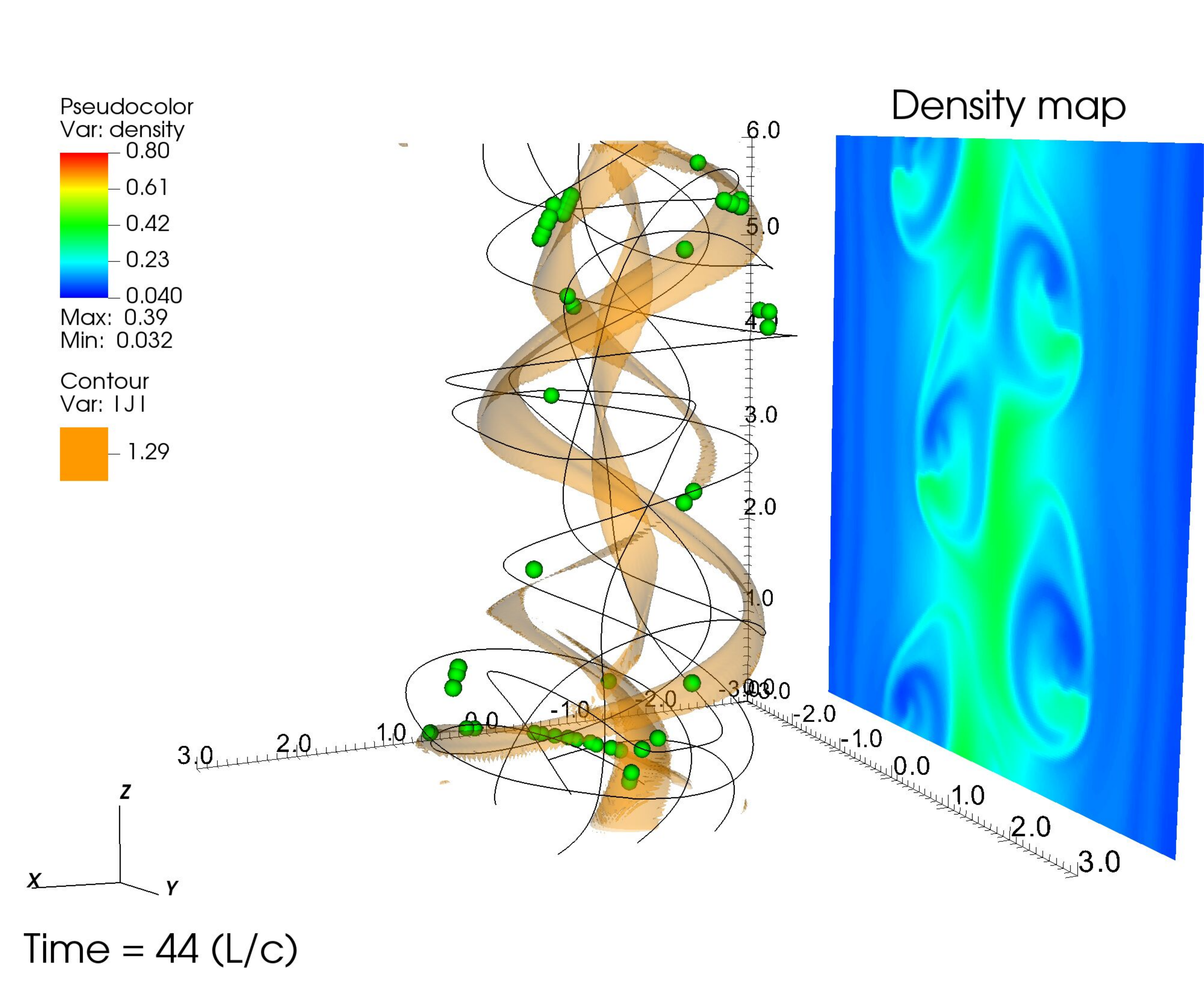}
   \includegraphics[scale=0.12]{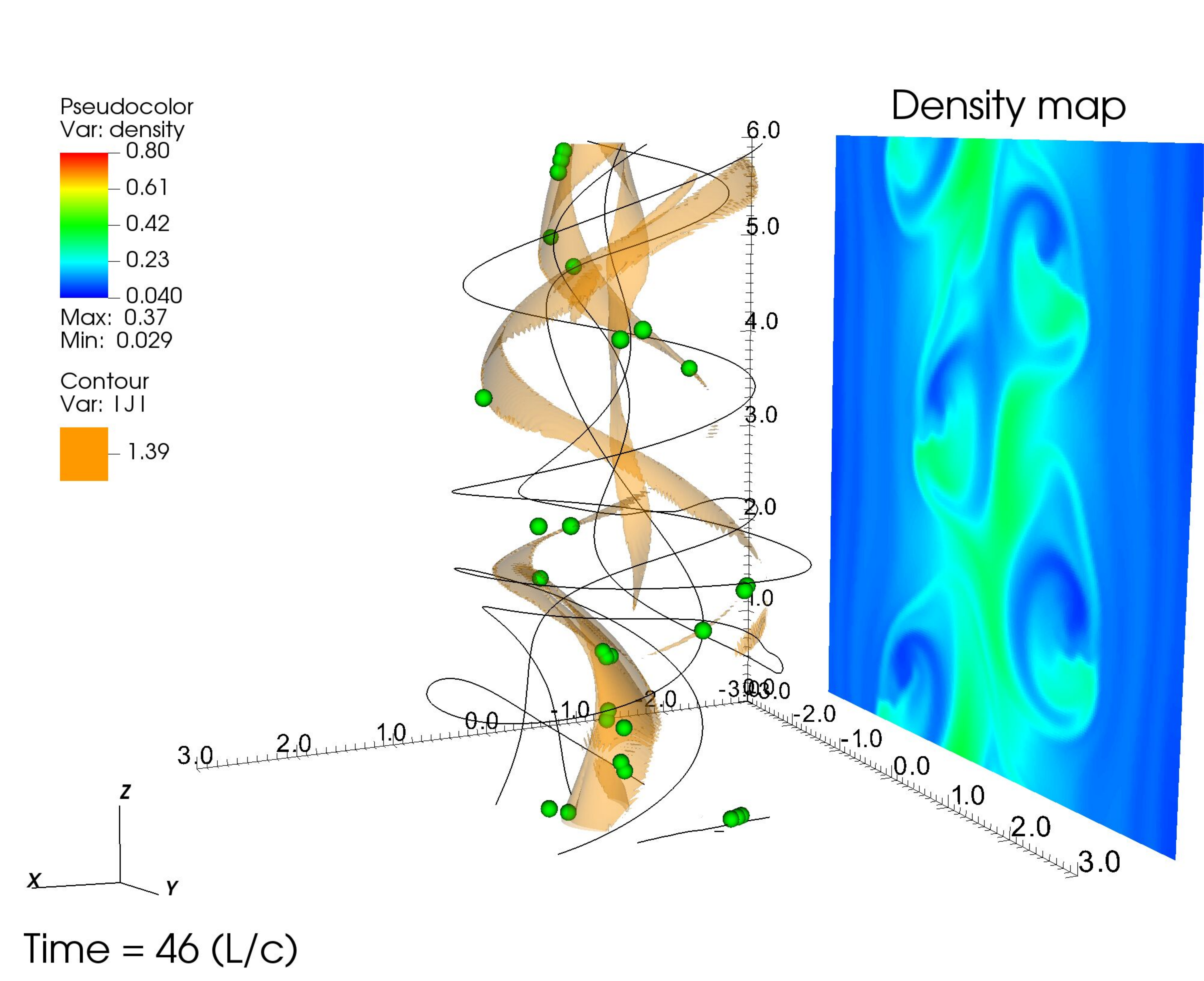}
   \includegraphics[scale=0.12]{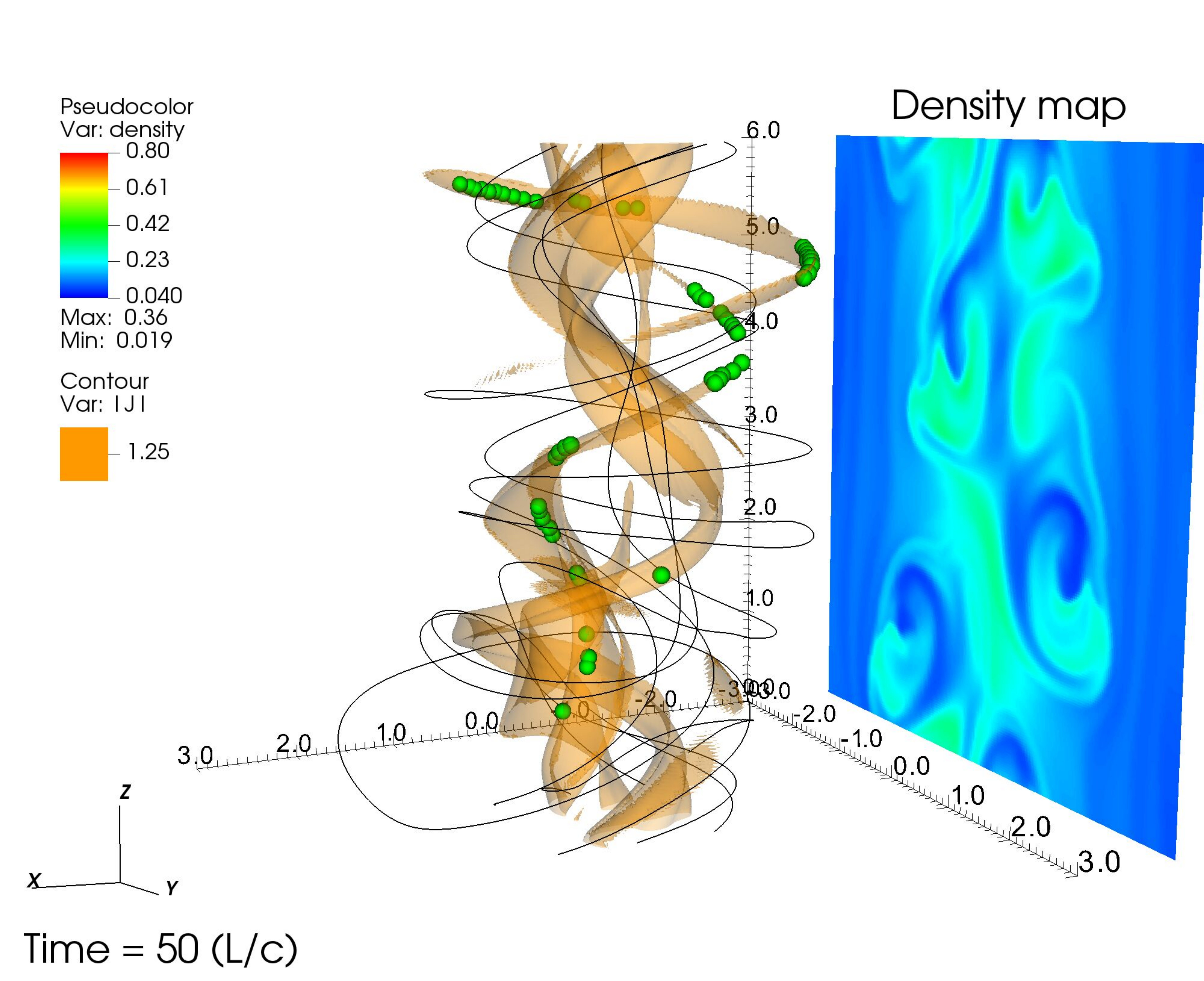}
\caption{
Time evolution of the tower jet at $t=25$, $30$, $40$, $44$, $46$ and $50$ $L/c$ (from top left to bottom right). 
The diagrams depict isosurfaces of the current density intensity at half maximum $|J|$ (orange color), the solid black lines correspond to the magnetic field lines, and the green circles correspond to the position of fast magnetic reconnection regions (with velocities larger than or equal to the average reconnection speed) identified with an algorithm described in \citet{kadowaki_etal_18a} and \citetalias{kadowaki_etal_2020}.  
At zy plane, it is shown the density map of the central slice of the jet (at $x=0$). The time $t$ is in units of $L/c$, 
the density in units of 
$\rho{_o}$, 
and the current density is also in code units.
}
\label{jet_points}
\end{figure*}

\begin{figure} 
\centering
    \includegraphics[scale=0.5]{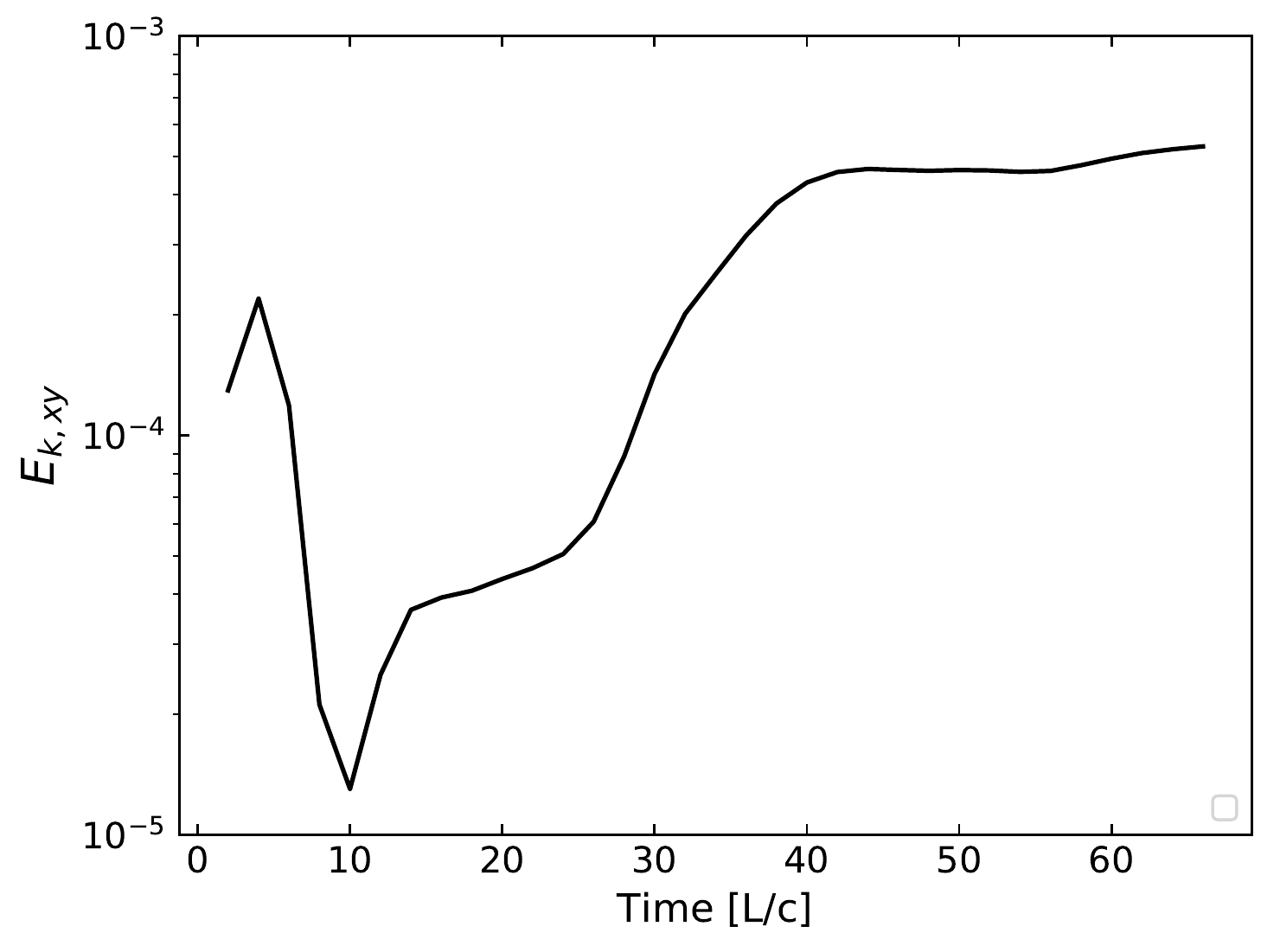}
    \includegraphics[scale=0.5]{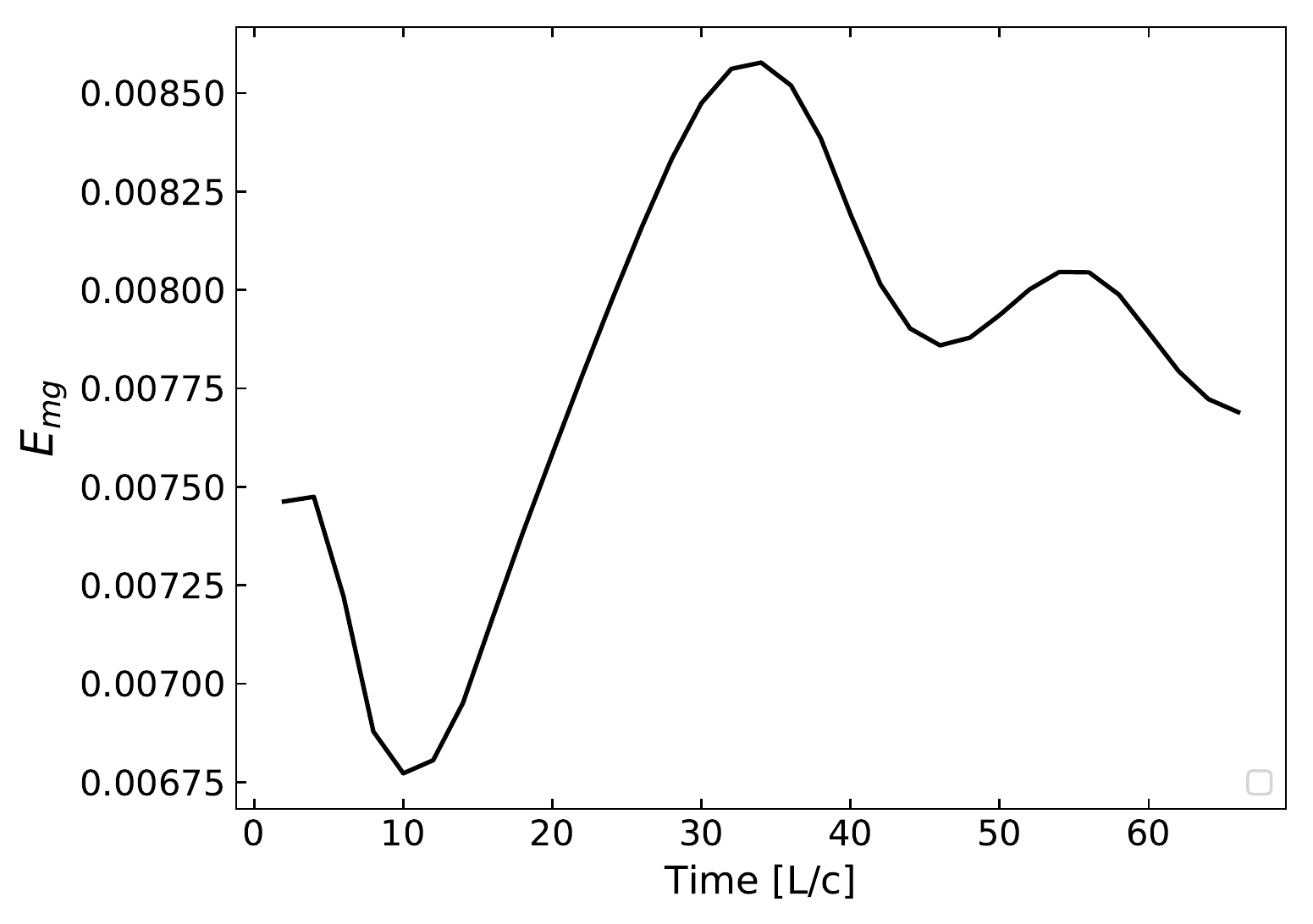}
\caption{Time evolution of the volume-averaged energy  within a cylinder of radius $R/L \leq 3.0$ for the simulation. On the top: kinetic energy transverse to the z-axis. The bottom: total electromagnetic energy. The kinetic energy density is in log scale, while the electromagnetic energy density is in linear scale.}\label{E240}
\end{figure}

We start by presenting the results of the 3D RMHD simulations of the relativistic jet and then we describe the results of the injection of test particles into this domain in different snapshots along the jet evolution.

\subsection{3D MHD simulation of the relativistic tower jet}

Figure~\ref{jet_points} shows the time evolution of the relativistic jet (model j240 in Table \ref{tablejet}) for snapshots at $t =$ 25, 30, 40, 44, 46, and 50 $L/c$.  
It depicts, in color scales, the isosurface of the magnetic current density intensity at half maximum in orange and the density map of the jet central slice (at the plane zy, and $x=0$), as well as the magnetic field lines in black. As time evolves and the CDK instability grows, the jet spine becomes increasingly deformed and disruptive, driving turbulence and magnetic reconnection. 

The resolution of our  simulated jet is large enough to allow us to  follow the growth of the turbulence and disentangle the reconnection structures.
As described in Section \ref{sec:intro}, in  a concomitant work  (\citetalias{kadowaki_etal_2020}), we have performed a systematic analysis of identification of all sites of magnetic reconnection in the evolving tower jet based on the algorithm developed in \citep*{kadowaki_etal_18a}
(see Figures 1 and 2 of \citetalias{kadowaki_etal_2020}).
2D cuts of the 3D reconnection sites have  evidenced  the presence of X points and extended chains of  magnetic islands (see Figures 3 and 4 in \citetalias{kadowaki_etal_2020}). 
It has been also found that the distribution of the reconnection rates follow a  log-normal   (see Figure 7 in \citetalias{kadowaki_etal_2020}), which is characteristic of  turbulent flows thus being  additional evidence that the turbulence  (induced by the CDKI) is driving the fast reconnection events  
\citep[in agreement with the theory of][]{lazarian_vishiniac_99} \citep[see also][]{santos-lima_etal20, kadowaki_etal_18a} along the relativistic jet.

In Figure \ref{jet_points}, the green dots  characterize all sites of $fast$ reconnection (identified with the search algorithm), that is, with reconnection velocities larger than or equal to the average reconnection rate obtained for each snapshot. 
Table \ref{tablevrec} presents this average and the maximum reconnection velocity values for each of the snapshots of Figure \ref{jet_points} (in units of local Alfv\'en speed),  as well as the number of reconnection sites (or counts) with reconnection velocities larger than the average value. 

We note that in $t=$ 25 $L/c$, there is no green dots (or reconnection regions), because turbulence and fast reconnection driven by  the  CDK instability have not developed yet, but in $t=$ 30 $L/c$, we identify already the break of symmetry of the plasma column and the appearance of a few fast reconnection sites at the jet  axis. We find that until near $t=40\,L/c$, there are still very few sites of fast reconnection, but in this snapshot and beyond, there are several sites  spread over all the jet domain, as we see in $t= 44$, $46$, and $50\,L/c$, in Figure \ref{jet_points}. It is interesting that in  the snapshot $t= 40\,L/c$, though it presents several sites, their reconnection velocities are all around the average, contrary to what we see in the more evolved snapshots where there are sites with very high reconnection speeds.   
In \citetalias{kadowaki_etal_2020}, these reconnection events are  discussed in more detail, and an average reconnection velocity $\langle V_{rec}/V_A \rangle \simeq  0.05 $ is derived when accounting for all reconnection events in all snapshots  \citep[see also][and further discussion below]{kadowaki_etal_2018b,dalpino_etal_2018,dalpino_etal_2020}.

Figure \ref{E240} shows the time evolution of the volume-averaged kinetic energy transverse to the $z$-axis \citep{mizuno_etal_09,mizuno_etal_11,mizuno_etal_12, singh_etal_16} within a cylinder of radius $R/L\leqslant 3.0$ around the jet axis,
\begin{equation}
E_{\mathrm{k},xy}=\frac{1}{V_b}\int_{V_b}\frac{\rho v_x^2 + \rho v_y^2}{2}dxdydz\, ,
\end{equation}
and the volume-averaged total relativistic electromagnetic energy
\begin{equation}
E_{\mathrm{em}}=\frac{1}{V_b}\int_{V_b}\frac{\mathbf{B}^2+[\mathbf{v}^2 \mathbf{B}^2-(\mathbf{v}\cdot \mathbf{B})^2]}{2}dxdydz\, ,
\end{equation}
where $V_b$ is the total volume where the average is calculated. 

As in \cite{mizuno_etal_12} and \cite{singh_etal_16}, we can use these diagrams to identify the growth of the CDK instability. As it develops, electromagnetic (EM) energy is converted into kinetic energy and this is a striking feature revealed by Figure \ref{E240}. 
Note that in this Figure, the EM energy is presented in linear scale, while the kinetic energy is in log scale.
The initial relaxation of the system to 
equilibrium leads to a hump in the kinetic and EM energy density curves (until $\sim 10$ $L/c$).\footnote{Though the centrifugal and pressure forces of our initial setup are small, they are not entirely negligible and thus the initial force-free magnetic configuration is not in real equilibrium. Therefore, a little relaxation occurs after a few times steps.}
The kink instability comes into play only after the relaxation finishes. 
There is  an initial  linear growth of the EM energy between $10$ and  $\sim 30$ $L/c$  due to the increasing wiggling distortion of the magnetic field structure in the jet spine in the initial increase of the CDK instability 
\citep[which  in a log scale would be harder to perceive; see e.g.][]{mizuno_etal_12}. 
After a slower increase, the kinetic energy undergoes an exponential growth from $t \sim 30$ $L/c$ to a maximum near $t\sim $ 40 $L/c$, after which it  approximately reaches a plateau while the magnetic energy decreases \citep[see more details in][]{singh_etal_16}. 
We note that this plateau time also coincides with the one after which we have detected an increase in the turbulence and the number of fast reconnection sites in Figure \ref{jet_points}. This plateau regime characterizes the achievement of non-linear saturation of the CDK instability and a nearly steady-state turbulent regime in the system. 
A similar trend has been also detected in the evolution of the average magnetic reconnection speeds,  which  nearly achieves a plateau around the same epoch (see Figure 8  in \citetalias{kadowaki_etal_2020}).


\subsection{Particle acceleration}

We have injected test particles in  different snapshots of the RMHD jet model j240, specifically, in the simulation time  $t = 25$, $30$, $40$, $44$, $46$, and $50 \,L/c$ (see also Figure \ref{jet_points} and Table \ref{tablep}).

\begin{figure*} 
\centering
\includegraphics[scale=0.33]{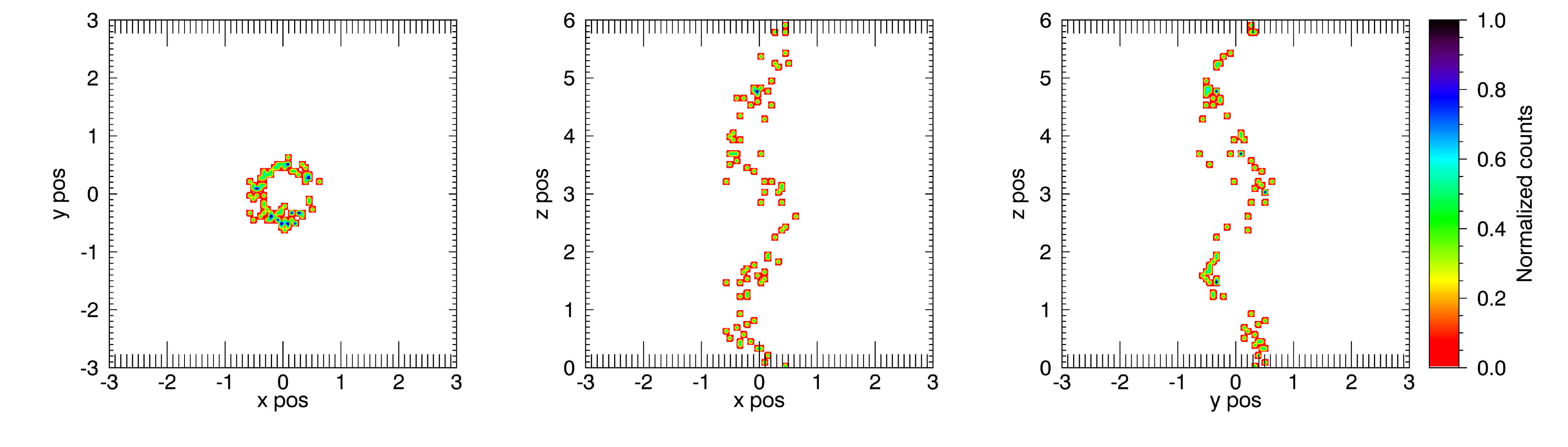}
\includegraphics[scale=0.33]{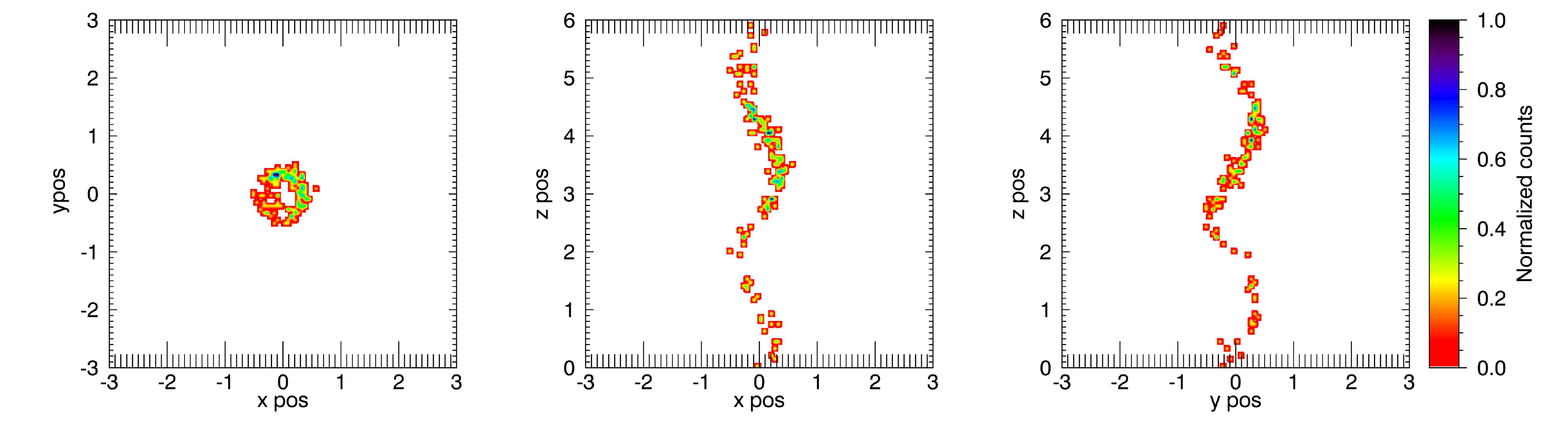}
\includegraphics[scale=0.33]{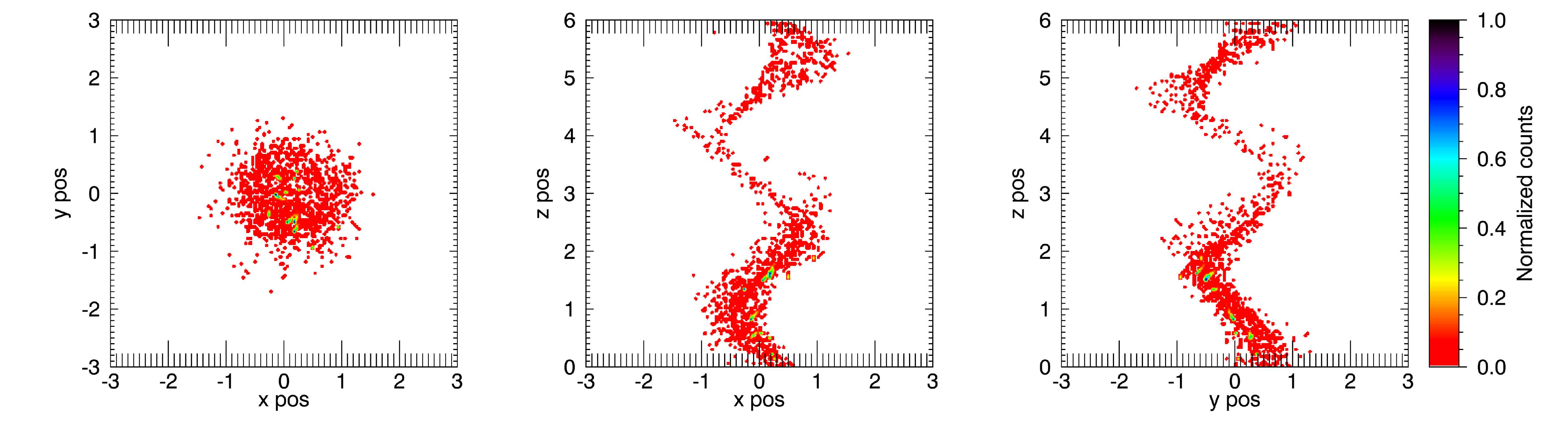}
\includegraphics[scale=0.33]{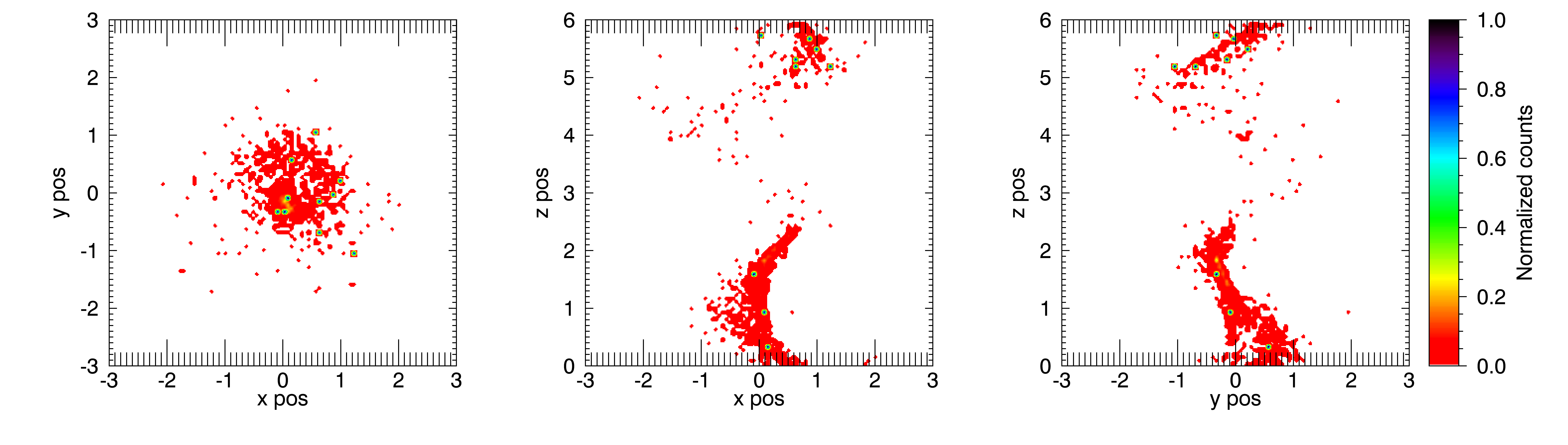}
\includegraphics[scale=0.33]{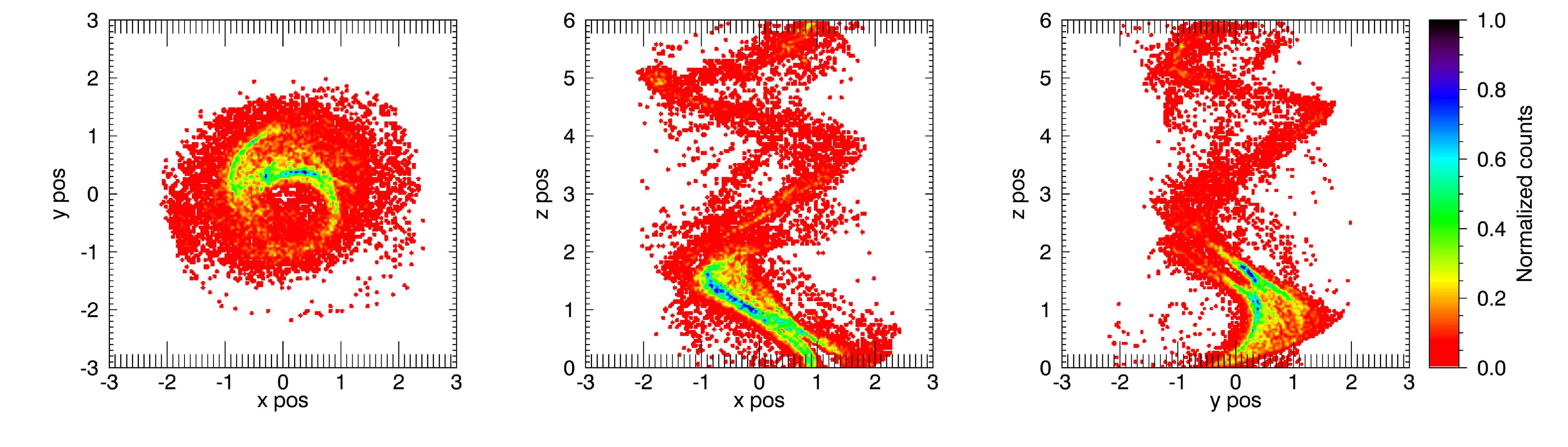}
\caption{
Two-dimensional histograms of particle positions for different snapshots of the model $j240$. From top to bottom, t = 25 $L/c$ ($ut25o$), t = 30  $L/c$ ($t30o$), t = 44 $L/c$ ($t44o$), $t =$ 46 $L/c$ ($t46o$), and  t = 50 $L/c$ ($t50o$), respectively. 
Each row,  from left to right,  shows histograms projected on xy, xz, and yz plans. To make visualization more clear, the histograms depict only particles  accelerating with increment $\Delta E_p/E_p > 0.4$, and with  energies larger than $10^2$ MeV 
($\sim 10^{- 1} m_p c^2$),
except for $t=25$ $L/c$, for which the minimum  energy depicted is $10^4$ MeV. The different colors indicate the (normalized) 
concentration of particles in each region of the jet 
(see text for details).
}
\label{posHisto}
\end{figure*}

Figure \ref{posHisto} shows two-dimensional (2D) histograms of the positions of  test particles in the snapshots $t =$ 25, 30, 44, 46 and 50 $L/c$, projected on planes xy, xz, and yz. All histograms show only the position of  particles accelerated with energy increment $ \Delta E_p / E_p> 0.4 $ (here $ E_p $ is the kinetic energy of the proton) and  energy larger than $10^2 $ MeV (or $ \sim 10^{- 1} m_p c^2 $), which is approximately the energy at which it starts an exponential acceleration growth (see Figure \ref{t44o}, in section \ref{sec:magrec} below). 
This condition applies to all snapshots, except to $t= 25$ $L/c$, for which particles start to undergo an exponential growth only for energies larger than $10^4$ MeV (or $\sim 10  m_p c^2 $). Also, only in this snapshot, the particles were injected with a monoenergetic spectrum ($\sim 10 m_p c^2$),  as discussed in section \ref{sec:t25}, while in the other snapshots, particles were launched with a Maxwellian distribution (section \ref{sec:setuptp}).

When compared to Figure \ref{jet_points}, Figure \ref{posHisto} indicates that  particles are mainly accelerated along the  wiggling jet spine for which the amplitude of the distortion increases as the CDK instability grows and turbulent disruption develops along with the appearance of fast reconnection regions. Particles are clearly accelerated in these regions where the strength of the current density is larger and, particularly for times larger than $t=40$ $L/c$, 
there are clearly several reconnection sites all over the wiggling structure. Note that while the CDK instability is still growing in the early times, in $t=25$ $L/c$ there are no reconnection regions and in $t=30$ $L/c$ only very few along the jet axis. Particle acceleration in these snapshots will be discussed in section \ref{sec:t25}.



\begin{figure} 
\centering
\includegraphics[scale=0.08]{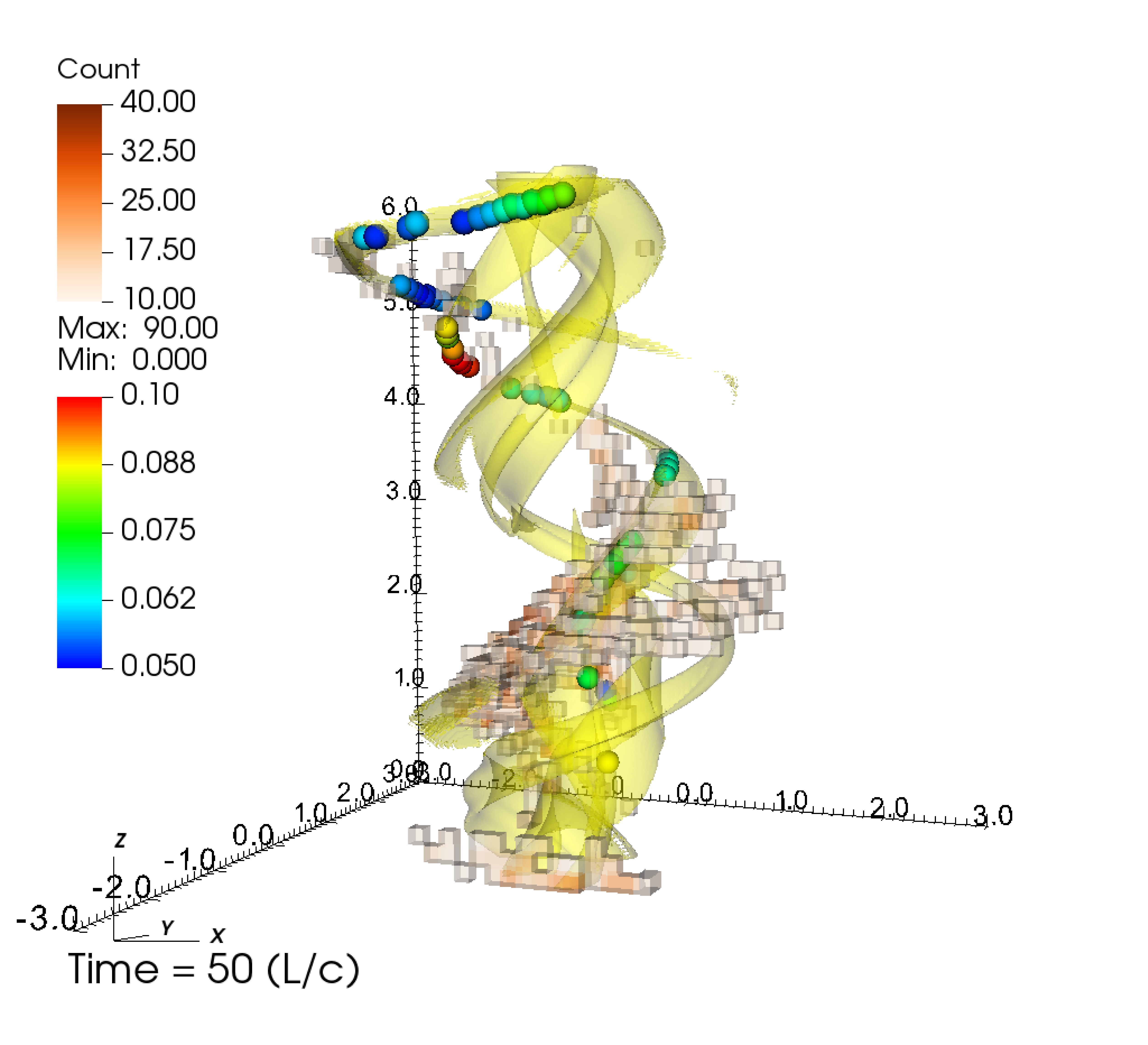}
\caption{
Three-dimensional histogram of accelerated particle positions (square symbols)  for the snapshot $t = 50\,L/c$ of jet model $j240$.  It was integrated over the particles acceleration time between 100 and 5000 hours, which corresponds to the exponential acceleration regime 
(see Figure \ref{t44o}, bottom panel). 
Only the positions of particles accelerated with energy increment $ \Delta E_p / E_p> 0.4 $ were included,  and energy between  $E_p > 10^{- 1} m_p c^2 $ 
and  the saturation energy 
$E_p\sim 10^{7} m_p c^2$. 
In order to improve  visualization, 
particles that accelerated to a maximum energy less than $10^{3}$ MeV were also removed,  
and to avoid boundary effects, the counts were constrained to the domain between $[ -2.5, 2.5]\, L$ in the  x and y directions, and $[0.5, 5.5]\, L$ in the z direction. 
The circles correspond to the positions of fast  magnetic reconnection sites  (with velocities larger than or equal to the average reconnection speed).
The isosurfaces of  the current density intensity at half maximum $|J|$ (yellow color) are also depicted ($J_{max}/2 \sim 1.25$). 
}
\label{pos3DHisto}
\end{figure}

Figure \ref{pos3DHisto} further elucidates these connections between particle acceleration and the sites of high current density and fast reconnection.
It depicts a 3D histogram of the accelerated particles for the jet snapshot $t = 50\,L/ c$. The histogram was integrated over the particle acceleration time interval between 100 hours and 5000 hours, which corresponds to the exponential acceleration regime (see Figure \ref{t44o} and section \ref{sec:magrec} below). As in Figure \ref{posHisto}, only the position of particles accelerated with energy increment $ \Delta E_p / E_p> 0.4 $, and  starting with  energy $ \sim 10^{- 1} m_p c^2 $ ($\sim 100$ MeV)
were included. The particles depicted are accelerated up to the saturation energy of the exponential regime,  $\sim 10^{7} m_p c^2 $ (Figure \ref{t44o}, bottom panel). 
This figure indicates a clear association  of the particles (orange square symbols) with the acceleration regions, being mostly confined within the wiggling configuration of the half-maximum current density iso-contours along the jet spine (shown in yellow). We also see a trend for a larger concentration of particles in regions of faster reconnection rates  (the green, yellow, and red circles), particularly in the heights between 
$1.0$ and $ 3.5\,L$, and between $4.0$ and $5.5\,L$ approximately. 
Besides, we note that there are also accelerated particles in the bottom 
of the domain. Since we have periodic boundary conditions in z-direction, these   particles  are  likely associated  with the reconnection sites seen in the top of the diagram. Finally, there are a few particles at the right edge around height $2.0$ $L$ which seem to be detached from the maximum current density layers. An analysis of their energies indicates that most of them have already achieved the saturation value and thus have already disconnected from the  reconnection regions.

In the next paragraphs we describe in detail the properties and the nature of the acceleration of the particles.


\subsubsection{Magnetic Reconnection acceleration} \label{sec:magrec}

\begin{figure} %
  \centering
  \begin{overpic}[scale=0.6]{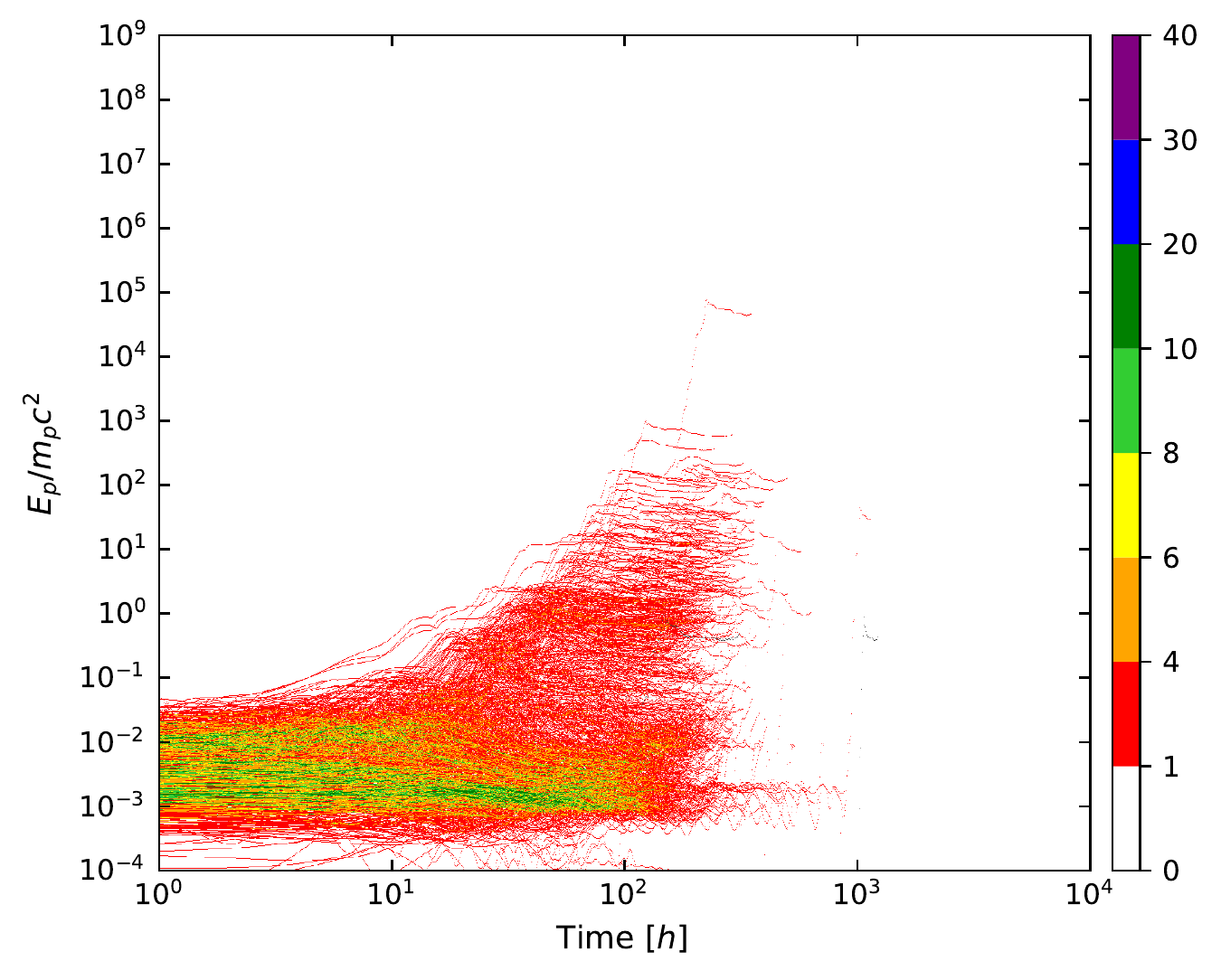}
    \put(14,46){\includegraphics[scale=0.21]{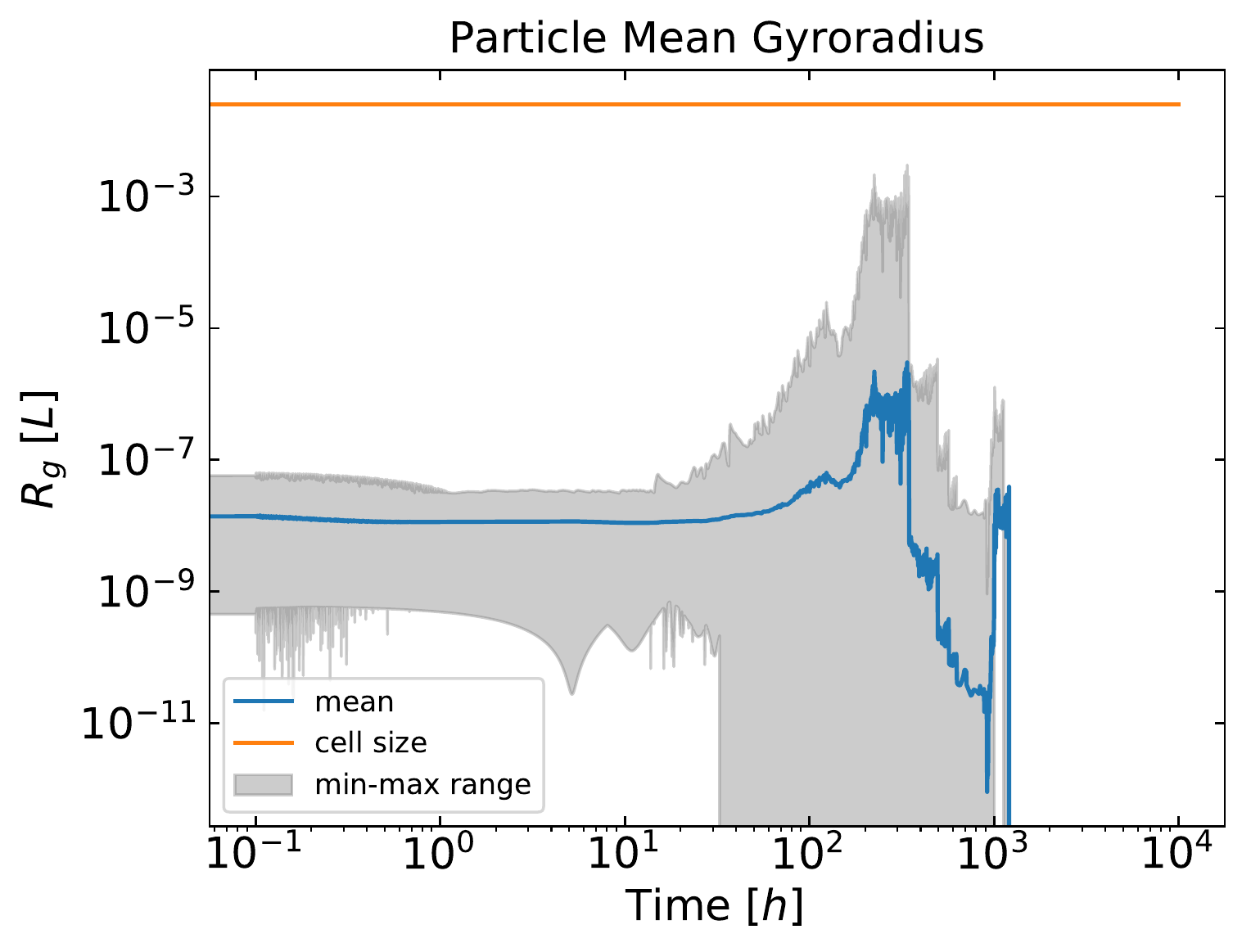}}
  \end{overpic}  
  \begin{overpic}[scale=0.37]{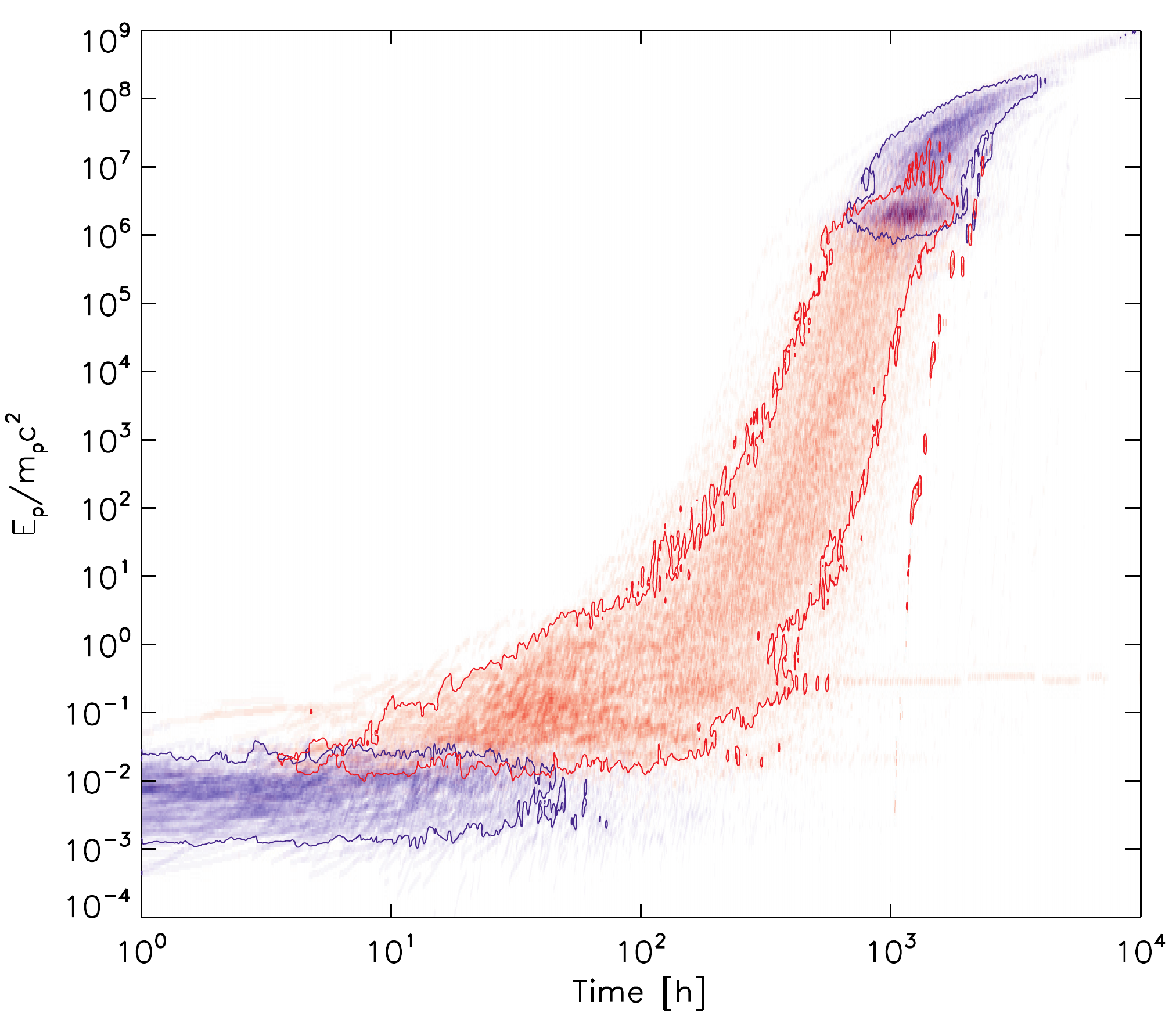}
    \put(14,50){\includegraphics[scale=0.21]{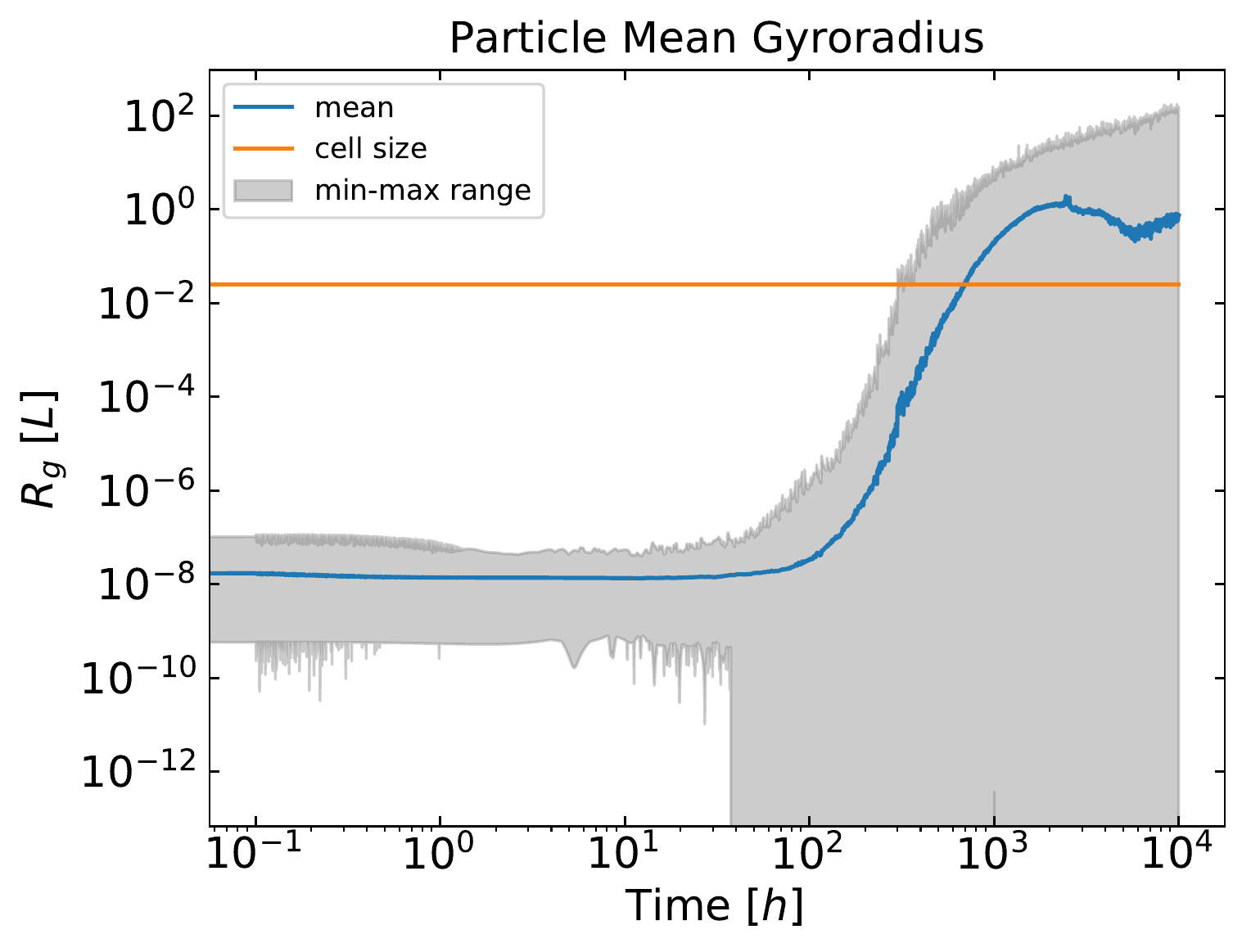}}
  \end{overpic}
  \begin{overpic}[scale=0.6]{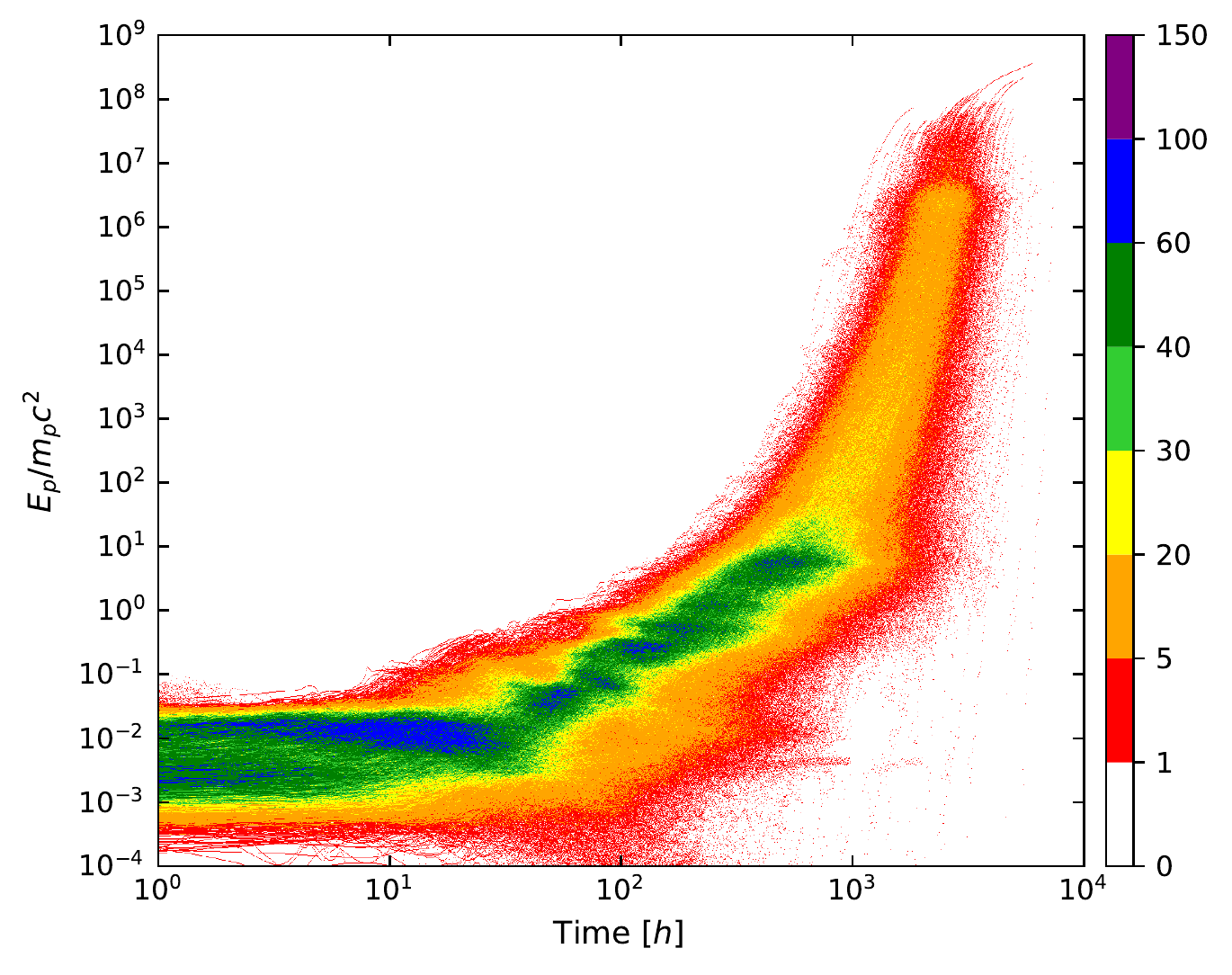}
    \put(14,46){\includegraphics[scale=0.21]{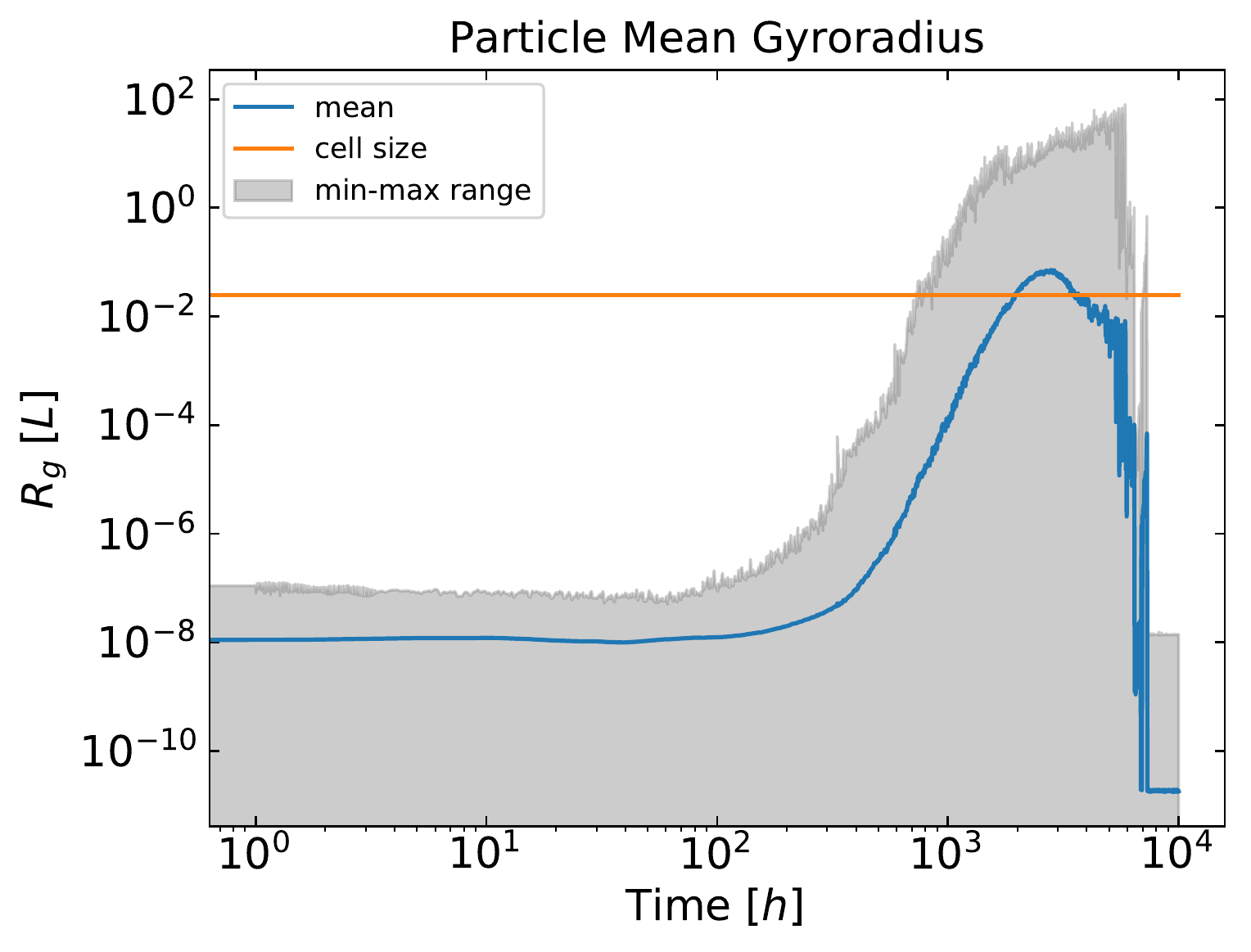}}
  \end{overpic}  
\caption{
Kinetic energy evolution, normalized by the proton rest mass energy, for the particles injected in the snapshots $t = 40$, $46$ and $50\,L/c$ of the jet model $j240$, from top to bottom, respectively. The color bars indicate the number of particles. 
In the middle panel, the colors indicate which velocity component is being accelerated (red or blue for parallel or perpendicular component to the local magnetic field, respectively).
The 
plots in the upper left inset of each panel show the time evolution of the particles gyro-radius. The horizontal orange line corresponds to the cell size of the simulated background jet, the grey color corresponds to the entire distribution of gyro-radius values, and the blue line gives the average value.}
\label{t44o}
\end{figure}

Let us first discuss in detail particle acceleration in the more evolved snapshots of the jet, after the CDK instability reaches the plateau, driving fully developed, near steady-state turbulence and fast reconnection all over the jet. 
Figure \ref{t44o} compares the kinetic energy
evolution of  test particles injected in the snapshots $t=40$ (the initial time of the plateau), $46$, and $50$ $L/c$.  The small plots in detail in each diagram show the evolution of the Larmor radius of the particles, with the orange line representing the cell size of the jet simulation. 

As in previous studies of test particles in single currents sheets \citep{kowal_etal_2011, kowal_etal_2012, delvalle_etal_16},  we clearly see that the injected particles, after an initial slow drift, undergo an exponential growth in their kinetic energy up to a maximum value around  $\sim 10^7 m_p c^2$ or $\sim 10^{10}$ MeV,
around $t\sim 10^3$ hr, for the snapshots $t= 46$, and $50\,L/c$. 
This is due to the stochastic Fermi-like acceleration in the current sheets, as described in section \ref{sec:intro} \citep[see also][]{dalpino_lazarian_2005, dalpino_kowal_15}. The maximum energy growth corresponds to a Larmor radius $E/(qB) \sim 4 L$, which is approximately equal to the jet diameter, above which the particles escape from the acceleration region \citep[][]{kowal_etal_2012,delvalle_etal_16}.  Beyond this value, the particles energy may still grow further, as we see for the $t= 46$ and $50\,L/c$ snapshots, but at a smaller rate.  As it is seen also in \cite{kowal_etal_2011,kowal_etal_2012}, this is due to further linear drift acceleration in the varying background large scale magnetic field of the system. 
The Larmor radius evolution plots indicate that initially it is very small compared to the size of the cell. When it approaches the cell size, particles then start to interact resonantly with the magnetic fluctuations of the background plasma,  undergoing exponential growth both in energy and Larmor radius. 

In the snapshot $t=46$ $L/c$ at the middle panel of  Figure \ref{t44o}, we depict which component of the velocity of the particles is being predominantly accelerated (red for the parallel, and blue for the perpendicular component to the direction of the local magnetic field).  We note that in the exponential regime, there is a clear dominance of the parallel component, characterizing an effective electric field mostly parallel to the reconnection layers, as expected, though the stochastic nature of the whole process also allows for the acceleration of the perpendicular component \citep[see also][]{kowal_etal_2012}. The slower drift acceleration regimes, both in the beginning and after the exponential growth regime, are dominated by the acceleration in the vertical direction. The same behaviour has been identified in the other snapshots.

We note that in the snapshot $t=40\,L/c$, which has not developed yet full turbulence with substantial number of very fast reconnection events (Figure \ref{jet_points} and Table \ref{tablevrec}), though particles also undergo exponential acceleration, most of them do not achieve the saturation energy, contrary to what happens in the more evolved snapshots where nearly steady-state, and fully developed turbulence has been already achieved.


\subsubsection{Particle acceleration in the earlier stages of the CDK instability} \label{sec:t25}

\begin{figure}
\centering
 \begin{overpic}[scale=0.6]{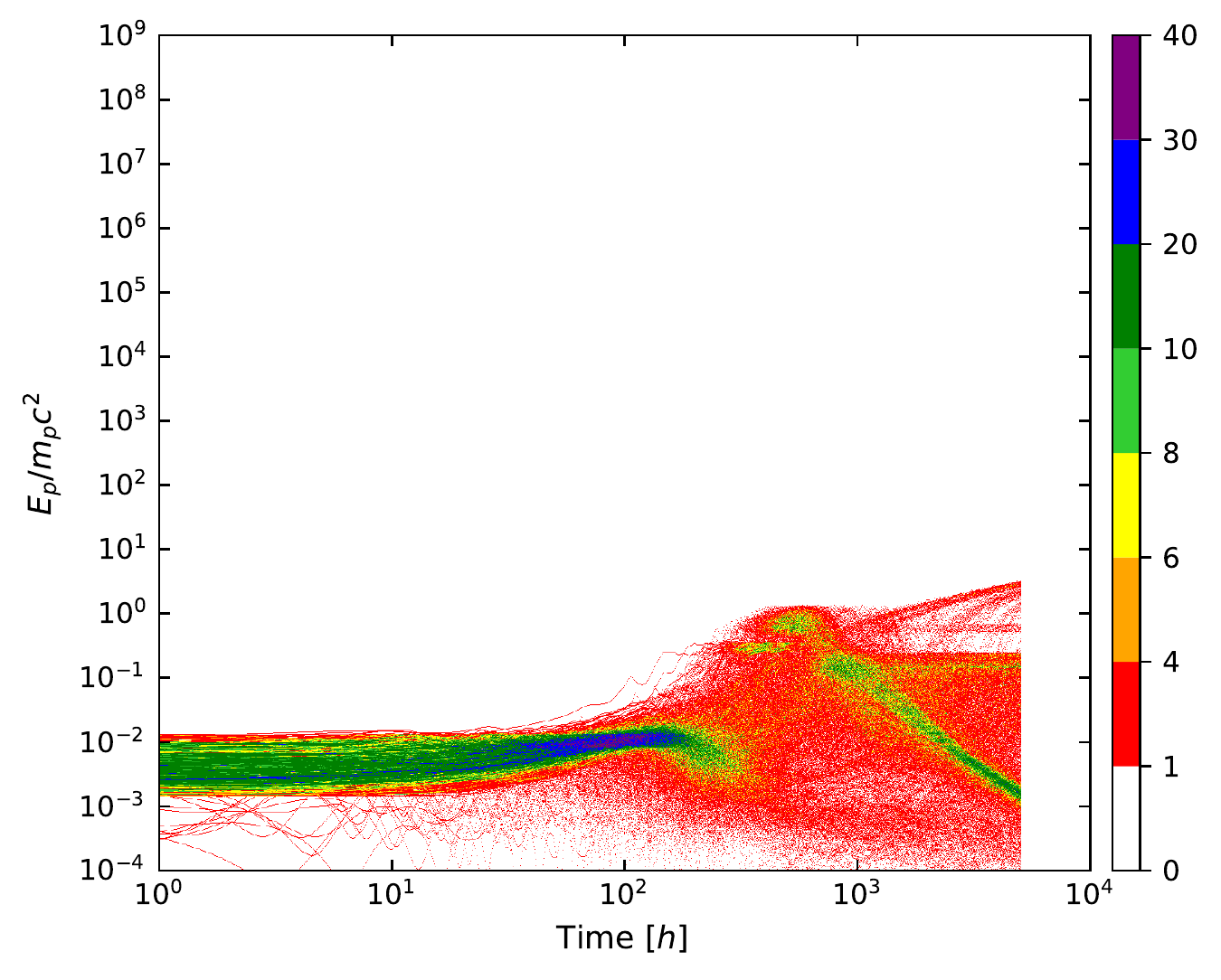}
 \put(14,46){\includegraphics[scale=0.21]{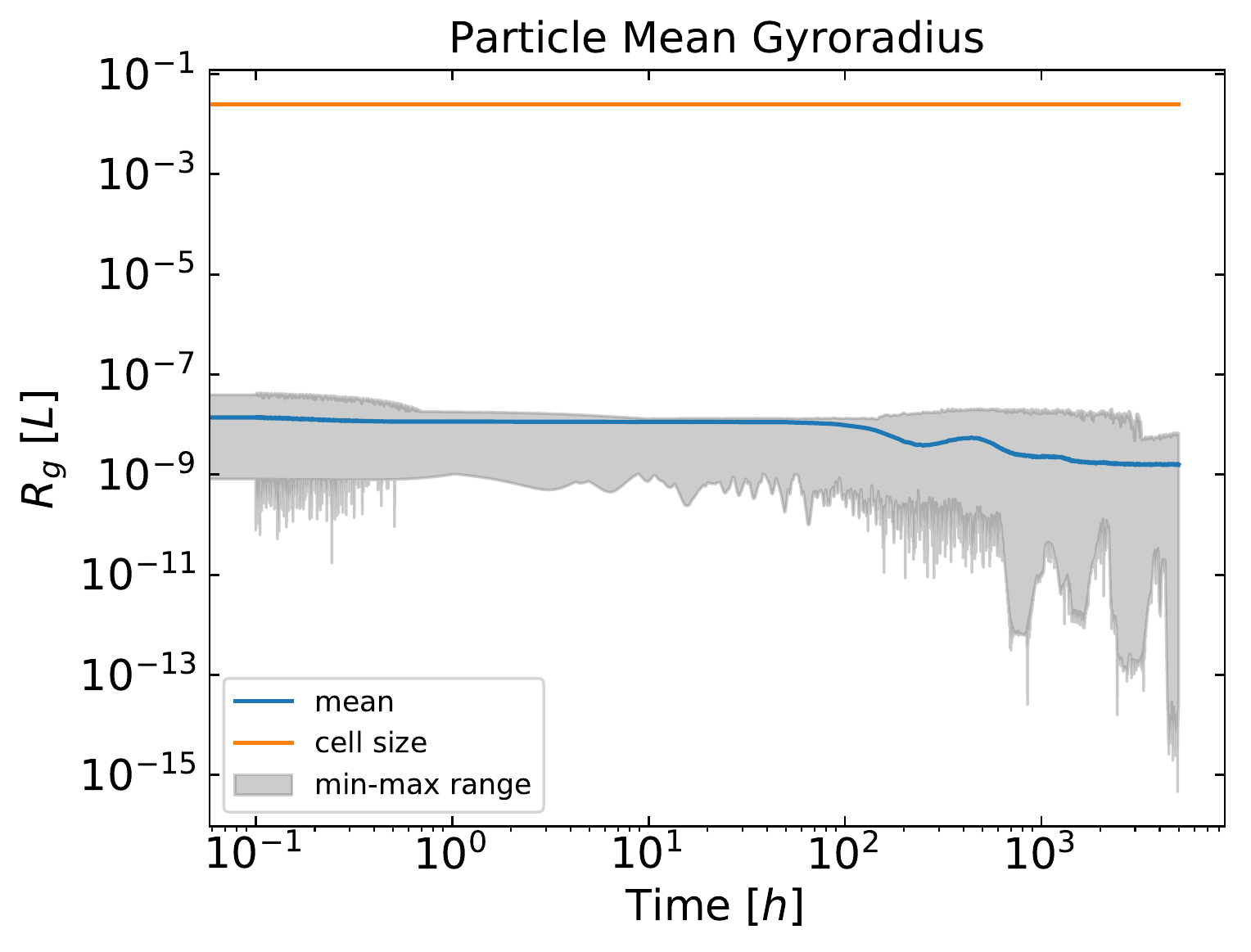}}
 \end{overpic}
 \begin{overpic}[scale=0.37]{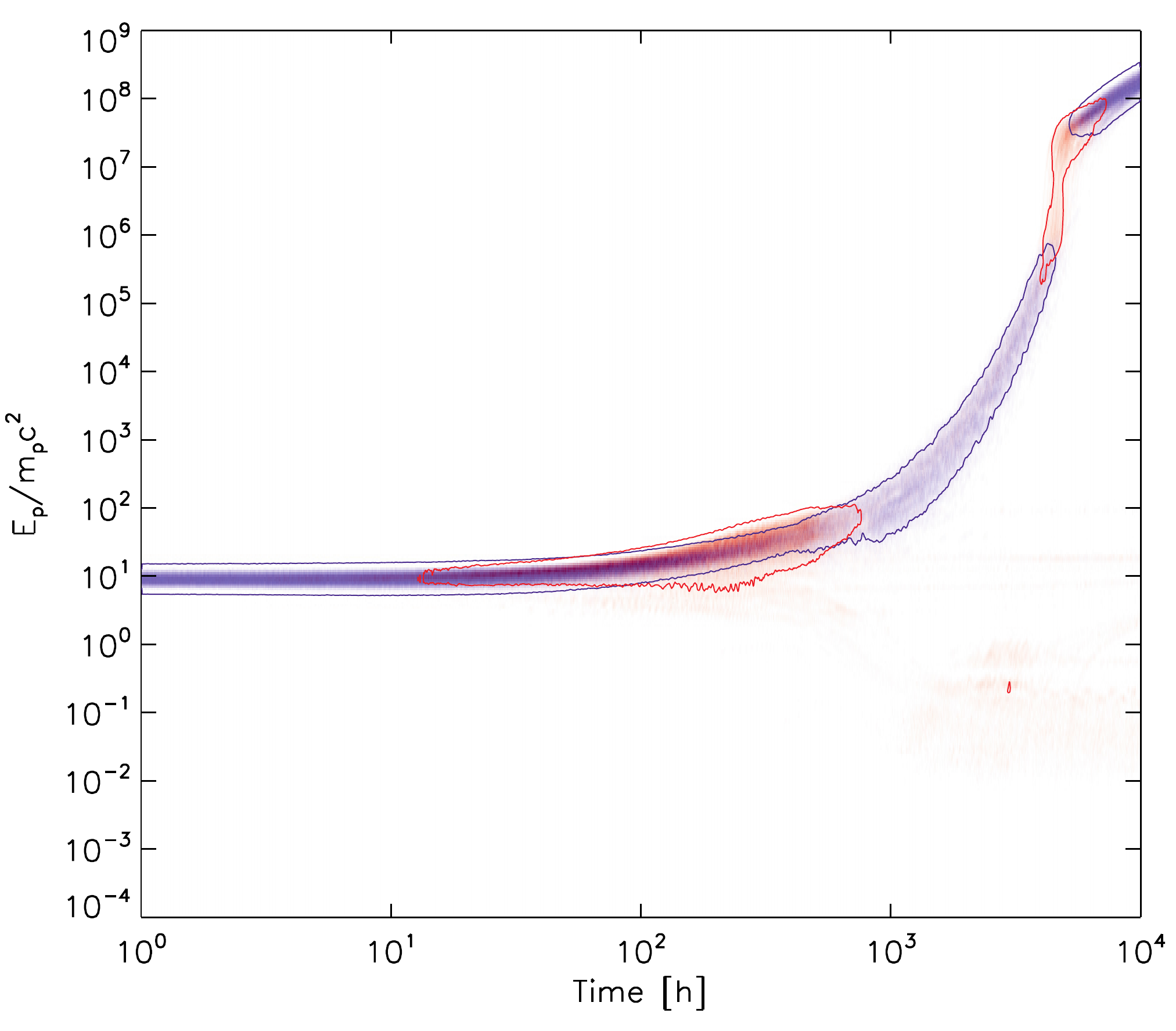}
 \put(14,50){\includegraphics[scale=0.21]{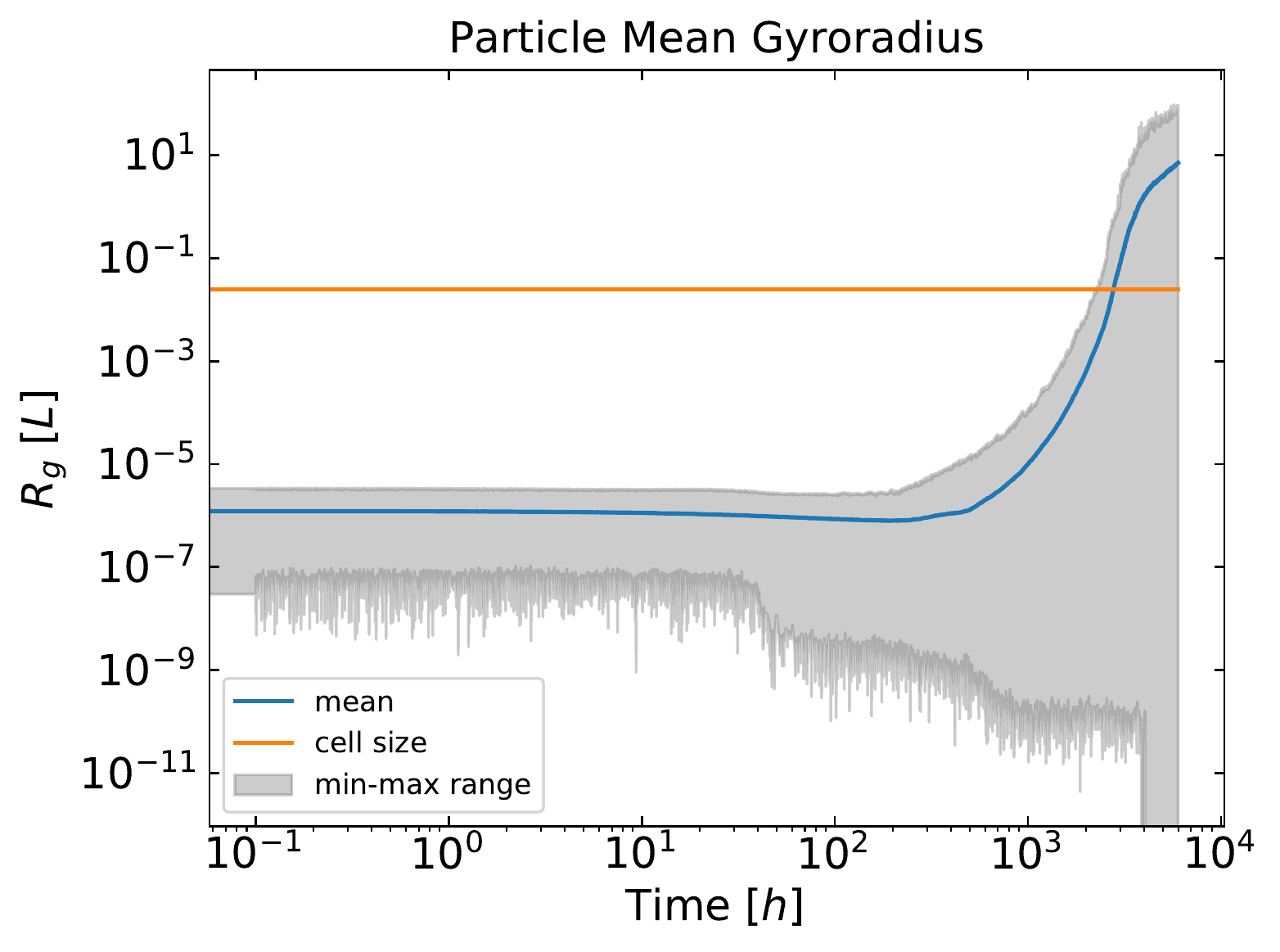}}
 \end{overpic}
\caption{Kinetic energy evolution for particles injected in the  snapshot  $t =$ 25 $L/c$ of the jet  model 
$j240$.
The top panel corresponds to  particles injected with a Maxwellian distribution and  $<E_p> \sim 10^{-3} m_p c^2$, as in Figure \ref{t44o}. The color bar indicates the number of particles. The  bottom panel corresponds to particles injected with a monoenergetic distribution with larger energy $E_p \sim 10 m_p c^2$. 
The colors in this diagram indicate which velocity component of the particles is being accelerated (red or blue for parallel or perpendicular component to the local magnetic field, respectively). The small plots in the detail depict  the evolution of the particles gyro-radius, the same as in Figure \ref{t44o}.} 

\label{t25}
\end{figure}

As remarked  previously, there is no reconnection 
 events in the jet snapshot $t=25$ $L/c$ (Figure \ref{jet_points}) and thus one would not expect any acceleration by magnetic reconnection. 
 Nevertheless, motivated by the recent work of \cite{alves_etal_2018} who claimed to detect magnetic curvature drift acceleration in a relativistic jet subject to the kink mode instability, we have also launched test particles in this snapshot in order to seek out for this process. The  results are presented in Figure \ref{t25}. The upper diagram shows particles launched with similar initial energy distribution and intensity as in the evolved snapshots of Figure \ref{t44o} (see test particle model $t25o$ in Table \ref{tablep}). In this case, we see that part of the particles undergo some acceleration, but saturate at an energy $E_p \sim m_p c^2$, which is much smaller than the values reached by the particles accelerated by reconnection in the evolved snapshots of Figure \ref{t44o}. Some particles even lose their energy. 
On the other hand, if we inject particles with an initial much larger energy $10^4$ MeV or $\sim 10 m_p c^2$ (see test particle model $ut25o$ in Table \ref{tablep}), they are accelerated as efficiently as in the evolved snapshots of Figure \ref{t44o}, though we find that part of the particles still lose their energy as in the run of the top diagram of Figure \ref{t25}.  A closer view into the plot of the Larmor radius evolution, indicates that this also increases exponentially once their value gets closer to the cell size of the background jet and exceeds it.  
 This exponential acceleration is similar to what \cite{alves_etal_2018} obtained in their PIC simulation of a tower jet. 

 Particle acceleration by magnetic curvature drift may also occur in helical jets subject to  kink mode instability \citep{alves_etal_2018}. 
 It may happen in the early stages of the development of the instability, before turbulence and reconnection break out. 
 As we have seen, the kink instability induces a growing helical modulation of the jet spine (see top panels of Figure \ref{jet_points}), and the  
 transverse motions excite an inductive electric field, $\mathbf{E}=-\mathbf{v} \times \mathbf{B}$. According to \cite{alves_etal_2018}, the axial component of this field $\left \langle E_z \right \rangle$ becomes strong and coherent throughout the jet spine  when  the transverse displacements excited by the tangling magnetic field of the jet become comparable to its radius, i.e., when the kink instability enters the non-linear regime. This leads to a potent acceleration that we see in the bottom panel of Figure \ref{t25}. However, we clearly see the difference from the comparison of the two diagrams of Figure \ref{t25} that, in order to the particles to obtain the effects of curvature drift they require some pre-acceleration.
 
 The bottom panel of Figure \ref{t25} also depicts  which component of the velocity of the particles is being accelerated (red for the parallel, and blue for the perpendicular component  to  the  direction  of  the  local  magentic field). In contrast  to  the middle panel of Figure \ref{t44o}, we see a dominance of the perpendicular component in the exponential regime and beyond that, thus confirming the dominance of the curvature drift acceleration in this case. 
 
\begin{figure}
  \centering
  \begin{overpic}[scale=0.37]{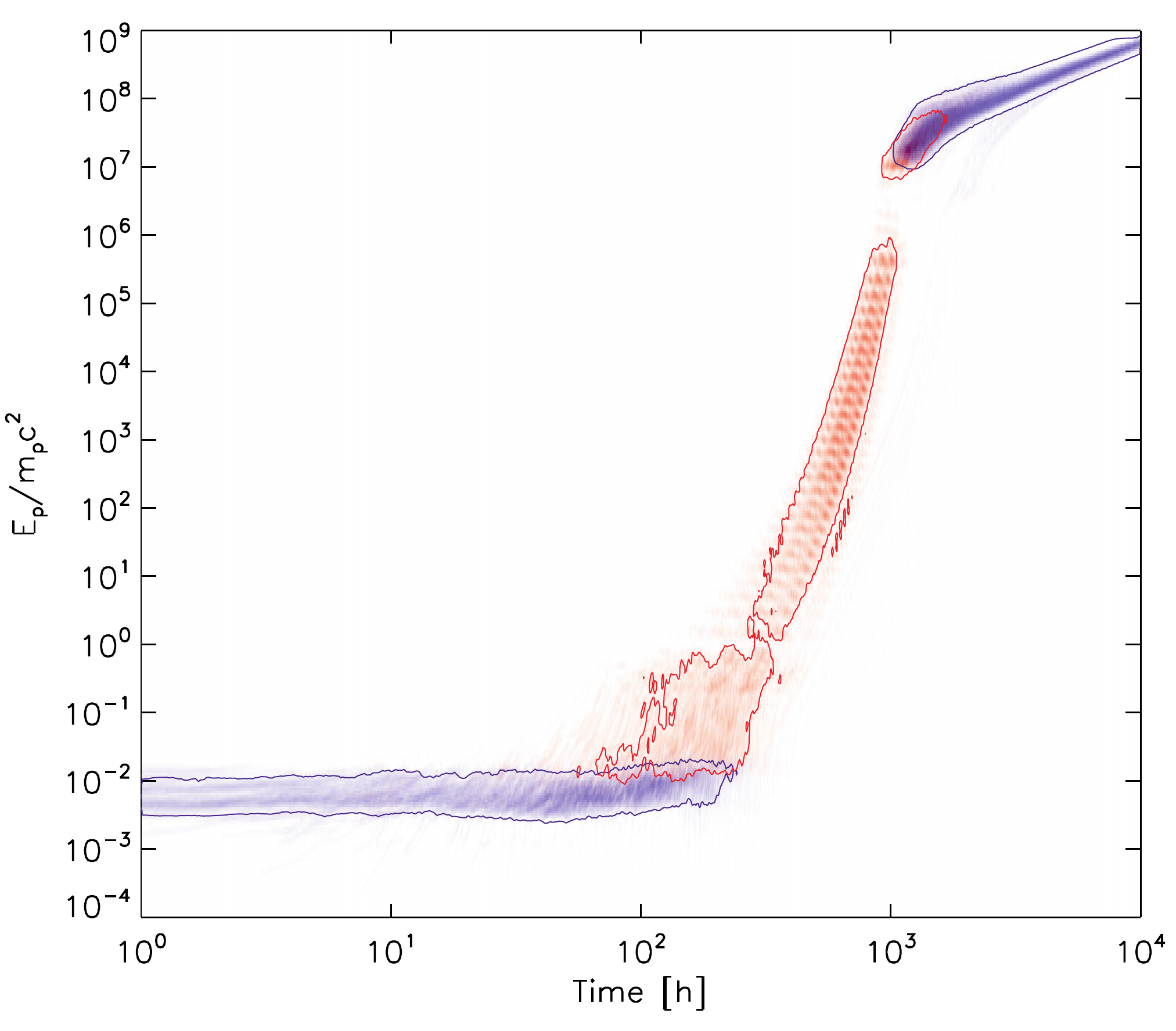}
     \put(14,50){\includegraphics[scale=0.21]{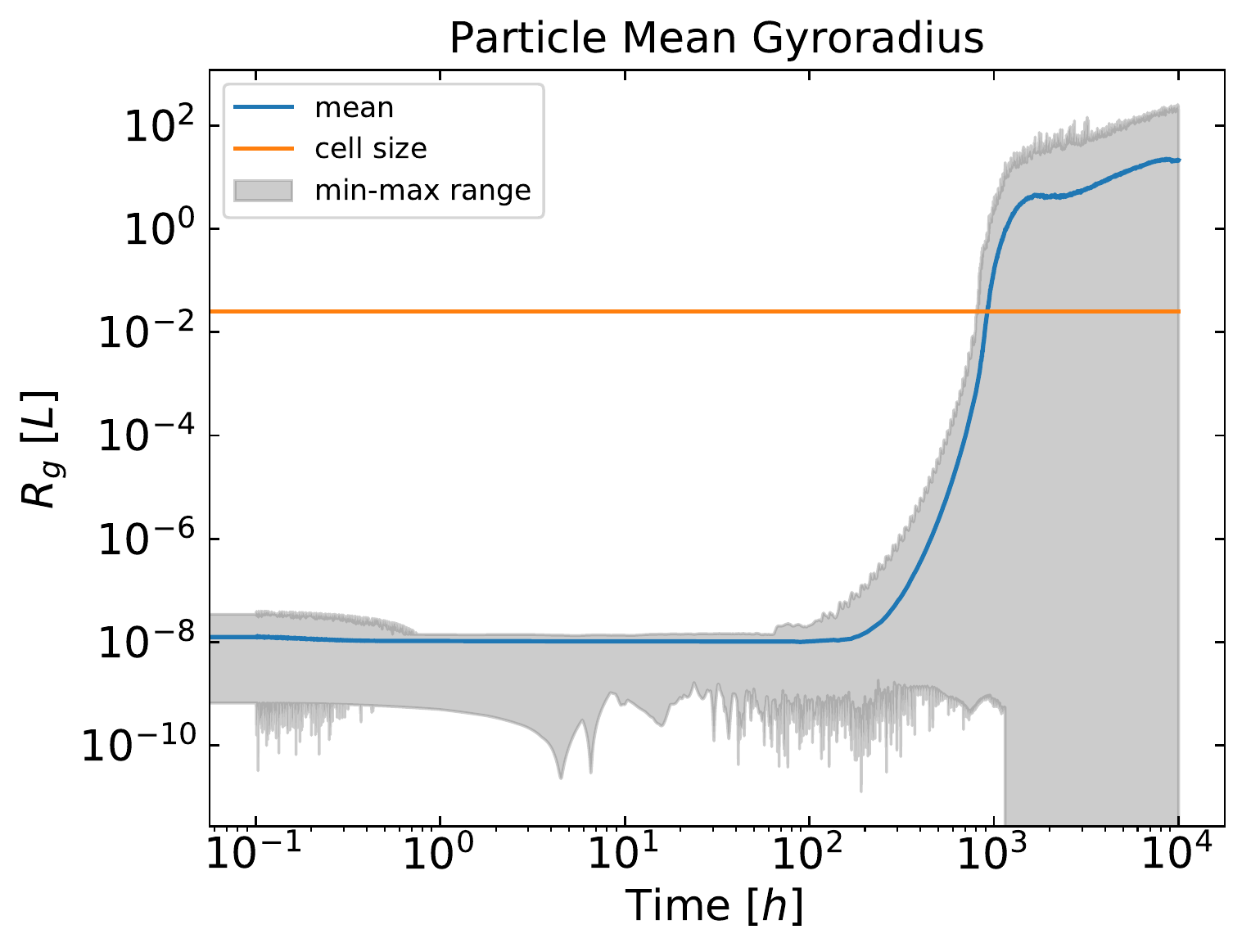}}
  \end{overpic}
\caption{
Kinetic energy evolution for particles injected in the  snapshot  $t =30$ $L/c$ of the jet  model 
$jet240$. As in Figure \ref{t44o},  particles  are injected with $<E_p> \sim 10^{-3} m_p c^2$ and a Maxwellian distribution. 
The colors  indicate which velocity component of the particles is being accelerated (red or blue for parallel or perpendicular component to the local magnetic field, respectively).
The small plot in the detail depicts the evolution of the particles gyro-radius.
}
\label{t30_240_1000_o}
\end{figure}

 Figure \ref{t30_240_1000_o}  depicts the kinetic energy evolution of test particles injected in the jet snapshot $t=30$ $L/c$ ($t30o$ in Table \ref{tablep}), for which  a few fast reconnection sites have been detected (see Figure \ref{jet_points}). This snapshot is also still in the non-linear growing phase of the CDK instability in the jet, before saturation (Figure \ref{E240}). The particles were injected  with the same initial  energy $\sim 1$ MeV ($\sim 10^{-3} m_p c^2$) as in the evolved jet snapshots of Figure \ref{t44o}, or the model of the top panel of Figure \ref{t25}.
 
 Interestingly, the particles now undergo 
  an exponential increase in the kinetic energy up to the same  maximum value of the more evolved snapshots (Figure \ref{t44o}, snapshots $t= 46$ and $50\,L/c$), even having only a few fast reconnection sites. 
  Furthermore, contrary to what we see in snapshot $t=25\,L/c$ (Figure  \ref{t25}), where particles could accelerate by magnetic curvature drift only starting with injection energy around $10^4$ MeV (or  $\sim 10 m_p c^2$), at $t= 30\,L/c$ they get accelerated starting  with energies well below ($\sim 10^{-3} m_p c^2$), as in the evolved snapshots. These results suggest  that the particles are also experiencing  magnetic reconnection acceleration in this case. 
  Moreover, it seems that when reconnection is present, particles do not require the pre-acceleration, as in $t=25\,L/c$ snapshot (Figure \ref{t25} bottom panel).  During the exponential growth, the fact that both processes are present, i.e., fast reconnection and a large amplitude tangled spine with a still coherent magnetic field, it is possible that both mechanisms, curvature drift and reconnection acceleration, are operating simultaneously. This combination may  also explain why we see a more efficient acceleration in this snapshot than around $t=40$ $L/c$, when CDK instability had just saturated and turbulent reconnection had just broken out, driven by  it. 
  
In fact, Figure \ref{t30_240_1000_o} indicates that the dominant velocity component accelerated in the exponential region is the parallel (as in Figure \ref{t44o}, middle panel), though the number of particles is smaller in this regime,  while  in the more extended  region of slower acceleration beyond that, where we see a much larger concentration of particles, the dominant component is the perpendicular one. The dominance of the parallel component in the exponential region is  suggestive of a predominance of reconnection acceleration, while the dominance of the perpendicular component in the second region is characteristic of  curvature and normal drift acceleration.


\subsubsection{Resolution Effects}\label{sec:resol}

\begin{figure} %
  \centering
   \begin{overpic}[scale=0.6]{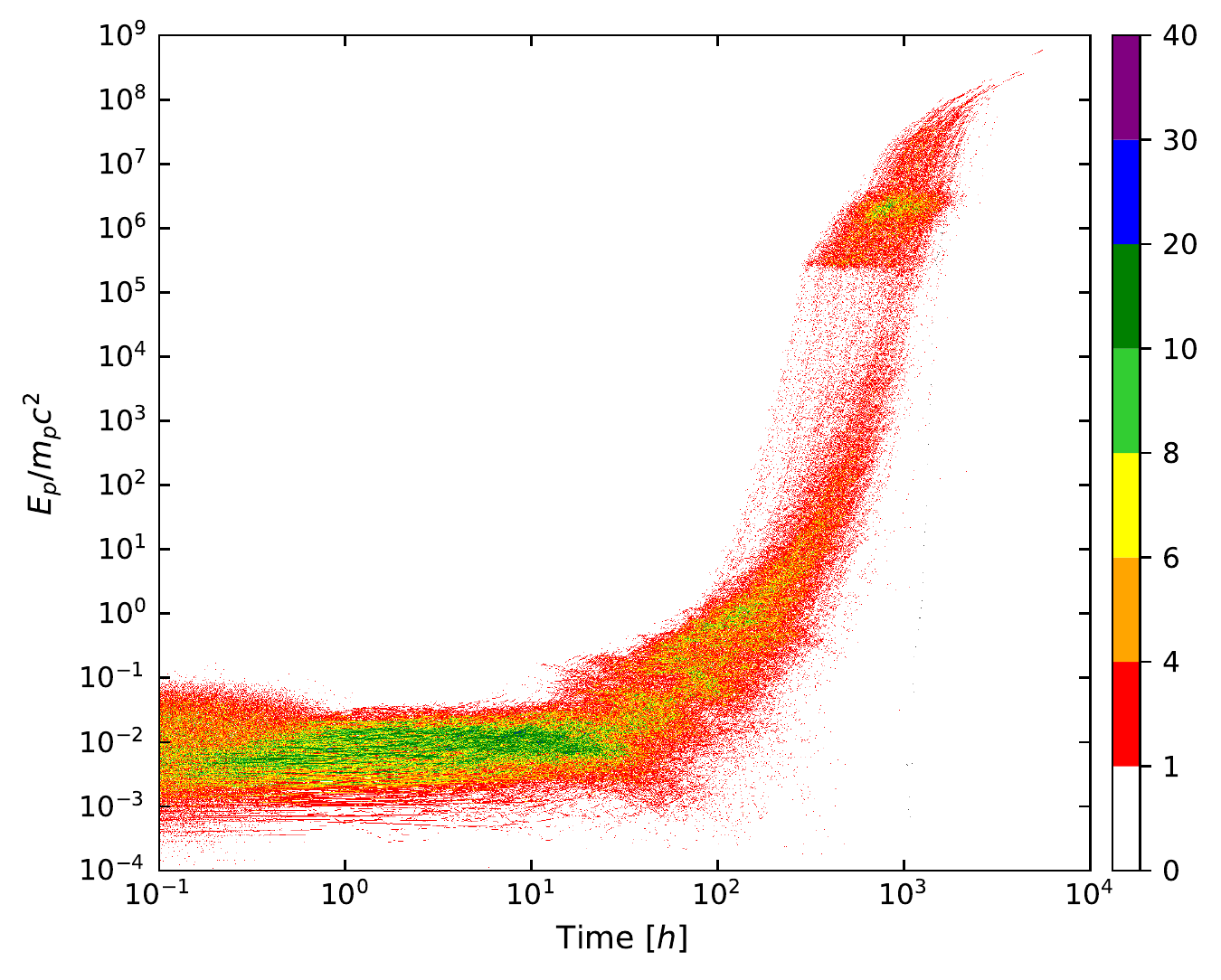}
     \put(14,46){\includegraphics[scale=0.21]{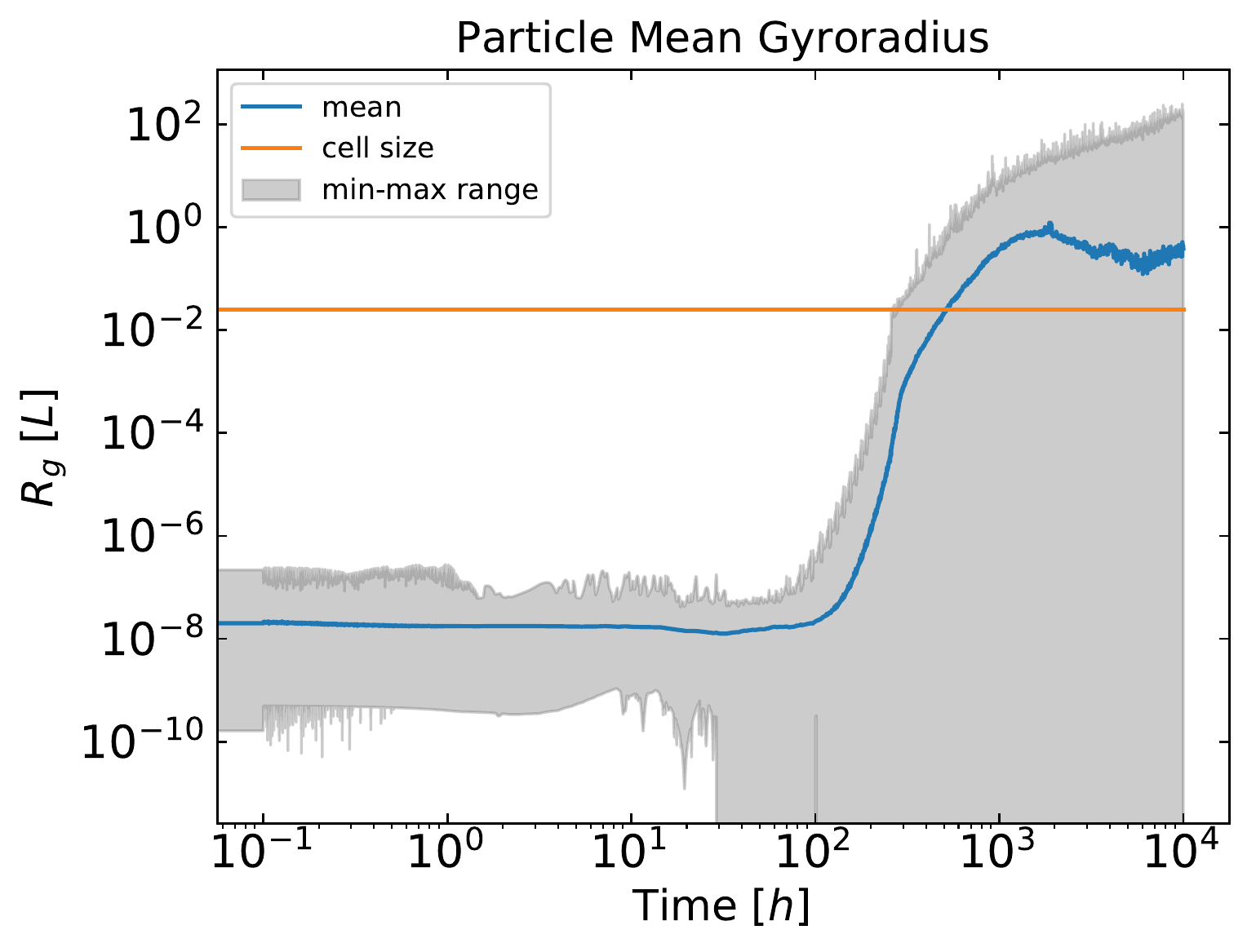}}
  \end{overpic}  
\caption{
Particle kinetic energy evolution, normalized by the rest mass energy, for 1,000  protons injected in $t=50\,L/c$ snapshot of the higher resolution jet model ($t480$, see Table \ref{tablejet}). The initial conditions for the test particles are the same as in the lower resolution test shown in Figure \ref{t44o} for the same time step (bottom panel), except that there 10,000 protons were used. The color bar indicates the number of particles. The small plot on upper left  shows the evolution of the particles gyro-radius.
}
\label{t50_240&480}
\end{figure}

In order to examine potential resolution effects, we have also injected test particles in a higher resolution background relativistic jet simulation (model $j480$ of Table \ref{tablejet}). 
Figure \ref{t50_240&480} shows the kinetic energy evolution for accelerated  particles launched at snapshot $t=$ 50 $L/c$ of this jet (test particle model $480t50o$ in Table \ref{tablep}). The initial conditions for this run are the same as in the corresponding lower resolution model at $t= 50$ $L/c$ (see bottom panel of Figure \ref{t44o}).
 
Comparing Figure \ref{t50_240&480} with its lower resolution counterpart in Figure \ref{t44o} (bottom panel), we clearly see that they are very similar. The only visible difference is due to the number of test particles used in each test. In the snapshot of the lower resolution jet model, we employed 10,000 particles, while in the high resolution jet model we injected only 1,000 particles, for being computationally much more time consuming and expensive. 
The energy growth rate is also very similar in both cases, as we will see in the next section (Figure \ref{rateacc}), though  the  particles in the higher resolution case reach the saturation energy of the exponential growth regime a little before $10^3$ hr, while in the low resolution jet, a little after $10^3$ hr, reflecting a slightly larger acceleration rate in the higher resolution case (see Figure \ref{rateacc}). 

This is due to the fact that in  the higher resolution jet (smaller cell size), more  regions of fast magnetic reconnection can be resolved at smaller scales, so that particles, starting with smaller Larmor radius, can interact more frequently with resonant  magnetic fluctuations, making the acceleration rate slightly more efficient. In fact, we have found that the number of reconnection sites is six  times larger than in the lower resolution jet. Nevertheless, since the change in the acceleration rates or the particles spectra are not substantial (see section \ref{sec:rates} and \ref{sec:spect}), we proceed our analysis considering the lower resolution jet model ($j240$), because the employment of the larger resolution counterpart for the entire analysis would be computationally rather long and  expensive.

We should note that we  have also repeated the test particle run for the snapshot $t =$ 25 $L/c$ (as in the top panel of Figure \ref{t25}, section \ref{sec:t25}), but employing the higher resolution jet model ($j480$), and we have obtained the same result, thus showing that that result is also not changed by the increase of  the background resolution.

\subsubsection{Particle acceleration rates}\label{sec:rates}

\begin{figure} %
\centering
\includegraphics[scale=0.5]{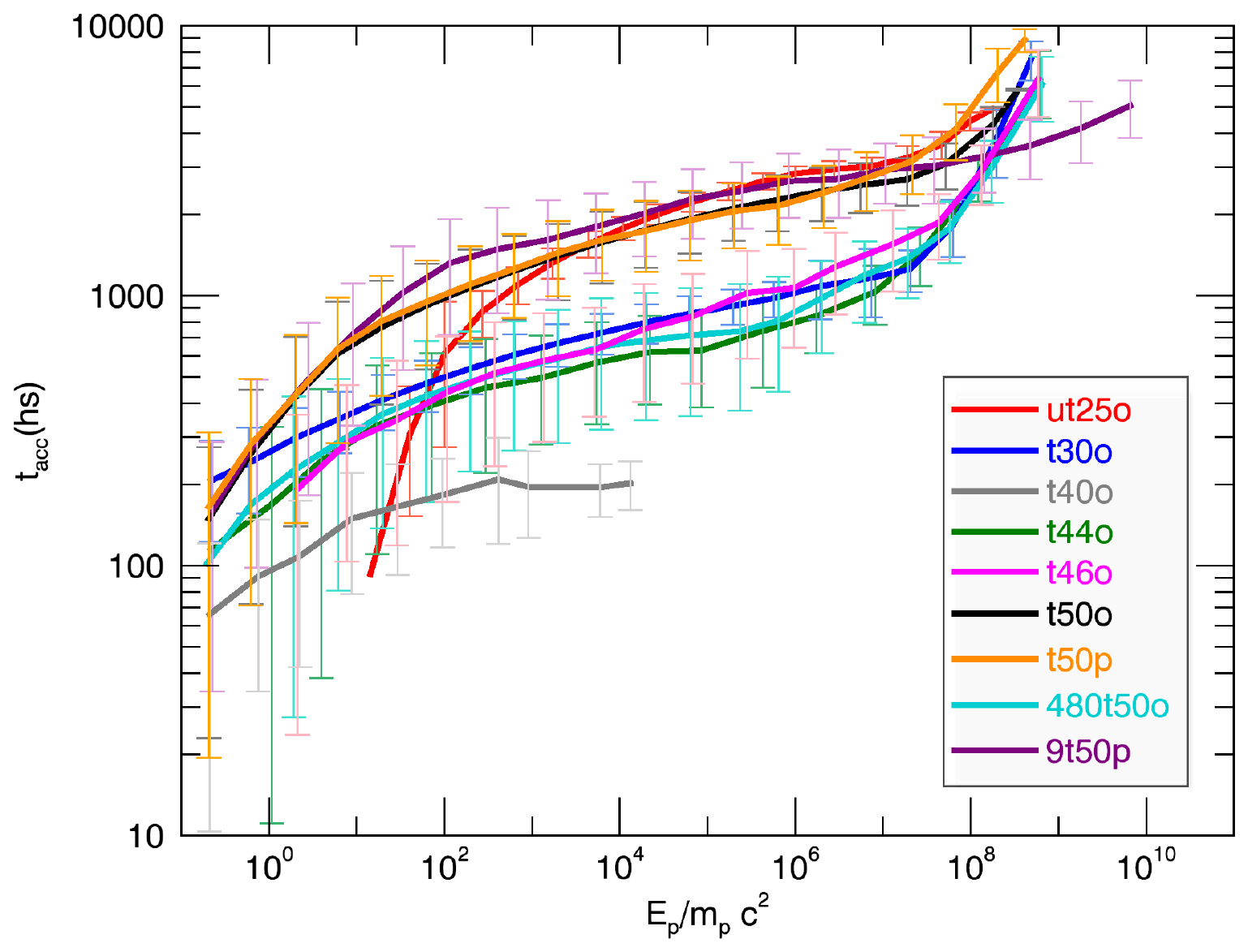}
\includegraphics[scale=0.5]{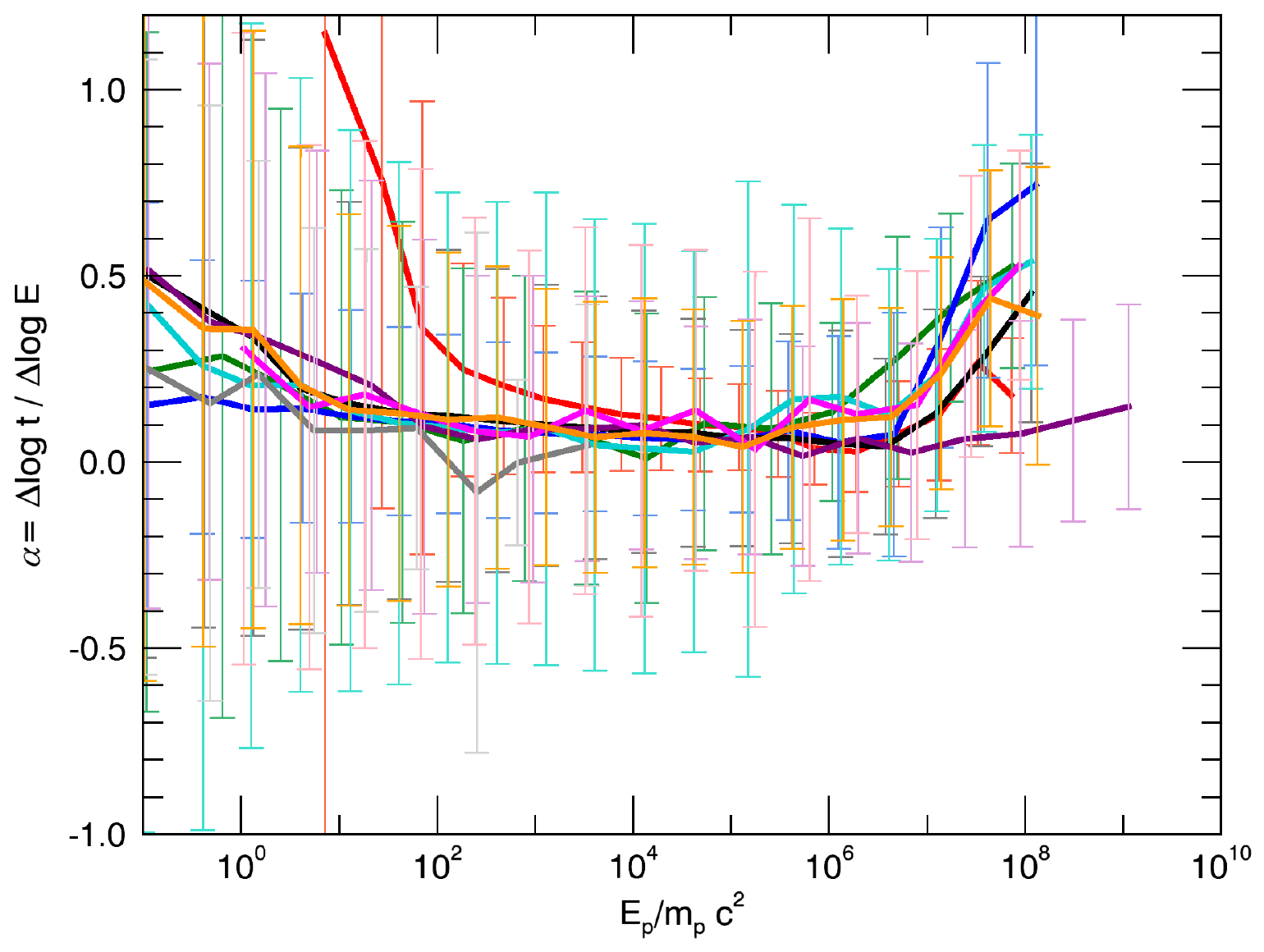}
\includegraphics[scale=0.34]{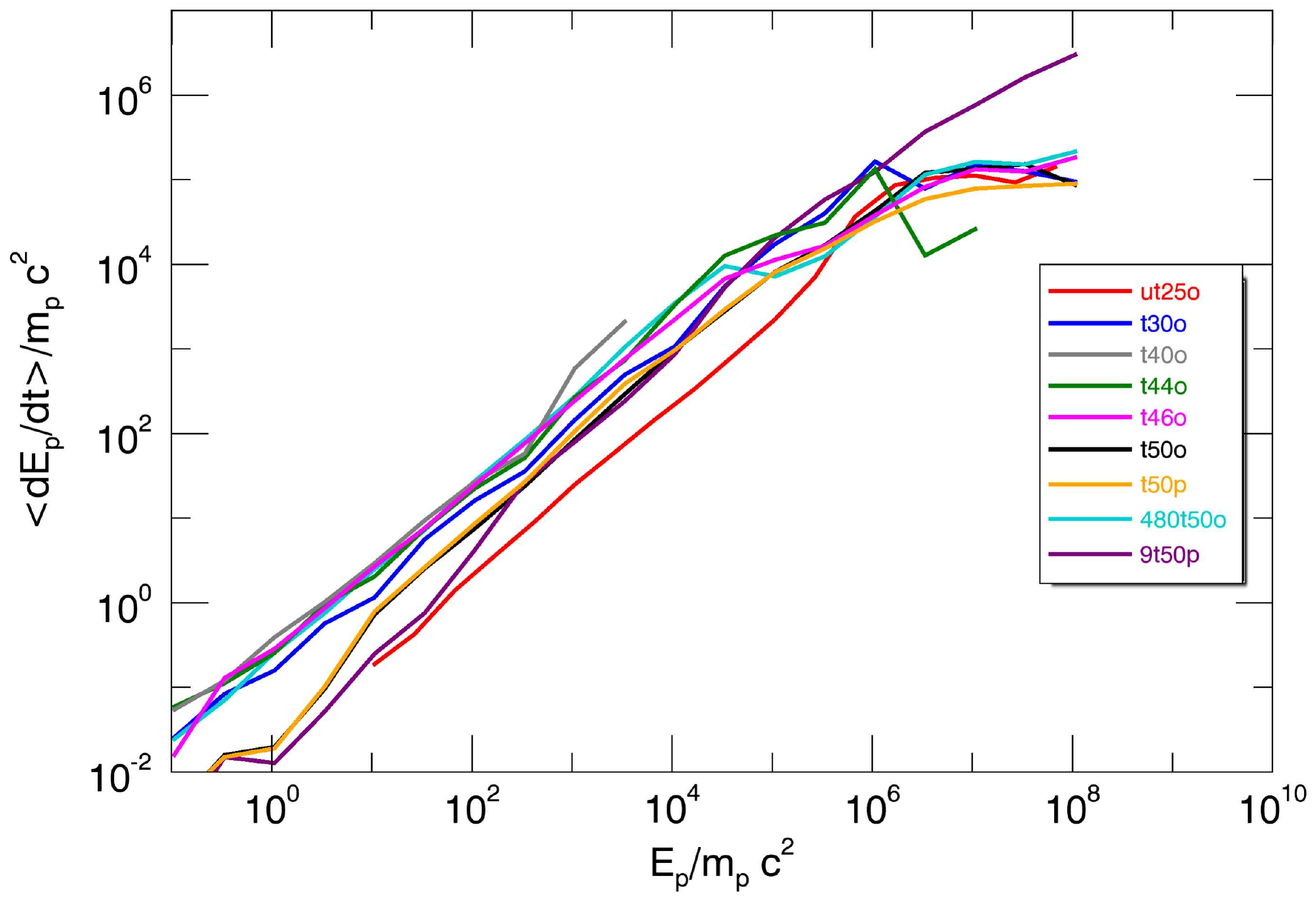}
\caption{ Acceleration time (top panel),  power law index of the acceleration time, $\alpha$ index (middle), and  kinetic energy growth rate (bottom) as  functions of the particle kinetic energy normalized by the proton rest mass energy for the different test particle models of Table \ref{tablep}. The error bars are obtained calculating the standard deviation. }
\label{rateacc}
\end{figure}

Magnetic reconnection acceleration, as in the Fermi process, predicts a dependence of the acceleration rate with the reconnection velocity and the particles energy \citep{dalpino_kowal_15,delvalle_etal_16,matthews_etal_2020}.  
Similarly as in \cite{delvalle_etal_16}, in order to quantify the effectiveness of the acceleration of the particles for each test particle model (Table \ref{tablep}), we have calculated the average time per energy interval that particles take to reach a certain energy, which gives the $acceleration$ $time$ as a function of the energy shown in  Figure \ref{rateacc} (top panel) for all the models. Using the same simulated data set, we depict in Figure \ref{rateacc} (middle panel), the power law index, $\alpha = \Delta \log (t_{acc}) / \Delta \log E_p$, of the the acceleration time dependence with particle energy, $t_{acc} \propto {E_p}^{\alpha}$.

In both, top and middle diagrams of Figure \ref{rateacc}, the particles in the different models enter the exponential growth regime of acceleration approximately around the energy $\sim 10^{-1} m_p c^2$ (in agreement with the diagrams of kinetic energy evolution of Figures \ref{t44o} to \ref{t50_240&480}), and end the exponential growth near $\sim 10^{7} m_p c^2$ (except for model $9t50p$ that we discuss below in section \ref{sec:magfield}). 
Before starting the exponential acceleration, the particles experience a slower growth in their energies which reflects in the larger $\alpha$ index, specially for $t= 25\,L/c$. 
The test corresponding to the snapshot $t=30 \,L/c$ has the smallest $\alpha$ 
 during this initial phase (dark blue line in Figure \ref{rateacc}), characterizing a smoother transition to the exponential acceleration regime. This is compatible with the previous analysis where we have seen that in this early snapshot, the particles are experiencing both, magnetic curvature drift and reconnection acceleration.  
During the exponential growth regime, 
 $\alpha$ decreases to  similar values around $\alpha \sim 0.09 \pm 0.05$ for all models and energies in this regime.\footnote{We note that the error in this determination was derived taking into account only the models having the jet  $j240$ as background, but the same value of $\alpha$ is derived for the particle models having the higher resolution jet as background.}

 Beyond the exponential regime, the acceleration time and the $\alpha$ index grow  a little further due to the slower drift acceleration that particles experience after leaving the reconnection (or curvature drift in the case of $t=$ 25 $L/c$) acceleration regions, as discussed in Figures \ref{t44o} to \ref{t50_240&480} (sections \ref{sec:magrec} and \ref{sec:t25}).

The kinetic energy growth rate as function of the particles energy depicted in the bottom diagram reflects the results of the upper panels. It increases with the energy, at the same rate in the exponential regime for all models. Interestingly, the only model that shows a slightly smaller rate (and slightly larger $\alpha$ index) is the one corresponding to snapshot $t= 25$ $L/c$, possibly due to the different acceleration process.
 
 A closer look into the acceleration time presented in the top diagram of Figure \ref{rateacc}, shows some slight differences between the models. Though these differences are approximately encompassed by the uncertainties of the numerical calculations (as we see from the error bars, which were calculated by the standard deviation method), the higher resolution jet model ($480t50o$) produces a slightly smaller acceleration time than its lower resolution test counterpart ($t50o$). This is compatible with the results found in Figures \ref{t50_240&480} and \ref{t44o} (bottom), discussed in section \ref{sec:resol}.). 
 
 Another interesting result is that for snapshot $t= 30$ $L/c$ ($t30o$ in  Figure \ref{t30_240_1000_o}) which also shows a smaller acceleration time, comparable to that of more evolved snapshots, like   $t= 44\,L/c$ ( $t44o$) and $t= 46\,L/c$ ($t46o$), for which we have detected a large number of fast reconnection sites with rates larger than the average value (see Table \ref{tablevrec}). Furthermore,  these tests have acceleration times comparable to the high resolution model ($480t50o$). Regarding the  models  $t44o$ and $t46o$, the larger efficiency can be attributed to the larger number of fast reconnection regions, while in $t30o$, this seems to be due to the combination of the two acceleration processes (as discussed in section \ref{sec:t25}).
 
More peculiar behaviour is found in the acceleration time for the snapshot $t= 40 \,L/c$ ($t40o$ in the top diagram), which shows the lowest values since the beginning. We remind  that this snapshot corresponds to the plateau of the CDK instability and the transition  of the jet from the laminar to the fully developed turbulence regime. Besides, it has already several sites of fast reconnection, but all with velocities near the average value (Figure \ref{jet_points} and Table \ref{tablevrec}). Moreover, Figure \ref{t44o} (top) has indicated that in this snapshot most of the particles enter the exponential regime of acceleration, but do not achieve the saturation energy. All these facts combined seem to  have favored this  slightly smaller acceleration time in the beginning of the acceleration for this snapshot in the transition regime.

\subsubsection{Particles Spectrum}\label{sec:spect}

\begin{figure} %
\centering
\includegraphics[scale=0.45]{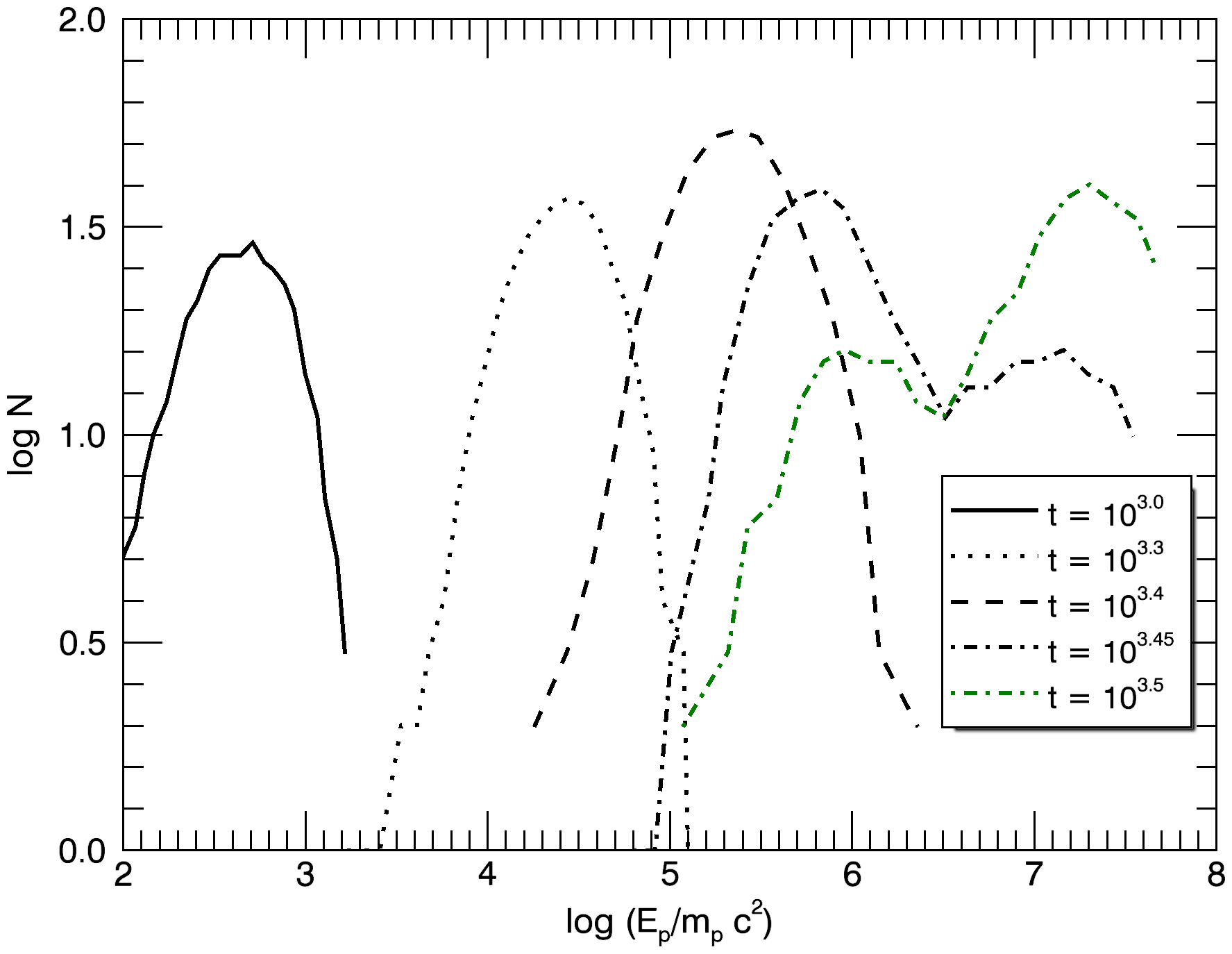}
\includegraphics[scale=0.45]{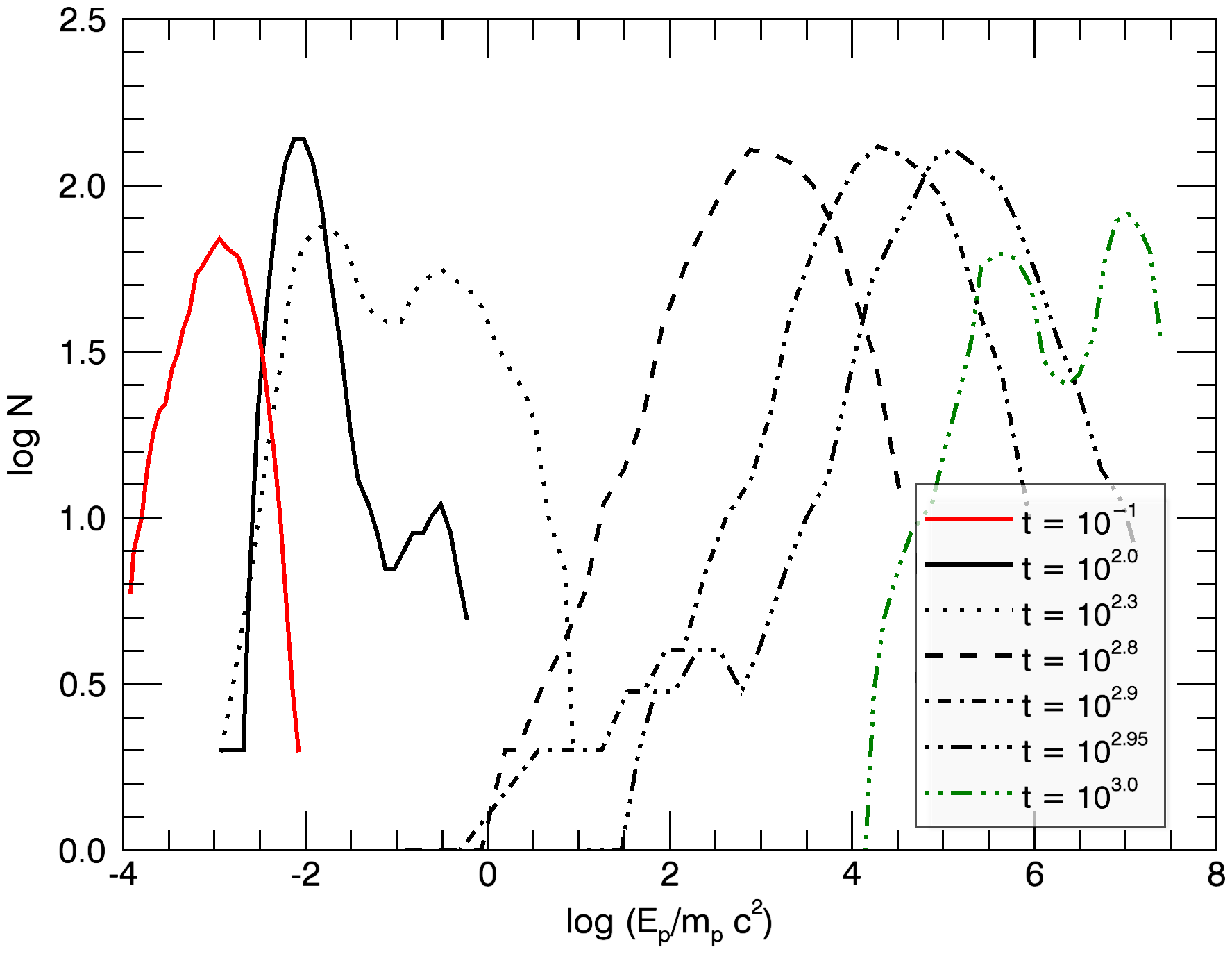}
\includegraphics[scale=0.45]{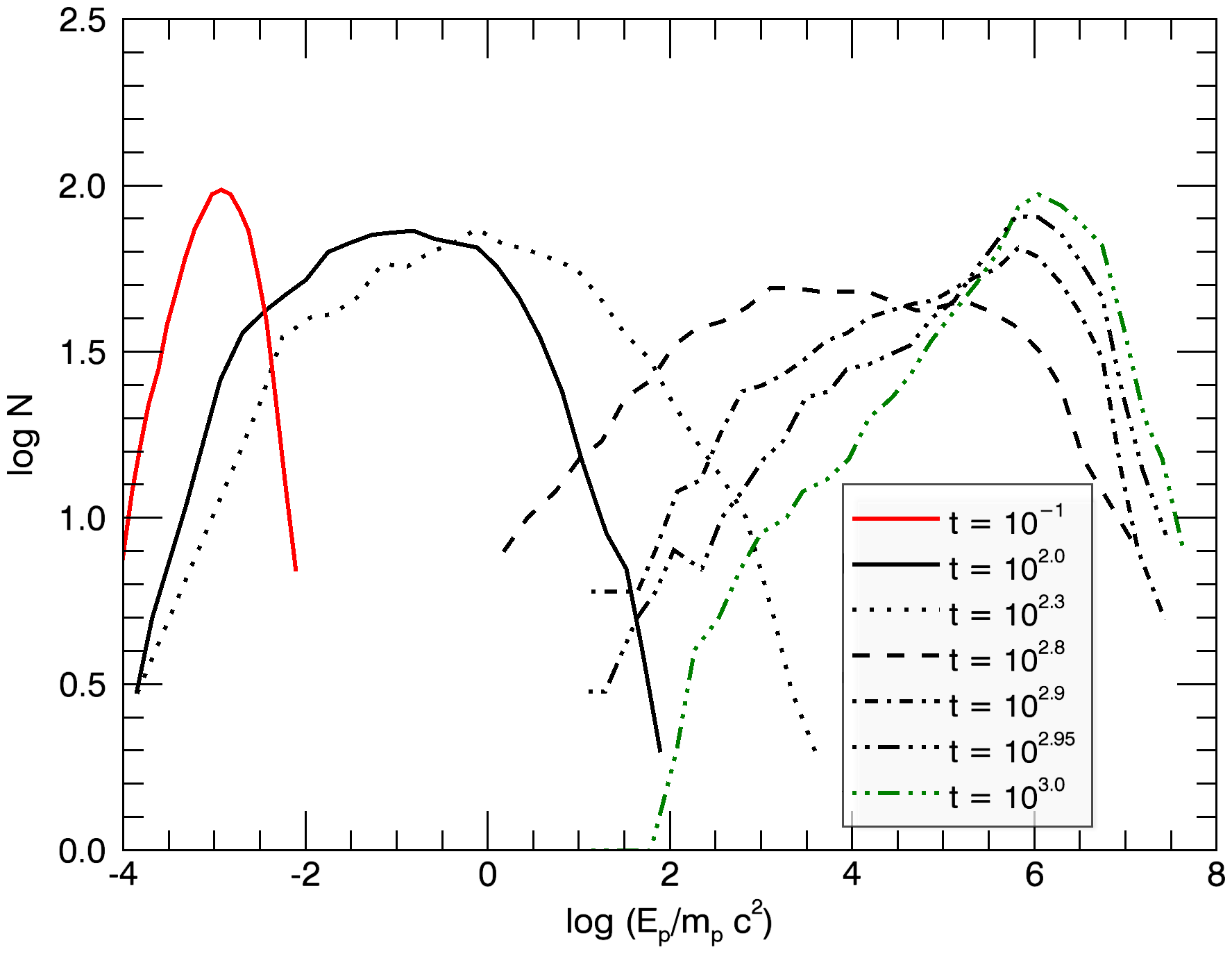}
\includegraphics[scale=0.45]{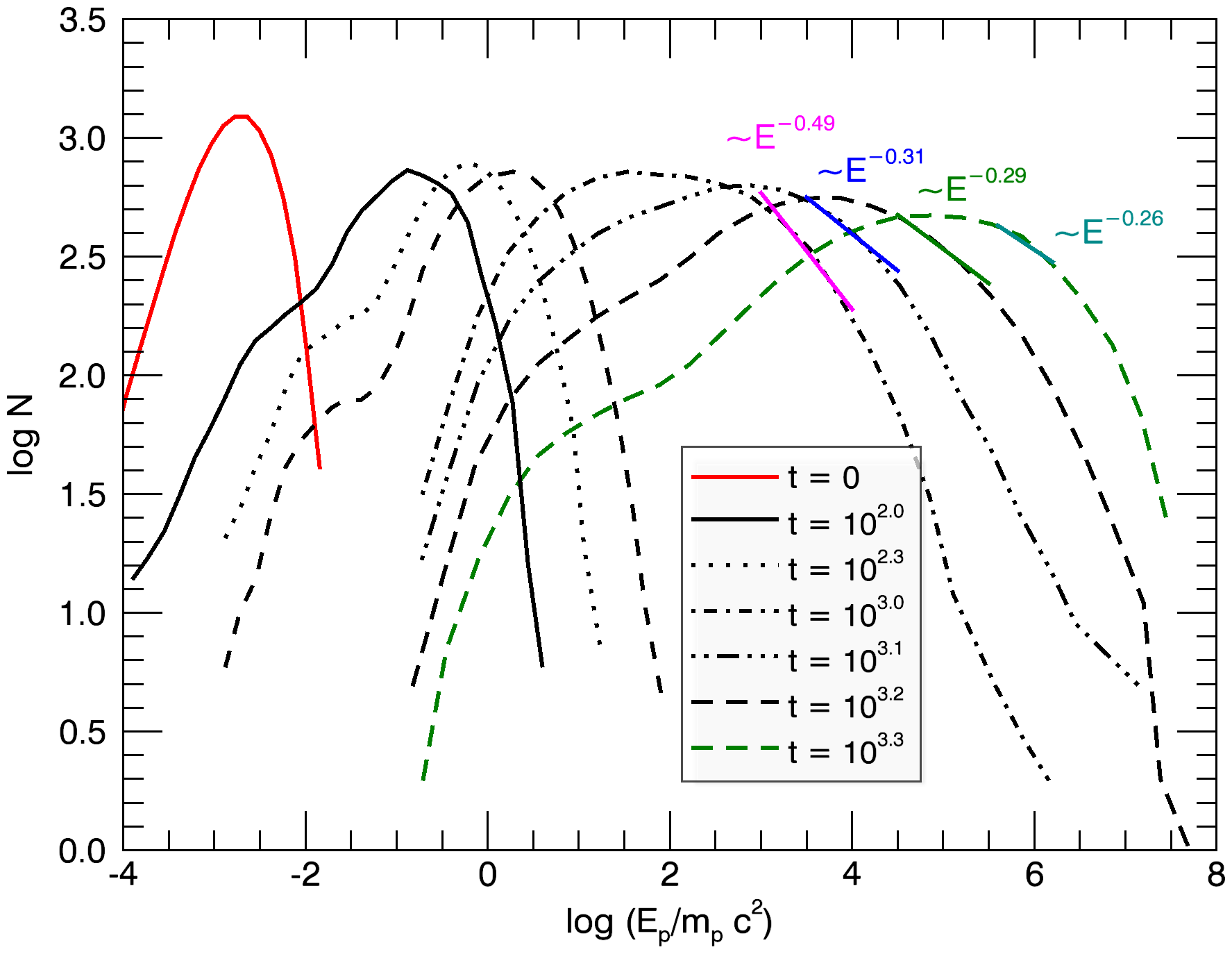}
\caption{Particles energy spectrum evolution  as a function of the normalized kinetic energy ($m_p c^2$)  in the jet snapshots, from top to bottom: $t=25$ (particle model $ut25o$); $30$ (particle model $t30o$), $46$ ($t46o$), and $50\,L/c$ (test particle model $t50o$). The red line in all panels but the top, corresponds to the initial Maxwellian  distribution of the particles. In the top panel particles are injected with a monoenergetic spectrum. The timesteps (in hours) of the acceleration are depicted in the detail of each panel.
}\label{NE}
\end{figure}

Figure~\ref{NE} shows the evolution of the energy spectrum of all particles (not only the accelerated ones) during the acceleration in the jet snapshots at $t=25\,L/c$ (particle model $ut25o$), $30$ ($t30o$), $46$ ($t46o$)  and $50\,L/c$ ($t50o$), from top to bottom panels, respectively. Several time intervals are depicted in each panel, in units of hour. Initially, the particles have a Maxwellian distribution (red line), except in the top panel for the jet snapshot $t= 25\,L/c$, where the injection spectrum is monoenergetic (with $\sim 10 m_p c^2$).  
As particles accelerate, they start to populate the higher energy tail of the distribution, which becomes flatter at these energies (see third and bottom panels, in particular). We should remember that in our numerical setup, particles are continuously re-injected into the system and therefore, they never stop being accelerated. For this reason, the distribution shifts to larger  and larger energies. Furthermore, even after the particles attain the maximum (saturated)  energy at the end of the exponential acceleration regime due to reconnection (or magnetic curvature in the case of the snapshot $t= 25\,L/c$, top panel), they continue to accelerate at a smaller rate  due to normal drift (as remarked in Section \ref{sec:magrec}). 

We should also remember that the maximum energy achievable by the stochastic  mechanism (at the saturation of the fast acceleration growth) occurs around $t \sim 10^3 $ hr for all the models depicted (see Figures \ref{t44o}, and \ref{t25} ), except in the top one, for which this occurs around $t \sim 10^{3.5} $ hr (see Figure \ref{t25}).
Interestingly, we see that for this model ($ut25o$),  and also model $t30o$ (second panel from top), around these times, there is a double hump in the distribution (green dot-dashed curve), with an accumulation of particles at energies above $10^7\,m_p c^2$  for both, thus  highlighting the transition from the exponential to the linear drift acceleration regimes. In the other models depicted, this transition is more smooth.
Also notable, is the double peak in the distribution that appears in model $t30o$ in the earlier times at $t\sim 10^2$ to  $10^{2.3}$ hr. This is possibly connected to the superposition of the two acceleration processes in this model, namely the magnetic curvature drift and the reconnection acceleration, as we discussed in section \ref{sec:t25}.


Perhaps, the most striking feature in all the diagrams of Figure \ref{NE} is that 
as particles reach very high energies, the distribution may even attain an almost zero power-law index tail in very evolved times, as we clearly see in the bottom diagram of the Figure.
In real systems, however, this acceleration process should be interrupted by the escape of the particles from the finite volume of the acceleration zone and also due to radiative losses.
As stressed above, since in our simulations the particles are continuously accelerated and there is no physical mechanism to allow them to escape, it is not possible to obtain the actual distribution of the accelerated particles. However, we can at least estimate the power-law index of the distribution soon after the particles start to populate the high-energy tail \citep[see e.g.,][]{delvalle_etal_16}. 

\begin{figure} %
\centering
\includegraphics[scale=0.5]{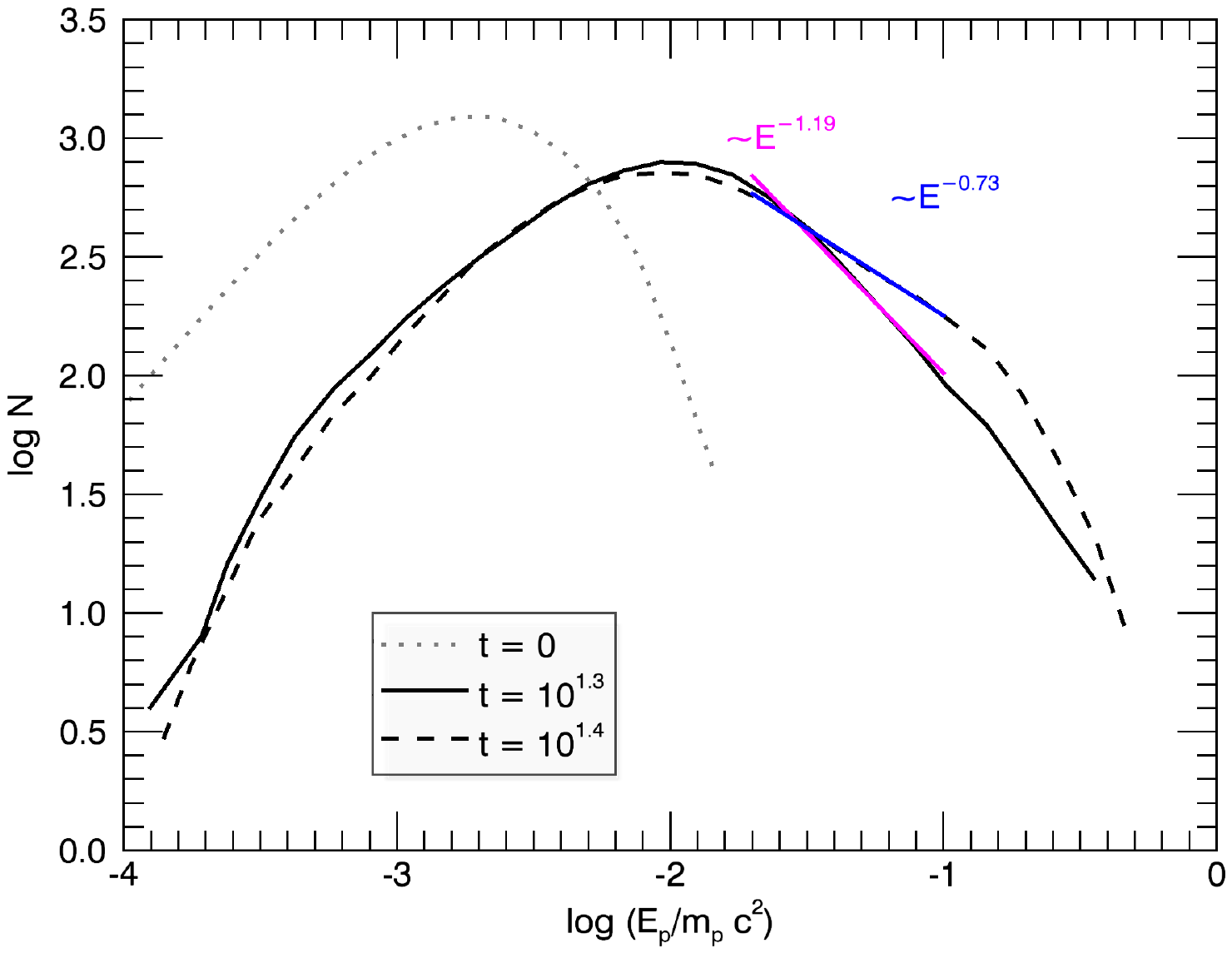}
\caption{
Particle energy spectrum  as a function of the normalized kinetic energy at two different early time steps of the acceleration (in hours)  for  the jet snapshot $t=50\,L/c$. The dotted gray line is  the initial Maxwellian  distribution. The high energy tail of each distribution is  fitted by a power-law.
}\label{NE50}
\end{figure}

In Figure \ref{NE50} we show the total number of particles as a function of energy for two different early  timesteps of the acceleration, for the jet snapshot $t=50$ $L/c$. The initial Maxwellian (normal) distribution 
is shown in gray dotted line. The earliest time step plotted corresponds to the approximate time when a high-energy power-law tail starts to form (i.e., when particles reach kinetic energies larger than $\sim 10^{-1}\,m_p c^2$, according to Figure \ref{t44o}, bottom panel); the second time corresponds to a little later time step. 
We see that the  power-law index at the earlier time can be fitted by $p= -1.19$. The second power-law at later time is flatter due to the effects discussed above and therefore, it must be taken only as illustrative of the limitations of the method. Of course, in realistic systems, the presence of physical particle escape from the acceleration zone, radiative losses and dynamical feedback of the accelerated particles into the plasma will result in steeper spectrum in the late times too ($|p|>1$).

\subsubsection{Boundary effects}\label{sec:bound}

\begin{figure} %
  \centering
   \begin{overpic}[scale=0.6]{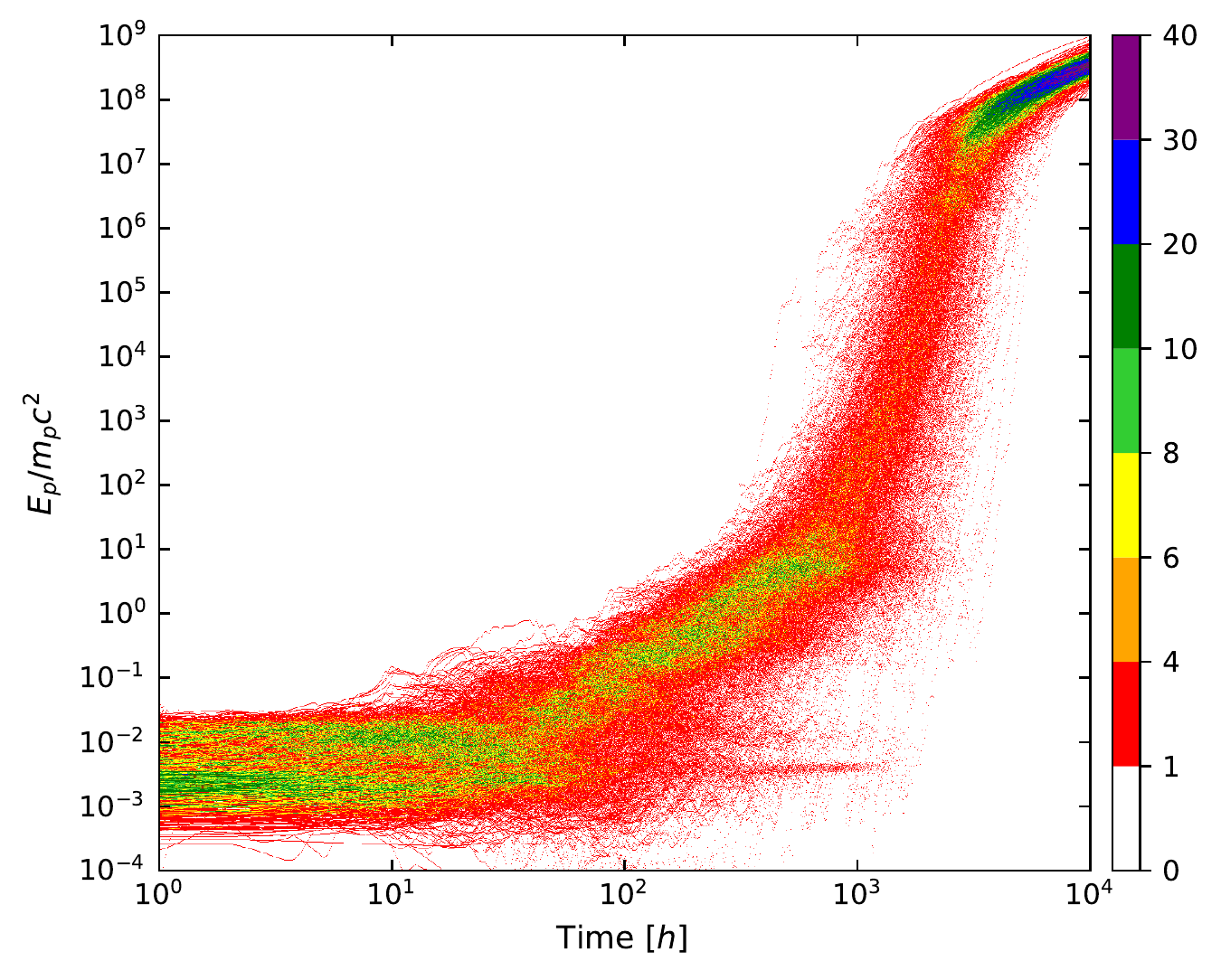}
     \put(14,46){\includegraphics[scale=0.21]{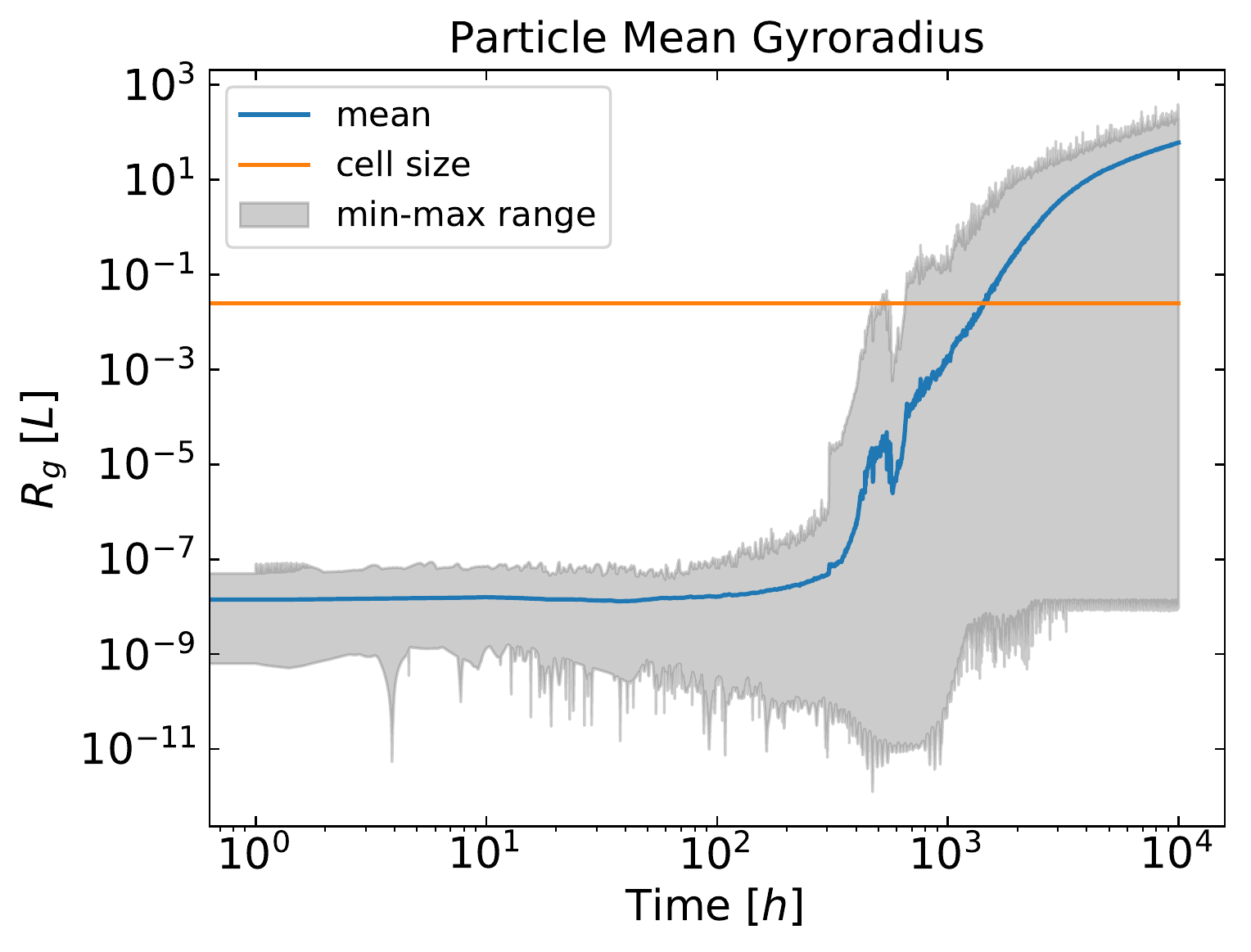}}
  \end{overpic}  
\caption{
Particle kinetic energy evolution, normalized by the rest mass energy, for  particles injected in $t=50\,L/c$ snapshot of the jet model ($j240$, see Table \ref{tablep}). This test is similar to that of the bottom diagram of Figure \ref{t44o}, except that here particles were periodically re-injected through all the boundaries of the jet system (see  model $t50p$ in Table \ref{tablep}). The color bar indicates the number of particles. The small plot on the upper left shows the evolution of the particles gyro-radius.
}
\label{t50p}
\end{figure}

As described in section \ref{sec:setuptp}, 
in most of the particle runs, we have allowed the particles to re-enter the system only through the jet periodic boundaries, along the $z$ direction. Nevertheless, we have also performed a few tests where we allowed the particles to be re-injected into the system through all the boundaries, i.e., also when crossing the jet outflow boundaries in the $x$ and $y$ directions, aiming at increasing the number of accelerated particles. In Table \ref{tablep}, these few tests are labeled with ``p". Figure \ref{t50p} shows one of these tests performed for the jet model $j240$ in the snapshot $t=$ 50 $L/c$ (model $t50p$), for which 1,000 particles were initially injected. It can be compared with its counterpart model shown in the bottom panel of Figure \ref{t44o}, in which 10,000 particles (rather than 1,000) were injected and allowed to re-enter the system only in the $z$ direction (test particle model $t50o$, Table \ref{tablep}). We note that both models have very similar behaviour, except for the amount of particles that are being accelerated along the system evolution. While in model $t50o$ (bottom panel of Figure \ref{t44o}), there are more particles in the beginning of the evolution, due to the much larger number of injected particles, in model $t50p$ (Figure \ref{t50p}), we see a larger number of particles that are accelerated up to the maximum energy at the exponential regime and beyond, due to the larger number of re-injected particles in the periodic boundaries in all directions.  
We also see in Figure \ref{rateacc} that both models have similar acceleration properties, i.e., acceleration rate, power-law index $\alpha$, and kinetic energy growth rate.


\subsubsection{Magnetic field effects}\label{sec:magfield}




\begin{figure} 
\centering
\begin{overpic}[scale=0.6]{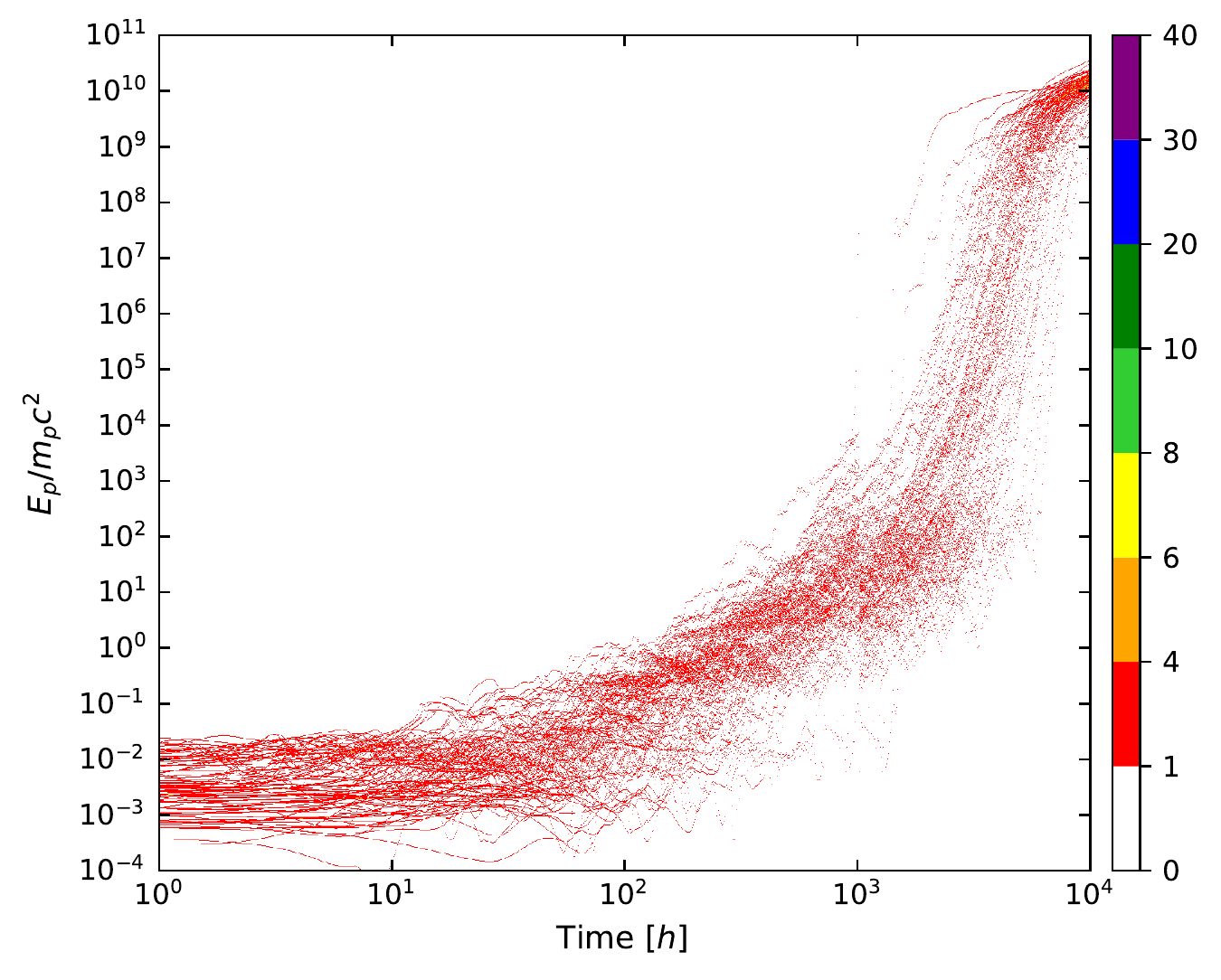}
 \put(14,46){\includegraphics[scale=0.21]{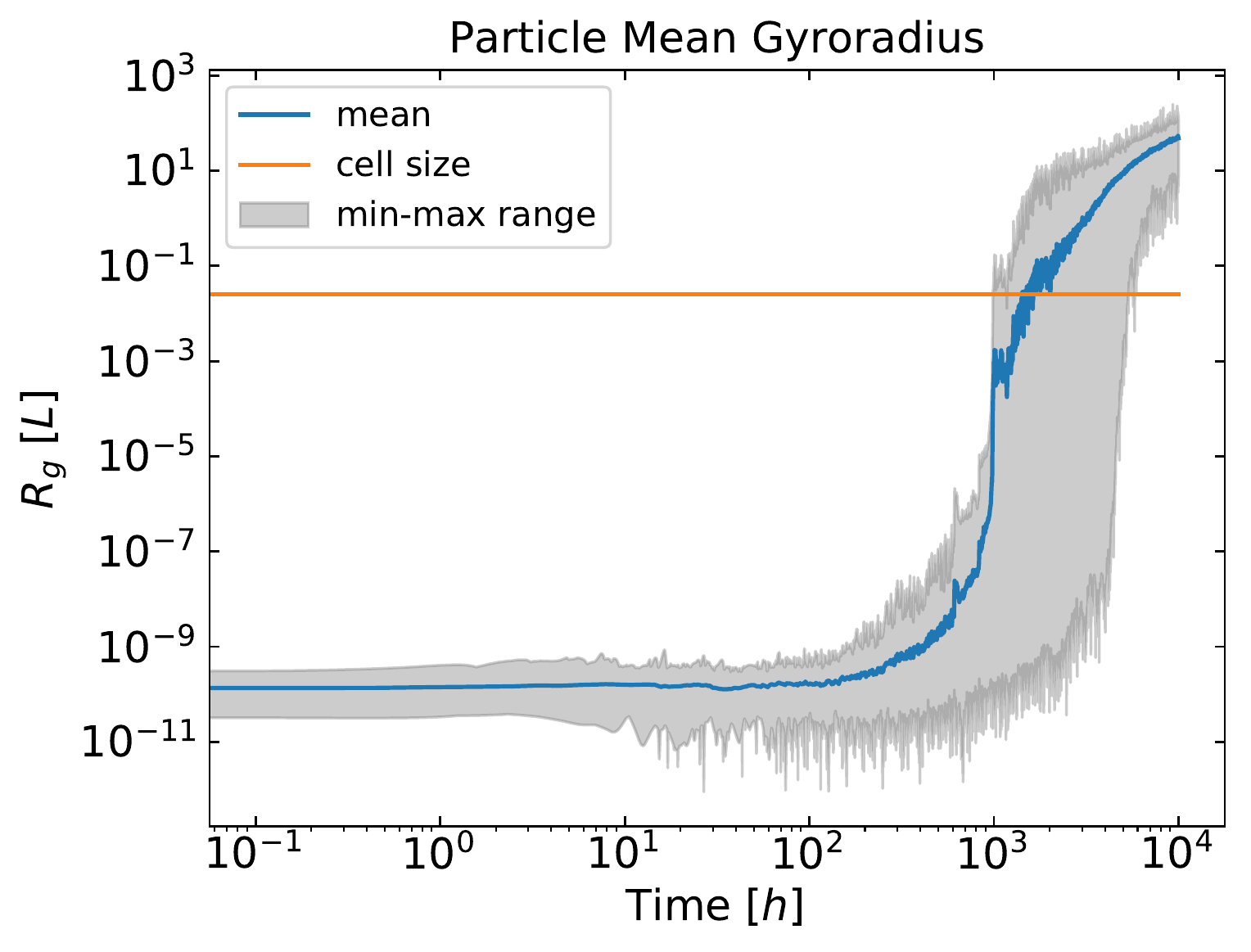}}
 \end{overpic}
\caption{
Kinetic energy evolution for particles injected at snapshot $t = 50\,L/c$ of the jet model $j240$ with an initial  background magnetic field at jet axis $B_0 =9.4$ G and a background density $\rho_0=10^4\,\mathrm{cm}^{-3}$  (see Table \ref{tablep}). The initial conditions are the same as in the test particle model ($t50o$) shown in the bottom panel of Figure \ref{t44o}, except that there $B_0 = 0.094\,\mathrm{G}$ and $\rho_0= 1\,\mathrm{cm}^{-3}$, leaving unaltered the Alfv\'en velocity in both tests (see text for details). The color bar indicates the number of particles and the small plot in the detail shows the evolution of the particles gyro-radius.}
\label{t50_B}
\end{figure}

As in Figure \ref{t44o} (bottom diagram, test particle model $t50o$), Figure \ref{t50_B} also shows the kinetic energy evolution for particles injected at snapshot $t = 50$ $L/c$ of the jet model $j240$, except that now the initial background magnetic field at the jet axis is 100 times larger ($B_0 = 9.4 \mathrm{G}$), corresponding to the test particle model $9t50p$ of Table \ref{tablep}.\footnote{We note that this test particle model $9t50p$ was run with periodic boundaries in all directions, while the counterpart model used for comparison with it, $t50o$ (Figure \ref{t50_B}), has periodic boundaries only in the $z$ direction. However, as we have seen in section \ref{sec:bound}, the employment of periodic boundaries in all directions produces results very similar to the corresponding model with periodic boundaries only in the z direction.} 
This change in the physical unit value of the magnetic field of the background jet was made in such a way that we have kept the scale invariance in the jet system. In other words, as stressed in section \ref{sec:numethod}, the magnetic field code unit in the RMHD jet simulation is given by  $\sqrt{4 \pi \rho_0 c^2}$. Thus, when increasing the physical unit of the magnetic field by a factor of 100, we had also to increase the background density physical unit by a factor $10^4$, in order to keep unaltered the magnetic field in code unit and thus, the corresponding Alfv\'en speed. This means that in this test particle model with larger physical magnetic fields and densities ($9t50$), the background reconnection velocities have also maintained the same as in model $t50o$.
Therefore, particles in this new model ($9t50$) should feel essentially the same acceleration rates of the counterpart test  model with smaller jet magnetic field and density (model $t50o$). This is what we see comparing these models in the diagrams of Figure \ref{rateacc}.

On the other hand, a closer view of Figure   \ref{t50_B} shows that the increase of the background magnetic field by a factor one hundred, causes an increase of the maximum energy achieved by the particles also by two orders of magnitude with respect to the counterpart model $t50o$ (see Figure \ref{t44o}, bottom panel). In other words, while the maximum energy that particles attain in the exponential regime in model t50o is $\sim 10^7 m_p c^2$, in model $9t50$, it is  $\sim 10^{9} m_p c^2$ (see also Figure \ref{rateacc}). 
The plot in the detail of Figure \ref{t50_B} shows that the corresponding maximum Larmor radius ($\propto{E_p/B}$) is the same as in model $t50o$, this because in both test particle models, the ratio of the maximum energy to the background magnetic field is the same, or in other words, the physical size of the acceleration region is the same in both cases.





~\\

\section{Discussion and Conclusions}\label{sec:discut}

In this work we have investigated 
the  acceleration of  particles injected  in several snapshots of a 3D Poynting flux dominated  jet with moderate magnetization ($\sigma \sim$ 1), subject to current driven kink instability (CDK) which drives turbulence and fast magnetic reconnection. Our results can be summarized as follows:

\begin{itemize}
 \item    
Once turbulence driven by the CDK instability is fully developed in the jet, achieving a nearly steady-state, the amplitude of the excited wiggles along the  jet spine also attains a maximum growth and gets disruptive with the formation fast magnetic reconnection in several sites. This occurs after the jet snapshot $t\sim 40 \,L/c$, when the CDK instability achieves a plateau.  
Injecting hundreds to thousands of protons in jet snapshots more evolved than this one, we find that, after about $10^2$ hr, the particles undergo an exponential acceleration up to a maximum energy. For a  background magnetic field around $B\sim 0.1$  G, this saturated kinetic energy is $\sim 10^7 m_p c^2$, or $\sim 10^{10}$ MeV (Figure \ref{t44o}), while for a magnetic strength one hundred times larger, $B\sim 1$ G, the maximum accelerated energy increases also by a factor one hundred, to  $\sim 10^{9} m_p c^2$, or $\sim 10^{12}$ MeV (Figure \ref{t50_B}). Beyond these values, the particles suffer further acceleration (to energies up to 100 times larger), but at a much slower rate due to drift in the varying magnetic field.  

\item
Particles achieving the saturation energy in the exponential regime of acceleration  attain a Larmor radius comparable to the size of the acceleration region of $\sim 4L$, which is of the order of the diameter of the perturbed wiggled jet. This regime of particle acceleration is very similar in all these evolved snapshots and lasts for several hundred hours until the saturation energy.

\item
In a companion work (\citetalias{kadowaki_etal_2020}), we have quantitatively identified the sites of reconnection over the entire jet and here we could correlate them with the accelerated particles (Figures \ref{posHisto} and \ref{pos3DHisto}). The results show a clear association with the regions of maximum current density and fast reconnection sites, indicating that particles are being mostly accelerated by magnetic reconnection, as detected in previous studies 
\citep[e.g.,][]{kowal_etal_2011,kowal_etal_2012,delvalle_etal_16}. 
The exponential acceleration in these sites is suggestive of a Fermi stochastic process \citep[e.g.,][]{dalpino_lazarian_2005, kowal_etal_2011, kowal_etal_2012, dalpino_kowal_15,matthews_etal_2020}. Furthermore, during the exponential  regime, we have found a predominance of acceleration of the parallel component of the particles velocity to the local magnetic field, which is characteristic of acceleration in reconnection domains. 

\item 
In the early stages of the development of the non-linear growth of the CDK instability, before this achieves the plateau and  the magnetic field lines start disruption, the jet spine oscillates
with growing amplitude. 
We find that during this early stage (jet snapshot $t= 25\,L/c$), there are no sites of fast reconnection, but the test particles are efficiently accelerated by  magnetic curvature drift, with a dominance of acceleration of the perpendicular component of the particles velocity to the local magnetic field, similarly as detected in the PIC simulations of \cite{alves_etal_2018}. However, in order to the particles to get accelerated by this process, they had to be injected with an initial energy much larger than that required for the particles to accelerate in the reconnection sites of the jet, in later snapshots.
While in the more evolved snapshots particles can be injected with energies 10$^{-3}$ $m_p c^2$ or less, in this early snapshot they have to be injected at least  with 10 $m_p c^2$ to be accelerated (four orders of magnitude larger).
This suggests that this mechanism requires pre-accelerated particles in order to work efficiently. This was confirmed by another test where we injected  particles in a later jet snapshot, $t= 30\,L/c$, where the wiggling amplitude of the magnetic field in the jet spine was still coherent, but a few sites of fast reconnection had already developed and, in such background conditions we find that the particles undergo an efficient acceleration starting with an injection energy of only 
$10^{-3} m_p c^2$. This occurs because in this case, particles are being accelerated from the beginning in the reconnection sites, and then further accelerated in the wiggling spine by curvature drift (Figures \ref{t25} and \ref{t30_240_1000_o}). 

 \item 
The acceleration time due to magnetic reconnection indicates a weak power law dependence with the particle energy given by $t_A \propto E^{\alpha}$, with $\alpha \sim $ 0.1,  obtained from all test particle models.


\item 
The energy spectrum of the accelerated particles develops a high energy tail that can be fitted by a power law index $p \sim$ -1.2 in the beginning of the acceleration,  which does not depend on the initial energy  of the injected particles, at least in the cases of the acceleration by magnetic reconnection.

\item
Particles injected in the background jet, assuming periodic conditions in all boundaries, or periodic boundaries only in the $z$ direction (along the jet axis) and outflow boundaries in the transverse direction, produce similar results. The only  remarkable difference is that the adoption of periodic boundaries in all directions allows for particles to re-enter the system more frequently thus increasing the number of accelerated particles. 

\end{itemize}{} 

The results above have important implications for particle acceleration and the associated non-thermal emission in relativistic jets, specially in their magnetically dominated regions.    
Though we have not taken into account particle losses, such as non-thermal radiation, electron-positron pair production, particle back reaction into the jet plasma, or particle diffusion, the  energies achieved by the particles, $\sim 10^{16}$  eV or $\sim 10^{18}$ eV (or even larger), depending on the strength of the background magnetic fields $\sim 0.1$ G or $\sim 10$ G,  are more than sufficient to explain energetic particles and even ultra-high-energy-cosmic-rays (UHECRs) in these sources. Protons with these energies could explain observed very high energy (VHE) emission, as well as the production of  neutrinos, out of interactions with the ambient photon and density fields in relativistic jets. This could be the case, for instance, of blazars like TXS 0506 +056 \citep{aartsen_etal_2018}, for which for the first time it has been observed  simultaneous TeV gamma-rays and neutrino emission. Although there might be other  possibilities \citep[see e.g.,][]{cerruti_2020}, this process should be explored in detail elsewhere. In our companion work, \citetalias{kadowaki_etal_2020}, we have applied  the results of  this study of magnetic reconnection in jets to the VHE light-curve of the blazar MRK421 \citep[e.g.,][]{kushwaha_etal_17} and found that the magnetic power and variability obtained from the reconnection events are compatible with the observed  emission.

Our results are comparable to those obtained from test particles injected in single non-relativistic current sheets in which forced turbulence was introduced to make reconnection fast \citep[e.g.,][]{kowal_etal_2011,kowal_etal_2012,delvalle_etal_16}. \citet{delvalle_etal_16}, for instance, have obtained an acceleration time with a similar weak energy dependence, with a power law index $\alpha \sim 0.2-0.6$ for a vast range of reconnection velocities. The slightly smaller values of $\alpha$ we obtained in this work are consistent with the fact that the jet has relativistic Alfv\'en velocities and thus intrinsically higher reconnection speeds $V_{rec} \simeq 0.05 V_A$ that  naturally make the process slightly more efficient. In \citet{delvalle_etal_16}, the particle spectrum power-law indices derived in the beginning of the acceleration process, are also compatible with our results. Moreover, our power-law indices are  remarkably similar to those obtained from  PIC simulations of single current sheets, in the kinetic scales of the plasma \citep[e.g.,][]{zenitani_H_2001,drake_etal_2013,sironi_spitkovsky_2014,guo_etal_2014,guo_etal_2015,li_etal_2015,werner_etal_2018}. 


These results are also consistent with the theoretical models of the Fermi acceleration process in  reconnection sites \citep[e.g.,][]{dalpino_lazarian_2005,drury2012,dalpino_kowal_15},  which predict an acceleration time similar to that of stochastic shock acceleration and approximately independent of the reconnection velocity.
Similarly as in these earlier studies,  particles achieve the maximum energy when their Larmor radius becomes comparable to the size of the acceleration zone. 

It is important to remark that in this work we have neglected potential effects of the dynamical variations in the background plasma on the acceleration of the particles. More specifically, we have neglected the betatron effect \citep[see e.g., ][]{kowal_etal_2012,dalpino_kowal_15}. 
When performing test particle in an MHD background system, in order to follow the particles interactions with it, we have to transform the background code units to physical units. As we have normalized the particles time unit in hours, this means that, for instance, from the background snapshot $t=40$ $L/c$ to $t =50$ $L/c$, 10 hours have elapsed.
Now, particles interact resonantly with the magnetic background fluctuations (according to eq. \ref{eqmov1}) when their Larmor radii are comparable to the wavelength of these fluctuations. When the turbulence in the jet attains a nearly steady state regime (beyond $t= 40$ $L/c$ in the jet), the background dynamical variations become statistically negligible. In other words, the particles  face a very similar background spectrum of fluctuations in the reconnection regions spread over the distorted magnetic field spine and outside them, in every snapshot of this nearly stationary regime. This is consistent with  results we obtain for the particle acceleration properties for the evolved snapshots $t = 46$ and $50$ $L/c$, which are very similar (see Figures \ref{t44o} and \ref{rateacc}). 
Earlier studies of  the betatron effect,  performing test particle simulations considering the background dynamical variations among the snapshots have shown that, in the case of pure turbulent environments (i.e., where the reconnection acceleration favors more a 2nd-order Fermi  than a 1st-order, and thus this effect is more important), it introduces only a factor 2 difference in the acceleration rate \citep[e.g.][and references therein]{dalpino_kowal_15}. 
Since particles attain energies which are more than 8 orders of magnitude larger than the energy at injection, this effect is almost negligible. It may have some impact (of a factor 2) only in the beginning of the acceleration when particles are still growing their energy linearly, essentially by drift acceleration. We have plans to perform studies of the particle back-reaction in the CDK unstable jet by  incorporating particles directly into the relativistic MHD code. This will  allow both, to compute the particles back-reaction in the fluid and perform a more accurate estimation of the betatron effect.
With regard to the earlier snapshots ($t= 25$, $30$  and $40$ $L/c$), when the CDK instability and the turbulence are still growing, and the system has not achieved the nearly-steady state yet, one may argue that the background dynamical effects might have implications on the stochastic reconnection acceleration evolution in these cases. However, at $t=25$ $L/c$, the system has no fast reconnection current sheets yet, and the particles experience only magnetic curvature drift acceleration, as discussed in Section \ref{sec:t25}.  Of course, this ideal situation may not sustain for long in the real jet since the CDK instability grows fast, but it illustrates what may happen with particles in the early stages of the growth of the instability and also serves to compare with other (PIC) works \citep[e.g.][see below]{alves_etal_2018}. The snapshot $30$ $L/c$, on the other hand, has already a few fast reconnection sites and, as discussed in Section \ref{sec:t25},  it combines the two processes of acceleration. The inclusion of dynamical effects of the background in this case would not affect much  the results since again we note that the results for particle acceleration are very similar to the more evolved snapshots (see Figures \ref{t30_240_1000_o}  and \ref{rateacc}). Finally, with regard to $t=40$ $L/c$, this snapshot which is  just before the  CDK  turbulence reaches the nearly steady-state regime, we see that though the exponential growth of the particles kinetic energy is similar to the evolved nearly stationary snapshots, the number of  particles that achieve the maximum energy is much smaller (Figure \ref{t44o}, top), which may be  reflecting the (minor) effects of neglecting the plasma background evolution in this case.

Other interesting implications arise from the conversion of the simulations   into physical units. 
The results from Figures \ref{t44o}, \ref{t50_240&480}, \ref{rateacc}, \ref{t50p} and \ref{t50_B} 
imply  a total acceleration  time (including  the  exponential acceleration regime by reconnection plus the slower final drift acceleration) of a few  $\sim 1000$  hr. During this time, particles have re-entered the system across the periodic boundaries in the longitudinal  direction (z) several times, traveling  a total length  of the order of   $\sim  10^{-1}$ pc along the jet.
If we consider only the time elapsed during the  acceleration exponential regime, the length scales are even smaller ($\sim  10^{-2}$ pc). 
 These  physical length scales
characterize the size of  the turbulent induced reconnection dissipation  region where particles are accelerated and, within these (time and length)  scales, the physical conditions in a real system are not expected to change substantially, except for the dissipation of the magnetic energy. It is interesting  to note that these scales are also compatible with  estimates of the size of the reconnection dissipation layer considered in recent blazar jet studies 
\citep[see e.g.][]{christie_etal_19,giannios2019}.

Besides the works above, other recent studies have also explored numerically acceleration by reconnection in relativistic jets, but considering PIC simulations \citep[e.g.,][]{nishikawa_etal_2020,davelaar_etal_2020}.  \citet{christie_etal_19} scaled the results of 2D PIC simulations of current sheets with the formation of plasmoids  (or magnetic islands) of different sizes, to the scales of  relativistic jets. Coupling these plasmoid simulations with a radiative transfer code, they reproduced light curves of blazar sources, showing the efficiency of reconnection acceleration to explain multi-scale variability in blazars across the entire electromagnetic spectrum. \citet{nishikawa_etal_2020}, on the other hand, performed 3D PIC simulations of a magnetically dominated relativistic  jet of electron-proton pairs, accounting for several  mechanisms driving  turbulence inside the jet and also found that magnetic reconnection should  be the dominant acceleration process. \citet{davelaar_etal_2020} achieved the same conclusion. However, none of these studies involving global jet simulations have derived the  properties  of the reconnection sites and their correlation with the accelerated particles, as in this work. Moreover, neither of them obtained the general properties of particle acceleration, like the acceleration rate, the size of the acceleration region and the saturation energy achieved by the particles. With regard to the spectrum, \citet{davelaar_etal_2020} obtained a much steeper power-law index than in this work, probably due to the higher magnetization of their jet model, and \citet{nishikawa_etal_2020} study did not obtain any power-law index due to the limited resolution. 

It is remarkable that 
 in our work, though we have considered a mild magnetization parameter  ($\sim1$ in the evolved jet),
 the results are very consistent even  with PIC studies with much higher magnetization
\citep[e.g.][]{comisso18, comisso19}. 
  For instance,
 the hardness of the power spectrum of the particles in our simulations, in spite of the intrinsic limitations of the test particle method, is similar to these studies.
In contrast, as remarked above, \cite{davelaar_etal_2020} 
 obtained much  steeper power spectrum (the larger the magnetization parameter the steeper the slope), which they interpreted   as due to an inhibition of the acceleration by the strong guide field of the plasma. 
\cite{comisso18, comisso19}, on the other hand,    have interpreted their  very hard spectrum (which improves with  increasing magnetization) as due to the high amplitude of the turbulent fluctuations that are accelerating the particles (though turbulence is decaying in their simulations) and in this way the underlying magnetic field is not sufficient to kill the stochastic process. Therefore, it is possible that one may obtain similar results when considering MHD relativistic jets with higher magnetization, as long as the amplitude of the magnetic fluctuations of the driving turbulence is maintained large enough \citep[i.e. $\delta B/B \sim 1$, as e.g. in][]{comisso18, comisso19}. This obviously requires further investigation since a critical difference of our study and that of \citet{comisso18, comisso19} is the absence in their case of an underlying strong large scale magnetic field, as we have in the jet \citep[see also][]{kowal_etal_2012,dalpino_kowal_15}.

As mentioned before, another recent study also explored particle acceleration in relativistic jets subject to the CDK instability by means of 3D PIC simulations of  electron-positron pairs \citep{alves_etal_2018}. These authors examined the early non-linear development of this instability and identified an acceleration of the particles due to curvature drift in  the  wiggling magnetic field structure of  growing amplitude  along the jet spine. 
According to their results, a  maximum  energy growth  rate  for the electrons  
$(\Delta E/\Delta t)/(m_e c ^2) \sim$ 12 $c/R $ is achieved at the maximum   energy to which the particles are accelerated,  $\sim$ 125 $m_e c^2$, where $R$ is the jet radius and the energy is normalized by the electron rest mass energy \citep[see Figure 3(d) in][]{alves_etal_2018}. This implies an acceleration time for the electrons,
$t_{acc,e} \simeq 10.4\,R/c$.
Considering our test particle model in the earlier jet snapshot where we also identified curvature drift acceleration, i.e.,  $t = 25\,L/c$ (model $ut25$), the Figure \ref{rateacc} (bottom diagram)  gives  for this model
a maximum energy growth rate for the protons
$(\Delta E/\Delta t)/(m_p c ^2) \sim 10^5$ hr$^{-1}$, which is achieved at the saturation energy $\sim 5 \times 10^6\,m_p c^2$. This implies an acceleration time for the  protons $t_{acc,p} \simeq 50$ hr.
In order to compare both rates, we need to estimate what would be the acceleration time  for the electrons in our simulation. 
 In the relativistic regime, the acceleration time for electrons is approximately given by   
$ t_{acc,e} \simeq t_{acc,p} ({m_e}/{m_p}) $ \citep[e.g.,][]{khiali_etal_15}. Thus, from our results we might expect an electron acceleration time up to the saturation energy due to curvature drift of the order of $t_{acc,e} \simeq 100$ s, which is comparable to the one obtained in \citet{alves_etal_2018} if one considers a jet radius $R \sim$ 10$^{-7}$ pc in their PIC simulation. However, this is only a rough estimate, since the scales implied in the two simulations are rather distinct.\footnote{
As remarked, considering the conversion in physical units adopted in our simulations where the time unit is $1$ hr, this implies a physical unit length in our jet  $L= c \, t \sim 10^{-4}$ pc.}
Moreover, in order to improve our estimate for  electrons we should perform also numerical simulations  for them. However,  
the numerical integration of the electron trajectories is much longer than for protons in MHD domains and computationally  expensive.

Finally, we should remark that $in$ $situ$  particle acceleration examined directly in real systems, like the 3D relativistic jets we considered here, is a very promising approach because it allows for testing the process under more realistic environmental  conditions, with turbulence and reconnection driven by natural physical processes, and allowing for direct applications to observed systems, and even including time dependence effects. The comparisons with former studies in single current sheets have further validated the process, confirming its ubiquitous nature. 
The similarity of our results with these earlier works which are based on local (higher resolution) simulations of current sheets (both in PIC and MHD+test particle simulations) have served not only to benchmark and validate our results, but also to highlight the stochastic and universal nature of the process across the scales. Furthermore, this method of injecting particles directly in  the collisional MHD  simulation of the relativistic jet, have allowed the  acceleration of the particles up to the real physical scales where they are accelerated, up to the observed ultra high energy values, without the need to extrapolate the much smaller energies that are achieved in the kinetic scales, to the large scales,  as it is required in PIC simulations. 
In forthcoming work we intend to include the radiative losses of the particles and apply to observed systems.
Moreover, since our MHD collisional approach
has limitations as only injected particles with Larmor radius close to the MHD scales can be effectively accelerated, and they allow  only for modest values of the magnetization parameter, future studies involving hybrid simulations combining PIC and MHD approaches, like those performed for single current sheets \citep[e.g.,][]{bai_etal_2015}, should be applied also to real systems, probing both the kinetic and the macroscopic scales of the process, and also accounting for the particles feedback in the system.



\acknowledgments
 TEMT acknowledges support from the Brazilian Agency CAPES, EMdGDP  from the Brazilian Funding Agencies FAPESP (grant 13/10559-5) and CNPq (grant 308643/2017-8), LK  from FAPESP (2016/12320-8), and G.K. from CNPq 
(grant 304891/2016-9) and FAPESP (grants 2013/10559-5 and 2019/03301-8). C.B.S. is supported by the National Natural Science Foundation of China under grant no. 12073021.
Y.M. is supported by the ERC Synergy Grant ``BlackHoleCam: Imaging the Event Horizon of Black Holes'' (Grant No. 610058). 
The  simulations presented in this work were performed in the cluster of the Group of Plasmas and High-Energy Astrophysics (GAPAE), acquired with support from  FAPESP (grant 2013/10559-5), in the Blue Gene/Q supercomputer supported by the Center for Research Computing (Rice University) and Superintend\^{e}ncia de Tecnologia da Informa\c{c}\~{a}o da Universidade de S\~{a}o Paulo (USP), and in the computing facilities of the Laboratory of Astroinformatics (IAG/USP, NAT/Unicsul), whose purchase was also made possible by FAPESP (grant 2009/54006-4) and the INCT-A.
The authors are also thankful to an anonymous referee whose comments have helped to improve the paper.

\bibliography{bibliography.bib}



\end{document}